\documentclass[12pt]{article}

\usepackage[authoryear]{natbib}
\usepackage{graphicx}
\usepackage{amssymb}
\usepackage{amsmath}
\usepackage{amsthm}
\usepackage{times}
\usepackage{bm, color}
\usepackage{subfigure}
\usepackage{multirow}
\usepackage{url}

\newcommand{\dd}{{\rm d}}
\newcommand{\Reais}{\mathbb{R}}
\newcommand{\diag}{\text{diag}}

\newcommand{\sinal}{\text{sign}}
\newcommand{\logit}{\text{logit}}
\newcommand{\norm}[1]{\left\lVert#1\right\rVert}

\newtheorem{definicao}{Definition} 

\title{Power logit regression for modeling bounded data}
\author{Francisco F.~Queiroz, Silvia L. P. Ferrari \footnote{Corresponding author: Silvia L.P. Ferrari, email silviaferrari@usp.br.}\\
{\small {\em Department of Statistics, University of S\~ao Paulo, Brazil}}}
\date{}

\begin{document}
\maketitle

\begin{abstract}

The main purpose of this paper is to introduce a new class of regression models for bounded continuous data, commonly encountered in applied research. The models, named the power logit regression models, assume that the response variable follows a distribution in a wide, flexible class of distributions with three parameters, namely the median, a dispersion parameter and a skewness parameter. The paper offers a comprehensive set of tools for likelihood inference and diagnostic analysis, and introduces the new R package PLreg. Applications with real and simulated data show the merits of the proposed models, the statistical tools, and the computational package.

\noindent {\it Keywords:} Continuous proportions; Fractional data; Power logit distributions.  
\end{abstract}

\section{Introduction}\label{Intro}

Bounded continuous data, particularly on the unit interval, appear in various research areas, including medicine, biology, sociology, psychology, economics, among many others. Some examples are the proportion of family income spent on health plans, mortality rate, percentage of body fat, fraction of the territory covered by treetops, loss given default, and efficiency scores calculated from data envelopment analysis (DEA). It is usually of interest to predict or explain the behavior of a continuous proportion from a set of other variables. Linear regression models may not be appropriate, as they may yield fitted values for the response variable that exceed its lower and upper bounds, below 0 and above 1. 
A natural strategy is to define a regression model where the dependent variable ($Y$) has a probability distribution function on the unit interval.

Different regression models for modeling data that are restricted to the interval $(0, 1)$ have been proposed. Beta regression models \citep{FerrariCribariNeto2004, Smithsonverk2006}, are widely used. Beta regression allows direct parameter interpretation, asymmetry and heteroscedasticity, while being reasonably flexible and having software available \citep{CribariNetoZeileis2010}. Several research papers have appeared in recent years regarding beta regression models for specific situations, such as errors-in-variables \citep{carrasco2014}, longitudinal data \citep{Wang2014, Brisco2020}, high-dimensional data \citep{schmid2013, Weinhold2020}, and time series \citep{rocha2009, pumi2021}. Other distributions have been proposed to model variables with support on the unit interval, such as the rectangular beta distribution \citep{bayesbazan2012}, the log-Lindley distribution \citep{gomezSordo2014}, the GJS class of distributions \citep{LemonteBazan2016}, and the CDF-quantile distributions \citep{SmithsonShou2017}. Regression models that accommodate observations at the extremes are found in \cite{OspinaFerrari2012} and \cite{queiroz2021}, among others.

Inference in beta regression models is usually based on maximum likelihood or Bayesian methods, for which the information from the data comes from the likelihood function. In either case, the inference can be highly influenced by atypical observations. To deal with this drawback,  the inference procedure may be replaced by a robust method \citep{RibeiroFerrari2020}. Alternatively, one may employ models based on distributions that are more flexible than the beta distribution. In this direction, \citet{LemonteBazan2016} propose the class of the GJS regression models, which is a generalization of the Johnson SB models \citep{Johnson1949}. The GJS distribution is defined from the transformation $X = [t(Y) - t(\mu)]/\sigma$, where $t(y) = \log[y/(1-y)]$, $y \in (0,1)$, $\mu \in (0,1)$, $\sigma>0$, assuming that $X$ has a standard symmetric distribution. Thus, the class of GJS distributions is constructed from symmetric distributions assigned to the logit of the variable that has support on $(0,1)$. However, it is known that the logit transformation is not able to bring common distributions of continuous fractions to symmetry; we illustrate this in Section \ref{PLdist}.

In this work, we develop a very flexible class of regression models for continuous data on the unit interval, named the power logit models. These models employ a new class of distributions with three parameters that represent median, dispersion and skewness. This class of models has the GJS regression models as a particular case, with the advantage that the skewness parameter provides extra flexibility and accommodates highly skewed data. The regression parameters are interpretable in terms of the median and dispersion of the response variable. Since the interest lies in skewed data, the median may be a more appropriate measure of central tendency than the mean. We also define diagnostic and influence measures and present applications that show the usefulness of the proposed regression models in practice. Additionally, we develop a new \texttt{R} package, named \texttt{PLreg}, for fitting both power logit and GJS models.

The outline for the remainder of this paper is as follows. Section \ref{PLdist} introduces the power logit class of distributions and some of its properties. Section \ref{PLreg} defines the class of power logit regression models. Section \ref{diagnostictools} presents some diagnostic tools and influence measures. Section \ref{package} gives a brief presentation of the \texttt{PLreg} package. Section \ref{app} presents and discusses real data applications. The paper closes with some discussions and directions for future extensions.

\section{The power logit distributions} \label{PLdist}

We say that a random variable $W$ has a distribution in the symmetric class of distributions if its probability density function is  
$$g(w)=g(w;\mu, \sigma)=\frac{1}{\sigma} r \left( \frac{(w- \mu)^2}{\sigma^2} \right), \quad w \in \Reais,$$
in which $\mu$ is the location parameter, $\sigma>0$ is the scale parameter, and $r(z)>0$, for $z \geq 0$, with $\int_0^\infty z^{-1/2} r(z) dz=1$, is the density generator function \citep{FangAnderson1990}.
We write $W \sim \text{S}(\mu, \sigma; r)$. The class of the symmetric
distributions has a number of well-known distributions as special cases depending on the choice of $r$. It includes the normal distribution as well as the Student-t, power exponential, type I and type II logistic, symmetric hyperbolic, sinh-normal and slash distributions, among others. Densities in this family have quite different tail behaviors, and some of them may have heavier or lighter tails than the normal distribution.

We now define a new class of distributions with support on the unit interval.
\begin{definicao}[Power logit distributions] \label{def1} Let $Y$ be a continuous random variable with support $(0,1)$ and let
\begin{align}\label{transfZ}
Z = h(Y; \mu, \sigma, \lambda) = \dfrac{1}{\sigma} \left[ \log \left( \dfrac{Y^\lambda}{1-Y^\lambda} \right) - \log \left( \dfrac{\mu^\lambda}{1-\mu^\lambda} \right)\right],
\end{align}
where $0< \mu<1$, $\sigma>0$, and $\lambda>0$. If $Z$ has a standard symmetric distribution, that is, $Z \sim \text{S}(0, 1; r)$, we say that $Y$ has a power logit distribution (PL) with parameters $\mu$, $\sigma$ and $\lambda$, and density generator function $r$. We write $Y \sim \mbox{PL} (\mu, \sigma, \lambda; r)$.
\end{definicao}

\begin{figure}[!htb]
\centering
\subfigure[][]{\includegraphics[scale=0.41]{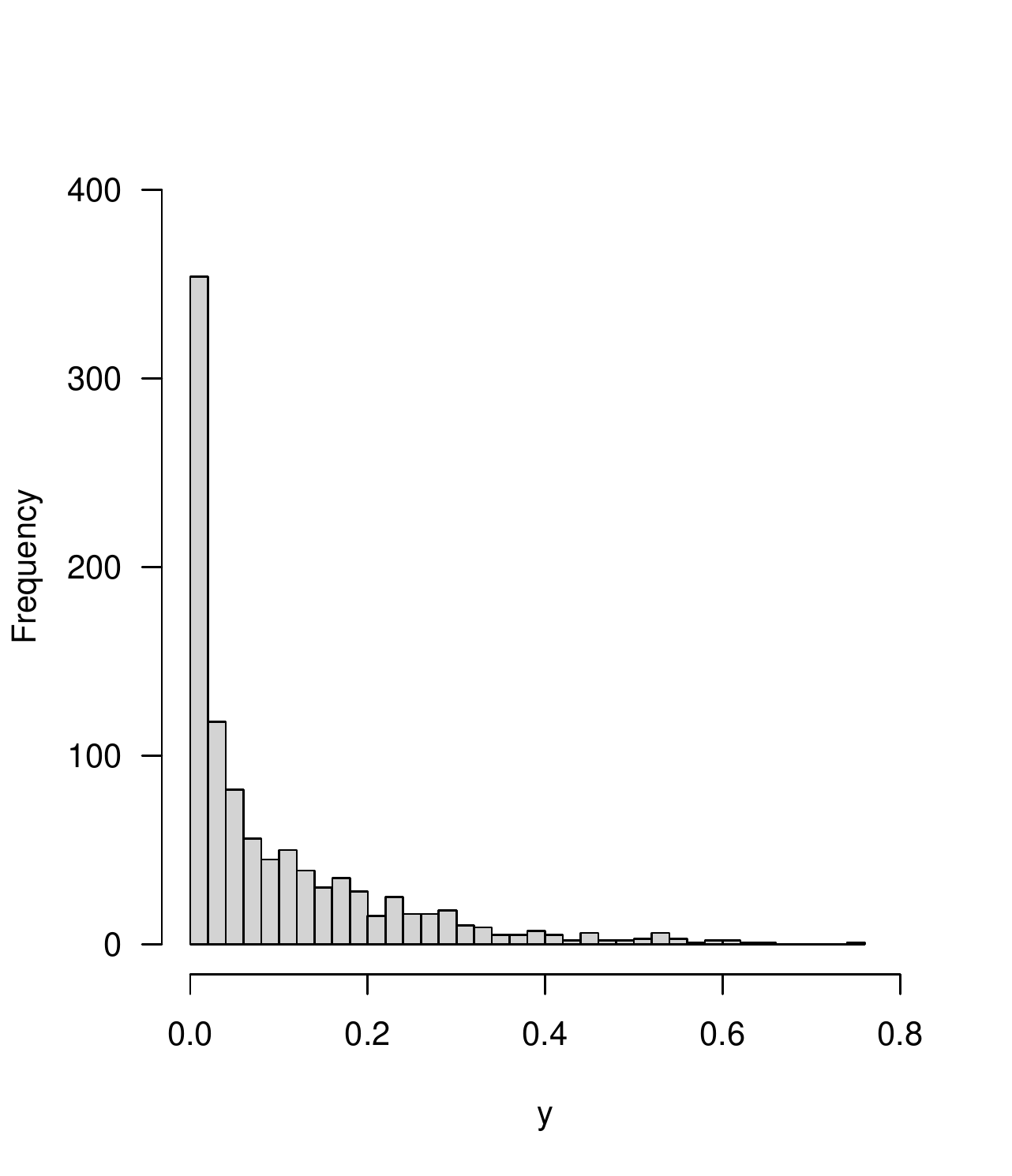}}
\subfigure[][]{\includegraphics[scale=0.41]{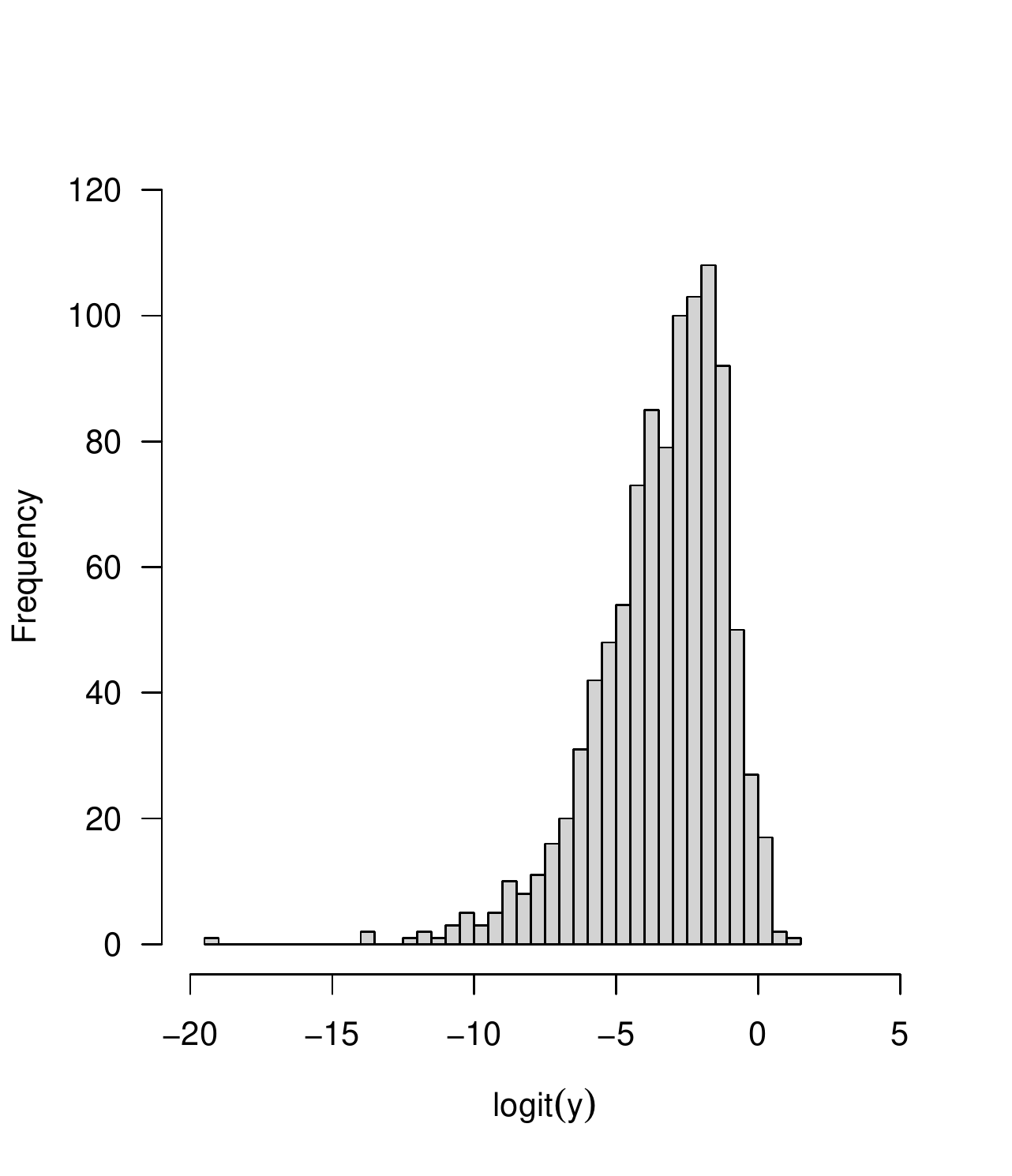}}
\subfigure[][]{\includegraphics[scale=0.41]{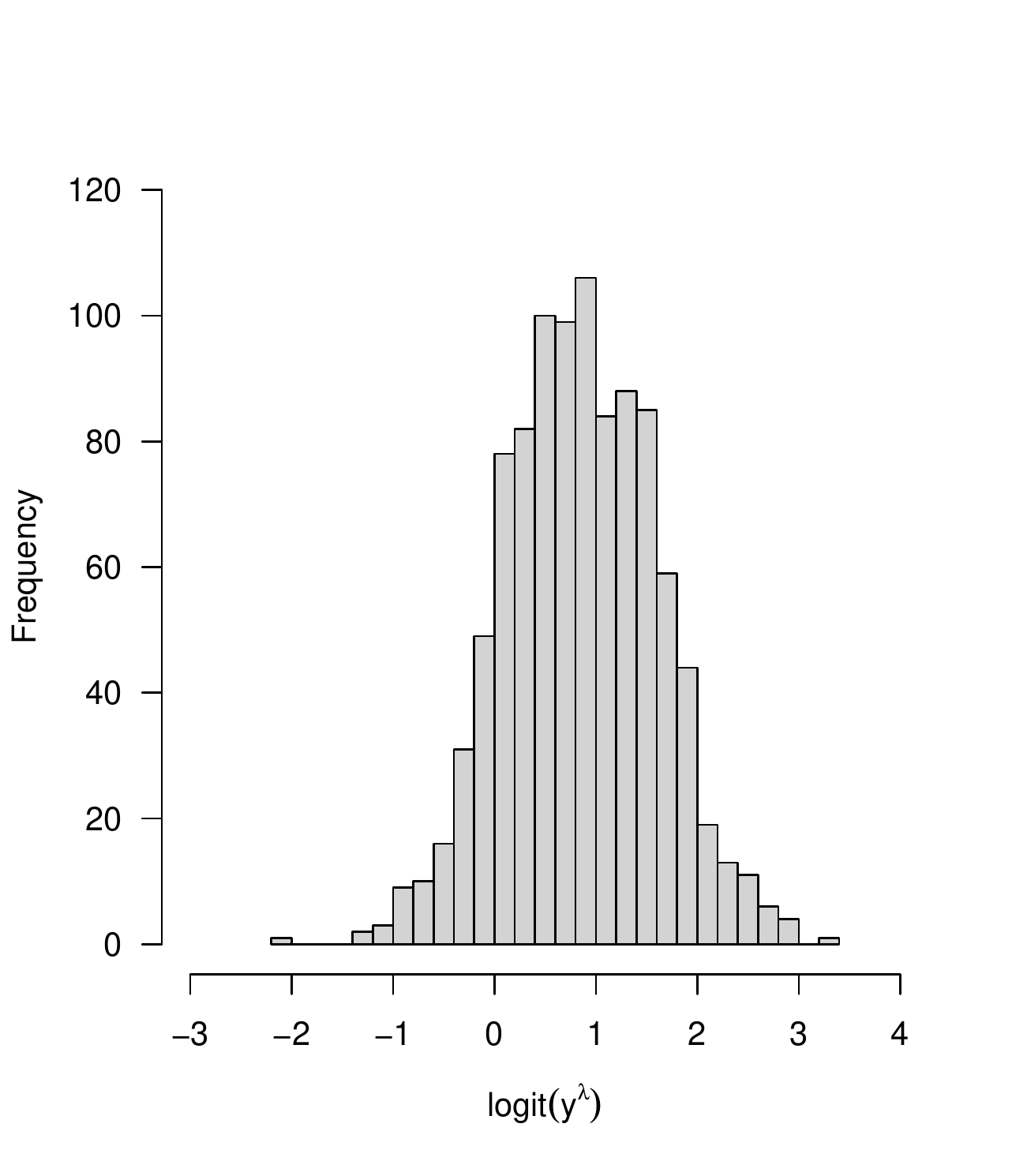}}
\caption{\small Histograms of $y$ (original data), $\logit(y)$ and $\logit(y^\lambda)$; $\lambda=0.11$.} \label{hist_transZ}
\end{figure}

The motivation for the power logit distributions is as follows. The logit transformation maps the interval $(0,1)$ to $(-\infty,+\infty)$ and, as such, is a candidate to define distributions with support in the unit interval from any distribution supported on the real line. In place of a particular distribution in $(-\infty,+\infty)$, we use the class of symmetric distributions. In addition, we add a power parameter in the logit transformation making it much more flexible. In fact, the power logit function, $\log[y^\lambda/(1-y^\lambda)]$, for $\lambda>0$, is successful in achieving symmetry in situations where the logit transformation ($\lambda=1$) fails. As an illustration, see  Figure \ref{hist_transZ}, which shows the histogram of a sample of $1000$ observations with a skewed distribution on $(0,1)$, along with the histograms of the observations transformed by the logit function and the power logit function with $\lambda=0.11$. It is clear that the histogram of the logit transformed data is far from symmetry unlike the histogram of the power logit transformed observations, which is nearly symmetric. The sample skewness coefficients for the original data and the transformed data are approximately $1.8$ (original), $-1.3$ (logit), and zero (power logit). The capacity of the power logit function to transform for symmetry is further illustrated in the Supplementary Material. Finally, Definition 1 implies that $\log[Y^\lambda/(1-Y^\lambda)]$ has a symmetric distribution with location parameter $\log[\mu^\lambda/(1-\mu^\lambda)]$ and scale parameter $\sigma$. This parametrization is convenient because $\mu$ represents the median of $Y$, while $\sigma$ and $\lambda$ are dispersion and skewness parameters, respectively, as we will show later.



The probability density function (pdf) of $Y$ is  
\begin{align}\label{pldist}
f_Y(y; \mu, \sigma, \lambda)= \frac{\lambda}{\sigma y(1-y^\lambda)} r (z^2), \quad y \in (0,1),
\end{align}
where 
\begin{align} \label{zfunction}
z=h(y; \mu, \sigma, \lambda) = \dfrac{1}{\sigma}  \left[ \log \left( \dfrac{y^\lambda}{1-y^\lambda} \right) - \log \left( \dfrac{\mu^\lambda}{1-\mu^\lambda} \right)\right].
\end{align}

The density generator $r(\cdot)$ may involve an extra parameter (or an extra parameter vector), which is denoted by $\zeta$. In addition, using the linear transformation $X = (b-a)Y + a$, we obtain the power logit distribution with support $(a,b)$, $a<b$.  

The power logit class of distributions reduces to the GJS class of distributions \citep{LemonteBazan2016} when $\lambda =1$ and we write $ Y \sim \text{GJS}(\mu, \sigma; r)$. Additionally, it leads to the logit normal distribution \citep{Johnson1949}, the L-Logistic distribution \citep{daPazetal2019}, and the logit slash distribution \citep{Korkmaz2020} by taking $\lambda =1$ and $Z$ as a standard normal, type II logistic, and slash random variable, respectively. The density generator function, $r(z)$, for $z \geq 0$, for the power logit normal (PL-N), power logit Student-t (PL-t$_{(\zeta)}$), power logit type I logistic (PL-LOI), power logit type II logistic (PL-LOII), power logit power exponential (PL-PE$_{(\zeta)}$), power logit slash (PL-slash$_{(\zeta)}$), power logit hyperbolic (PL-Hyp$_{(\zeta)}$), and power logit sinh-normal (PL-SN$_{(\zeta)}$) follow.

\begin{itemize}
\item normal: $r(z) = (2\pi)^{-1/2}\exp(-z/2)$;
\item Student-t: $r(z) = \zeta^{\zeta/2}B(1/2,\zeta/2)^{-1}(\zeta + z)^{-(\zeta+1)/2}$, $\zeta>0$ and $B(\cdot,\cdot)$ is the beta function;
\item type I logistic: $r(z) = c \exp\{-z\}(1 + \exp\{-z\})^2$, where $c \approx 1.484300029$ is the normalizing constant; 
\item type II logistic: $r(z) = \exp\{ - z^{1/2} \}(1 + \exp\{ - z^{1/2} \})^{-2}$;
\item power exponential: $r(z) = \zeta/[p(\zeta) 2^{1 + 1/\zeta} \Gamma(1/\zeta)]  \exp\left\{-z^{\zeta/2}/(2 p(\zeta)^\zeta) \right\}$, $\zeta>0$ and $p(\zeta)^2 = 2^{-2/\zeta} \Gamma(1/\zeta)/\Gamma(3/\zeta)$;
\item slash: $r(z) = (\zeta/\sqrt{2\pi}) \left( z/2 \right)^{-(\zeta + 1/2  )} G\left( \zeta + 1/2, z/2 \right)$, for $z>0$, and $r(z) = 2\zeta/[(2\zeta+1)\sqrt{2\pi}]$, for $z=0$, where $\zeta>0$ and $G(a,x) = \int_0^x t^{a-1}e^{-t}dt$ is the lower incomplete gamma function. When $\zeta = 1$ the slash distribution coincides with the canonical slash distribution;
\item hyperbolic: $r(z) = \exp\left\{ - \zeta \sqrt{1+z} \right\}/(2 \zeta K_1(\zeta))$, with $K_s(\zeta) = \int_0^\infty \frac{x^{s-1}}{2} \exp\{ -\frac{\zeta}{2} \left(x + \frac{1}{x} \right) \} \dd x$, is the modified Bessel function of third-order and index $s$.
\item sinh-normal: $r(z) = 1/(\zeta \sqrt{2 \pi}) \cosh(z^{1/2}) \exp\left[ - 2/\zeta^2 \sinh^2 (z^{1/2}) \right]$, where $\zeta>0$ and $\sinh(\cdot)$ and $\cosh(\cdot)$ represent the hyperbolic sine and cosine functions, respectively.
\end{itemize}

\begin{figure}[!htb]
\centering
\subfigure[][]{\includegraphics[scale=0.44]{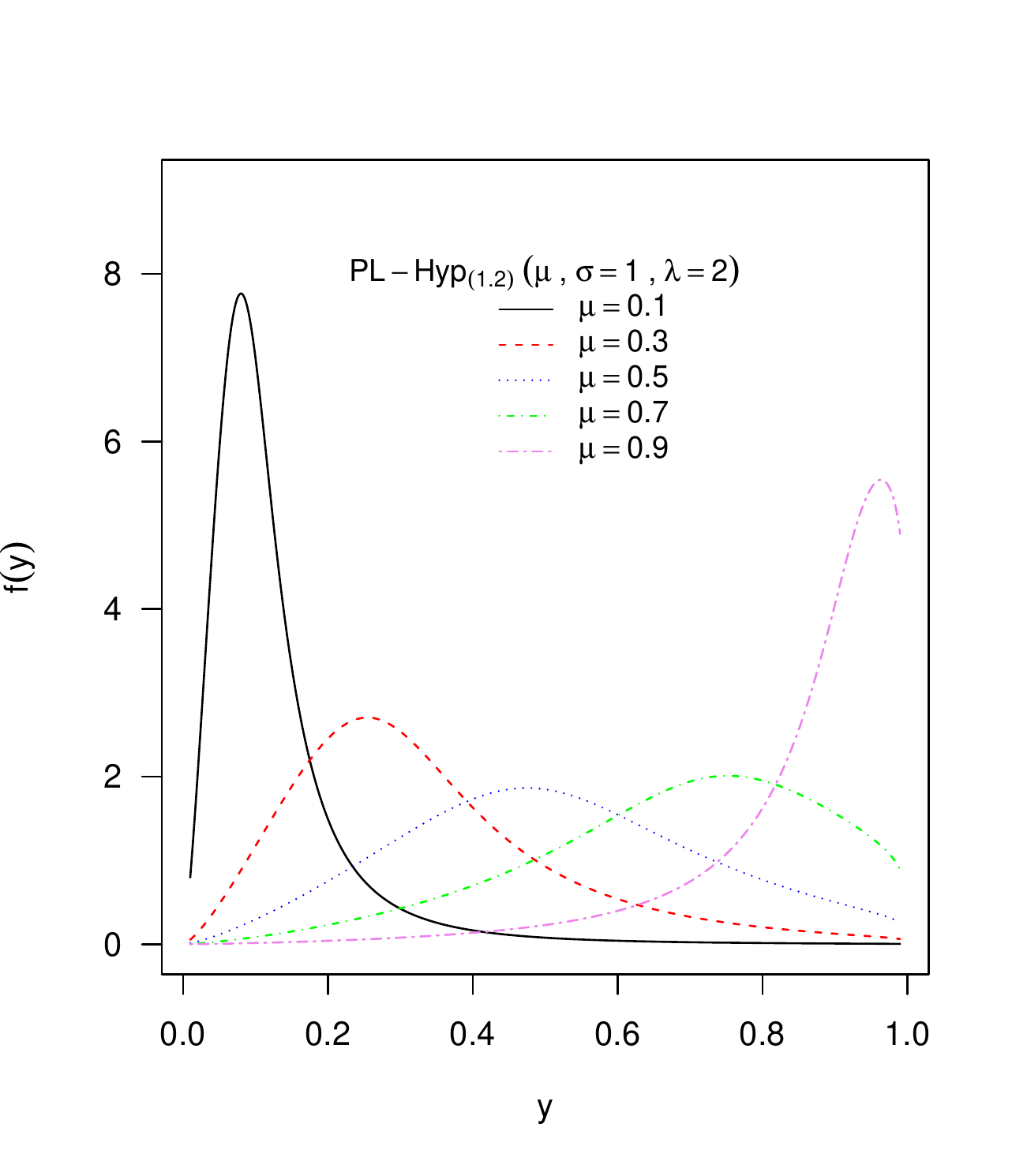}\label{Hyp1}}
\subfigure[][]{\includegraphics[scale=0.44]{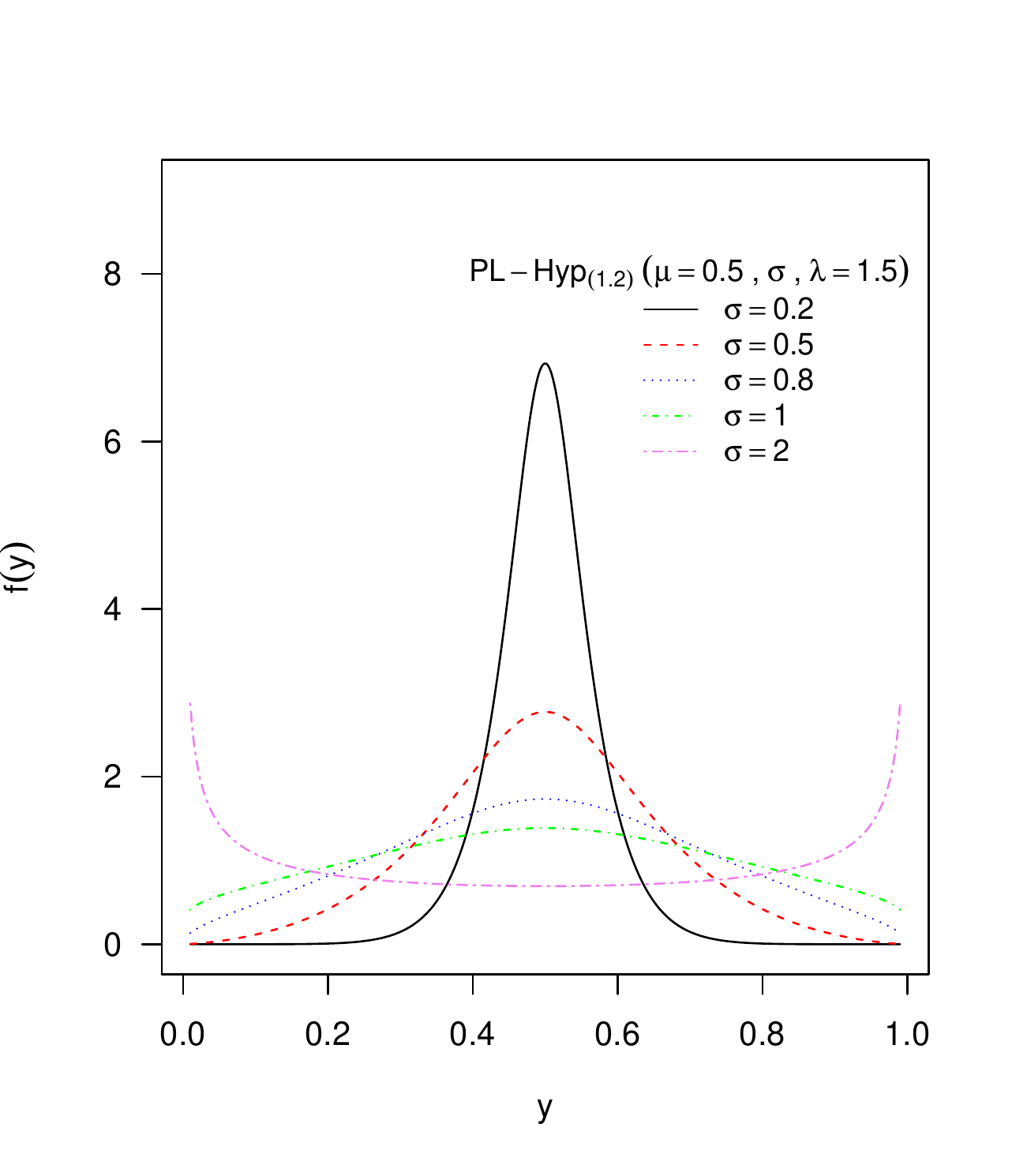}\label{Hyp2}}\\[-1ex]
\subfigure[][]{\includegraphics[scale=0.44]{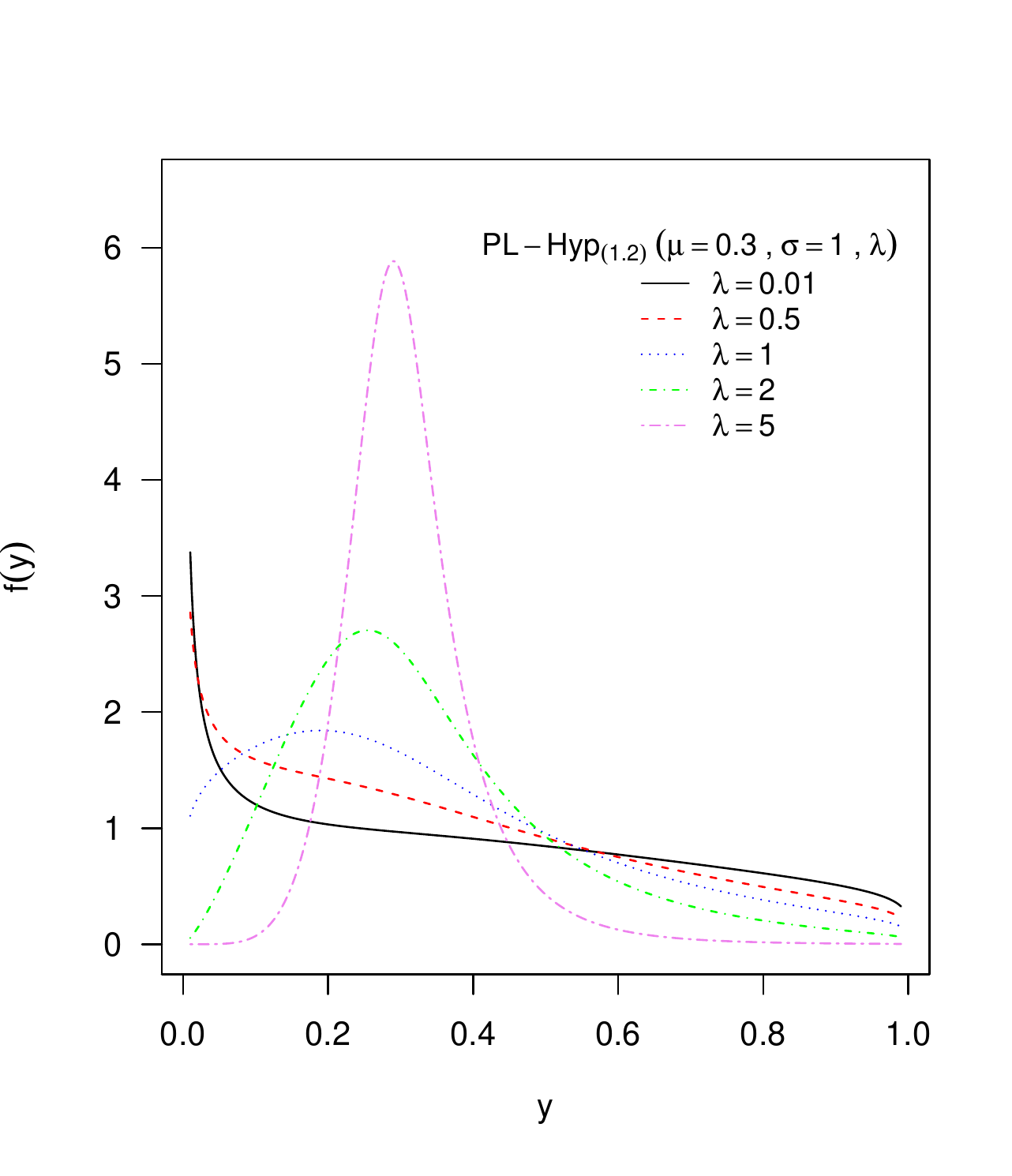}\label{Hyp3}}
\subfigure[][]{\includegraphics[scale=0.44]{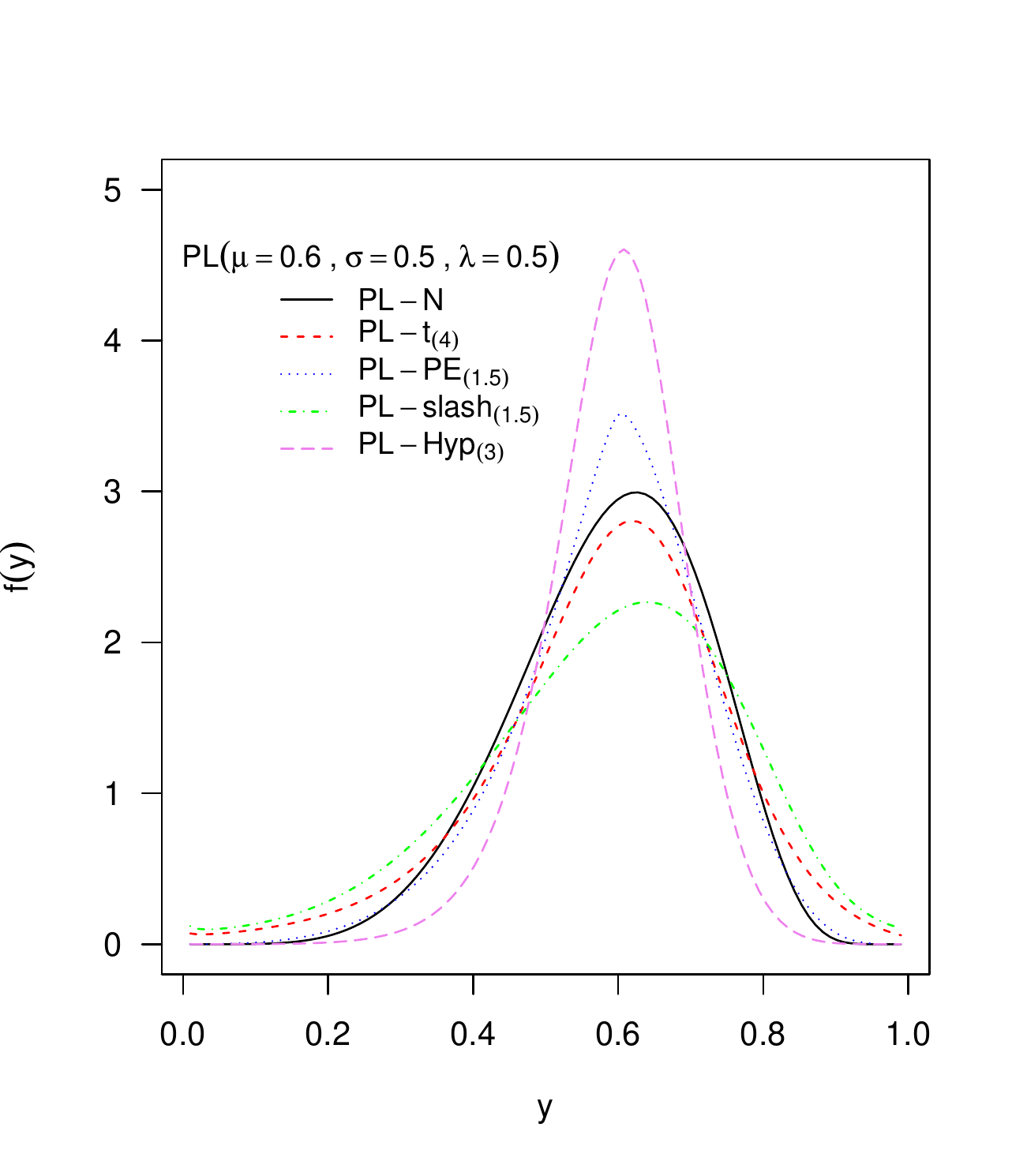}\label{pdff}}\\[-1ex]
\caption{\small Plots of the pdf of some power logit distributions.} \label{pdf}
\end{figure}

The power logit density can assume different shapes depending on the combination of parameter values and density generator functions; see Figure 2. Figures \ref{Hyp1}-\ref{Hyp3} present the pdf of the PL-Hyp$_{(1.2)}$ distribution. 
Figure \ref{Hyp1} suggests that $\mu$ is a parameter of central tendency. For fixed values of $\mu$, $\lambda$, and $\zeta$, the dispersion of the distribution increases as $\sigma$ increases, suggesting that $\sigma$ is a parameter that governs the dispersion of the distribution; see Figure \ref{Hyp2}. For fixed $\mu$, $\sigma$, and $\zeta$, the pdf becomes closer to symmetry around $\mu$ as $\lambda$ increases, indicating that $\lambda$ acts as a skewness parameter; see Figure \ref{Hyp2}. In the Supplementary Material, we show that the parameter $\sigma$ is in fact a dispersion parameter and that $\lambda$ can be regarded as a skewness parameter. Finally, Figure \ref{pdff} shows the pdf of the PL-N, PL-t$_{(4)}$, PL-PE$_{(1.5)}$, PL-slash$_{(1.5)}$ and PL-Hyp$_{(3)}$ distributions for a particular choice of the parameters. Note that the PL-t$_{(4)}$ and PL-slash$_{(1.5)}$ distributions display heavier tails than the other distributions.


We now give some properties of the power logit distributions. If $Y \sim \mbox{PL} (\mu, \sigma, \lambda; r)$ then, the following properties hold, most of which are immediate consequences of Definition \ref{def1}.

\begin{enumerate}
\item[(P1)] The cumulative distribution function (cdf) of $Y$ is $F_Y(y; \mu, \sigma, \lambda) = R(z)$, where $R(\cdot)$ is the cdf of $Z \sim \text{S}(0, 1; r)$ and $z$ is given in \eqref{zfunction}.
\item[(P2)] The $100u\%$ quantile of $Y$ is $y_u= \mu\left[ e^{\sigma z_u}/(1- \mu^\lambda (1- e^{\sigma z_u})) \right]^{1/\lambda}$, for $u \in (0,1)$, where $z_u$ is the $100u\%$ quantile of $Z \sim \mbox{S}(0,1;r)$.
\item[(P3)] $\mu$ is the median of $Y$. 
\item[(P4)] $\sigma$ is a dispersion parameter in the sense of the quantile spread order \citep{Townsend2005}. 
\item[(P5)] For $\lambda = 1$,
\begin{enumerate}
\item[1.] $1-Y \sim \mbox{PL}(1-\mu, ~\sigma,~ \lambda = 1;~r)$;
\item[2.] $Y \sim \mbox{GJS} (\mu, \sigma; r)$;
\item[3.] if $\mu = 0.5$, the power logit density function is symmetric around $y = 0.5$.
\end{enumerate}
\item[(P6)] $Y^\lambda \sim \mbox{GJS} (\mu^\lambda, \sigma; r)$.
\item[(P7)] $Y^c \sim \mbox{PL}(\mu^c, ~\sigma,~ \lambda/c;~r)$, for all $c>0$.
\item[(P8)] Let $W \sim \mbox{S}( -\log(-\log \mu ) , \sigma; r)$, then
\[
 -\log (- \log Y)   \stackrel{\mathcal{D}}{\longrightarrow} W, \quad \mbox{when} \quad \lambda \rightarrow 0^+,
\]
where $\stackrel{\mathcal{D}}{\longrightarrow}$ denotes convergence in distribution.
\end{enumerate}
Property P8 states that the power logit distributions have as a limiting case, when $\lambda \rightarrow 0^+$, the family of log-log distributions, as defined below.

\begin{definicao}[log-log distributions] We say that a continuous random variable $Y$ with support $(0,1)$ has a log-log distribution with parameters $\mu \in (0,1)$ and $\sigma>0$ and density generator function $r$ if 
\[
\dfrac{1}{\sigma} \{ - \log (-\log Y) - [ -\log (-\log \mu)] \}   \sim \text{S}(0,1;r).
\]
We write $Y \sim \mbox{log-log}(\mu, \sigma; r)$. The pdf of $Y$ is
\begin{align*} 
f(y; \mu, \sigma) = \dfrac{1}{\sigma y (-\log y)} r(z^2), \quad y \in (0,1),
\end{align*}
with $z = \sigma^{-1} \{ - \log (-\log y) - [ -\log (-\log \mu)] \}   $.
\end{definicao}
This class of distributions has some interesting properties, for instance: the $100u\%$ quantile of $Y$ is $y_u= \mu^{\exp\{-\sigma z_u\}}$, for $u \in (0,1)$, where $z_u$ is the $100u\%$ quantile of $Z \sim \mbox{S}(0,1;r)$; $\mu$ and $\sigma$ are the median and the dispersion of $Y$, respectively; if $Y \sim \mbox{log-log}(\mu, \sigma; r)$, then $Y^c \sim \mbox{log-log}(\mu^c, \sigma; r)$, for $c>0$.

\section{Power logit regression models} \label{PLreg}

\subsection{Definition}
We define a class of power logit regression models as follows. Let $Y_1,\ldots,Y_n$ be $n$ independent random variables, where $Y_i \sim \text{PL}(\mu_i, \sigma_i, \lambda; r)$, for $i=1,\ldots,n$, and
\begin{align}\label{link}
\begin{split}
d_1 (\mu_i) &= \bm{x_i}^{\top} \bm{\beta} = \eta_{1i}, \\ 
d_2 (\sigma_i) &= \bm{s_i}^{\top} \bm{\tau} = \eta_{2i},\\
\end{split} 
\end{align}
where $\bm{\beta}=(\beta_1, \ldots, \beta_p)^{\top} \in \Reais^p$, $\bm{\tau}=(\tau_1, \ldots, \tau_q)^{\top} \in \Reais^q$ and $\lambda>0$ are the unknown parameters, which are assumed to be functionally
independent and $p+q+1<n$; $\eta_{1i}$ and $\eta_{2i}$ are the linear predictors; $\bm{x}_i = (x_{i1}, \ldots, x_{ip})^{\top}$ and $\bm{s}_i = (s_{i1}, \ldots, s_{iq})^{\top}$ are observations on $p$ and $q$ known independent variables. We assume that the model matrices $\textbf{X}=[\bm{x}_1,\ldots,\bm{x}_n]^\top$ and $\textbf{S}=[\bm{s}_1,\ldots,\bm{s}_n]^\top$ have column rank $p$ and $q$, respectively. In addition, we assume that the link functions $d_1:(0,1)\to\Reais$ and
$d_2:(0,\infty)\to\Reais$ are strictly monotonic and twice differentiable. Some examples of link functions for the median submodel are: $d_1(\mu) = \log \{\mu/(1-\mu)\}$ (logit); $d_1(\mu) = \Phi^{-1}(\mu)$ (probit), where $\Phi^{-1}(\cdot)$ is the cdf of a standard normal random variable; $d_1(\mu) = - \log \{ - \log \mu \}$ (log-log); and $d_1(\mu) = \log \{ - \log(1-\mu)\}$ (complementary log-log). For the dispersion submodel, the log link, $d_2(\sigma) = \log \sigma$, is the natural choice.

The power logit regression models generalize some models: the GJS regression model \citep{LemonteBazan2016} is obtained by taking $\lambda = 1$; if $Y_i \sim \text{PL-LOII}(\mu_i, \sigma_i, 1)$ we have the L-logistic regression model \citep{daPazetal2019}. Additionally, the model parameters are interpreted in terms of the median, dispersion and skewness of the response variable. Also, the introduction of the skewness parameter $\lambda$ allows better fits for highly skewed data. 

Finally, the power logit regression models have the log-log regression models as a limiting case, when $\lambda \rightarrow 0^+$. In practice, the log-log regression models may be regarded as a parsimonious alternative to the power logit regression models when the estimated $\lambda$ is close to zero.\footnote{For brevity, the log-log regression model will not be studied in the following sections, but to obtain the quantities related to it, take the limit when $\lambda \rightarrow 0^+$ of those defined for the power logit regression model.}

\subsection{Parameter estimation}

Let $y_1,\ldots,y_n$ be $n$ observed responses from a power logit regression model. The log-likelihood function for $\bm{\theta}=(\bm{\beta}^{\top},\bm{\tau}^{\top}, \lambda)^{\top}$ is 
\begin{equation}
\ell (\bm{\theta}) = \sum_{i=1}^{n} \ell_i(\mu_i, \sigma_i, \lambda), \label{logver}
\end{equation}
where $\ell_i(\mu_i, \sigma_i, \lambda) = \log \lambda - \log \sigma_i - \log\{1-y_i^\lambda \} + \log\{ r(z_i^2)\} + c$, 
$z_i = h(y_i; \mu_i, \sigma_i, \lambda)$, and $c$ is constant with respect to $\bm{\theta}$. Note that, from \eqref{link}, $\mu_i$ and $\sigma_i$ are defined as functions of $\bm{\beta}$ and $\bm{\tau}$, respectively; that is, $\mu_i=d_1^{-1}(\eta_{1i})$, $\sigma_i=d_2^{-1}(\eta_{2i})$, for $i=1,\ldots,n$. The score function (see the Supplementary Material) is given by the $(p+q+1)$-vector $\textbf{U}(\bm{\theta}) = (\textbf{U}_{\bm{\beta}}^\top, \textbf{U}_{\bm{\tau}}^\top, \text{U}_{\lambda})^\top$, with 
\[
\textbf{U}_\beta = \textbf{X}^{\top}\textbf{W}\textbf{T}_1\bm{\mu}^*, \quad \textbf{U}_\tau = \textbf{S}^{\top}\textbf{T}_2\bm{\sigma}^*, \quad \text{U}_\lambda = \bm{1}_n^{\top}\bm{\lambda}^*,
\]
\sloppy where $\textbf{T}_1= \diag \{ 1/ \dot{d}_1(\mu_1), \ldots, 1/ \dot{d}_1(\mu_n)  \}$, $\textbf{T}_2= \diag \{ 1/ \dot{d}_2(\sigma_1), \ldots, 1/ \dot{d}_2(\sigma_n) \}$, $\textbf{W}= \diag \{z_1 v(z_1), \ldots, z_n v(z_n) \}$, $\bm{\mu}^* = (\mu_1^*, \ldots, \mu_n^*)^{\top}$, $\bm{\sigma}^* = (\sigma_1^*, \ldots, \sigma_n^*)^{\top}$, $\bm{\lambda}^* = (\lambda_1^*, \ldots, \lambda_n^*)^{\top}$, $\bm{1}_n$ denotes a $n$-dimensional vector of ones, $\dot{d}_j(t) =\dd d_j(t)/\dd t$, for $j=1,2$, $v(t)=-2 r'(t^2)/ r(t^2)$, $r'(u)=\dd r(u)/\dd u$, $\mu_i^* = \lambda/[\sigma_i \mu_i (1- \mu_i^\lambda)]$, $\sigma_i^* = \sigma^{-1}[z_i^2 v(z_i) -1]$,
and 
\[
\lambda_i^* = \dfrac{1}{\lambda} +\dfrac{y_i^\lambda \log y_i}{1- y_i^\lambda} - \dfrac{1}{\sigma_i} z_i v(z_i) \left[ \dfrac{\log y_i}{ (1-y_i^\lambda)} - \dfrac{\log \mu_i}{ (1-\mu_i^\lambda)} \right].
\]

The Hessian matrix for $\bm{\theta}$ is given by 
\begin{equation}\label{hessian}
\textbf{J}_n(\bm{\beta},\bm{\tau}, \lambda) =  \left[
\begin{array}{cccc}
\textbf{J}_{\bm{\beta} \bm{\beta}} & \textbf{J}_{\bm{\beta} \bm{\tau}} & \textbf{J}_{\bm{\beta} \lambda} \\
\textbf{J}_{\bm{\beta} \bm{\tau}}^\top & \textbf{J}_{\bm{\tau},\bm{\tau}} & \textbf{J}_{\bm{\tau} \lambda} \\
\textbf{J}_{\bm{\beta} \lambda}^\top & \textbf{J}_{\bm{\tau} \lambda}^\top & \text{J}_{\lambda \lambda}\\
\end{array}
\right], 
\end{equation}
\sloppy where $\textbf{J}_{\bm{\beta} \bm{\beta}}=\textbf{X}^\top \textbf{W}_1 \textbf{T}_1 \textbf{X}$, $\textbf{J}_{\bm{\tau} \bm{\tau}}=\textbf{S}^\top \textbf{W}_2 \textbf{T}_2 \textbf{S}$, $\text{J}_{\lambda \lambda}=\bm{1}_n^\top \textbf{W}_3 \bm{1}_n$, $\textbf{J}_{\bm{\beta} \bm{\tau}}=\textbf{X}^\top \textbf{W}_4 \textbf{T}_1 \textbf{T}_2 \textbf{S}$, $\textbf{J}_{\bm{\beta} \lambda}=\textbf{X}^\top \textbf{W}_5 \textbf{T}_1 \bm{1}_n$, $\textbf{J}_{\bm{\tau} \lambda}=\textbf{S}^\top \textbf{W}_6 \textbf{T}_1 \bm{1}_n$, with $\textbf{W}_j= \diag \{ w_1^{(j)}, \ldots, w_n^{(j)} \}$,
\begin{align*}
w_i^{(1)}&= \dfrac{\lambda}{[\sigma_i \mu_i (1- \mu_i^\lambda)]^2} \left\{ \sigma_i [1-\mu_i^\lambda (1+\lambda)] z_i v(z_i) + \lambda [ v(z_i) + z_i v'(z_i)] \right\} \dot{d}_1(\xi_i)^{-1} \\
& + \dfrac{\lambda}{\sigma_i \mu_i (1- \mu_i^\lambda)}  z_i v(z_i) \frac{\ddot{d}_1(\mu_i)}{\dot{d}_1(\mu_i)^2},\\
w_i^{(2)}&= - \left\{ \dfrac{1}{\sigma_i^2} - \dfrac{1}{\sigma_i^2} z_i^2[3 v(z_i) + z_i v'(z_i)] \right\} \dot{d}_2(\sigma_i)^{-1}  + \dfrac{1}{\sigma_i} [z_i^2 v(z_i) -1] \frac{\ddot{d}_2(\sigma_i)}{\dot{d}_2(\sigma_i)^2},\\
w_i^{(3)}&= \dfrac{1}{\lambda^2} - \dfrac{y_i^\lambda \log^2 y_i}{(1- y_i^\lambda)^2} + \dfrac{1}{\sigma_i^2} [ v(z_i) + z_i v'(z_i)] \left( \dfrac{\log y_i}{ 1-y_i^\lambda} - \dfrac{\log \mu_i}{ 1-\mu_i^\lambda} \right)^2\\
& + \dfrac{1}{\sigma_i} z_i v(z_i)\left[ \dfrac{y_i^\lambda \log^2 y_i}{ (1-y_i^\lambda)^2} - \dfrac{\mu_i^\lambda \log^2 \mu_i}{ (1-\mu_i^\lambda)^2} \right],\\
w_i^{(4)} & =  \dfrac{\lambda}{\sigma_i^2 \mu_i (1- \mu_i^\lambda)} z_i [2v(z_i) + z_i v'(z_i)],\\
w_i^{(5)} &= \dfrac{ - \lambda}{\sigma_i \mu_i (1- \mu_i^\lambda)} \left\{ \dfrac{1-\mu_i^\lambda (1-\lambda \log \mu_i)}{\lambda (1-\mu_i^\lambda)} z_i v(z_i) + \dfrac{1}{\sigma_i} \left( \dfrac{\log y_i}{ 1-y_i^\lambda} - \dfrac{\log \mu_i}{ 1-\mu_i^\lambda} \right) [v(z_i) + z_i v'(z_i)]  \right\},\\
w_i^{(6)} & = - \dfrac{1}{\sigma_i^2} \left( \dfrac{\log y_i}{ 1-y_i^\lambda} - \dfrac{\log \mu_i}{ 1-\mu_i^\lambda} \right) z_i [2v(z_i) + z_i v'(z_i)].
\end{align*}
where $\ddot{d_j}(z) =\dd\dot{d_j}(z)/dz$, for $j=1,2$.

The maximum likelihood estimate (mle) of $\bm{\theta}$ can be obtained by solving simultaneously the nonlinear system of equations $\textbf{U}(\bm{\theta}) = \bm{0}_{p+q+1},$ which does not have closed form, and $\bm{0}_{p+q+1}$ denotes a $(p+q+1)$-dimensional vector of zeros. As we can see from the score function $\textbf{U}_\beta$, $v(\cdot)$ acts as a weighting function, i.e., observations with small value for $v(\cdot)$ are downweighted for estimating $\bm{\beta}$.

The choice of $r(\cdot)$ may induce a function $v(z)$ that decreases as $z$ departs from zero, and hence some power logit distributions produce robust estimation against outliers. For the PL-N, PL-t$_{(\zeta)}$, PL-PE$_{(\zeta)}$ and PL-Hyp$_{(\zeta)}$ distributions one has, respectively, $v(z) = 1$, $v(z) = (\zeta +1)/(\zeta + z^2)$, $v(z) = \zeta (z^2)^{\zeta/2 - 1}/(2 p(\zeta)^\zeta)$, and $v(z) = \zeta/\sqrt{1+z^2}$. For the PL-t$_{(\zeta)}$, PL-PE$_{(\zeta)}$ (with $0<\zeta < 2$) and PL-Hyp$_{(\zeta)}$ distributions, $v(z)$ is decreasing in $z^2$ and hence extreme observations tend to have small weights in the estimation process. 

\begin{figure}[!htb]
\centering
\subfigure[][]{\includegraphics[scale=0.43]{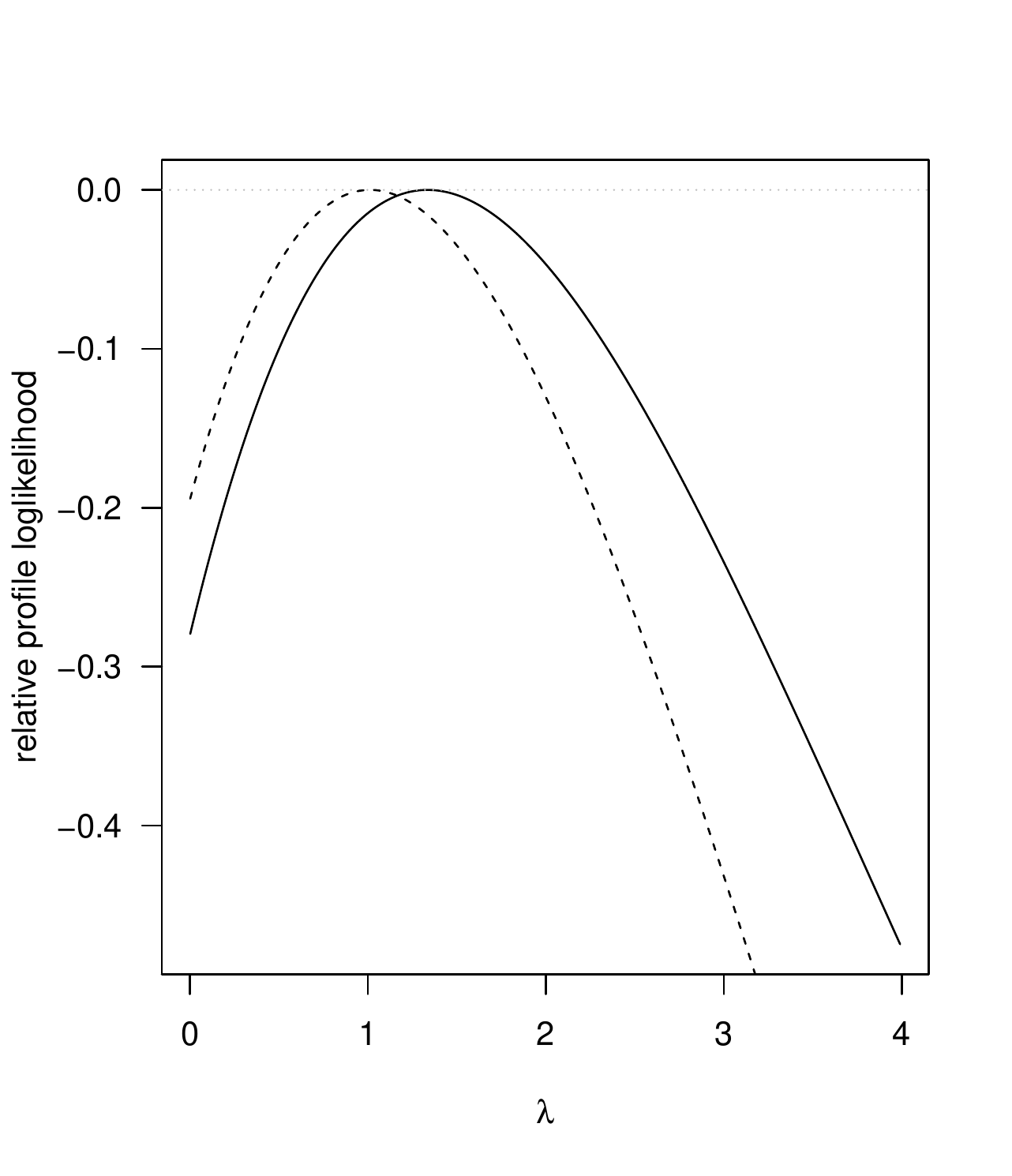}\label{samplePEok}}
\subfigure[][]{\includegraphics[scale=0.43]{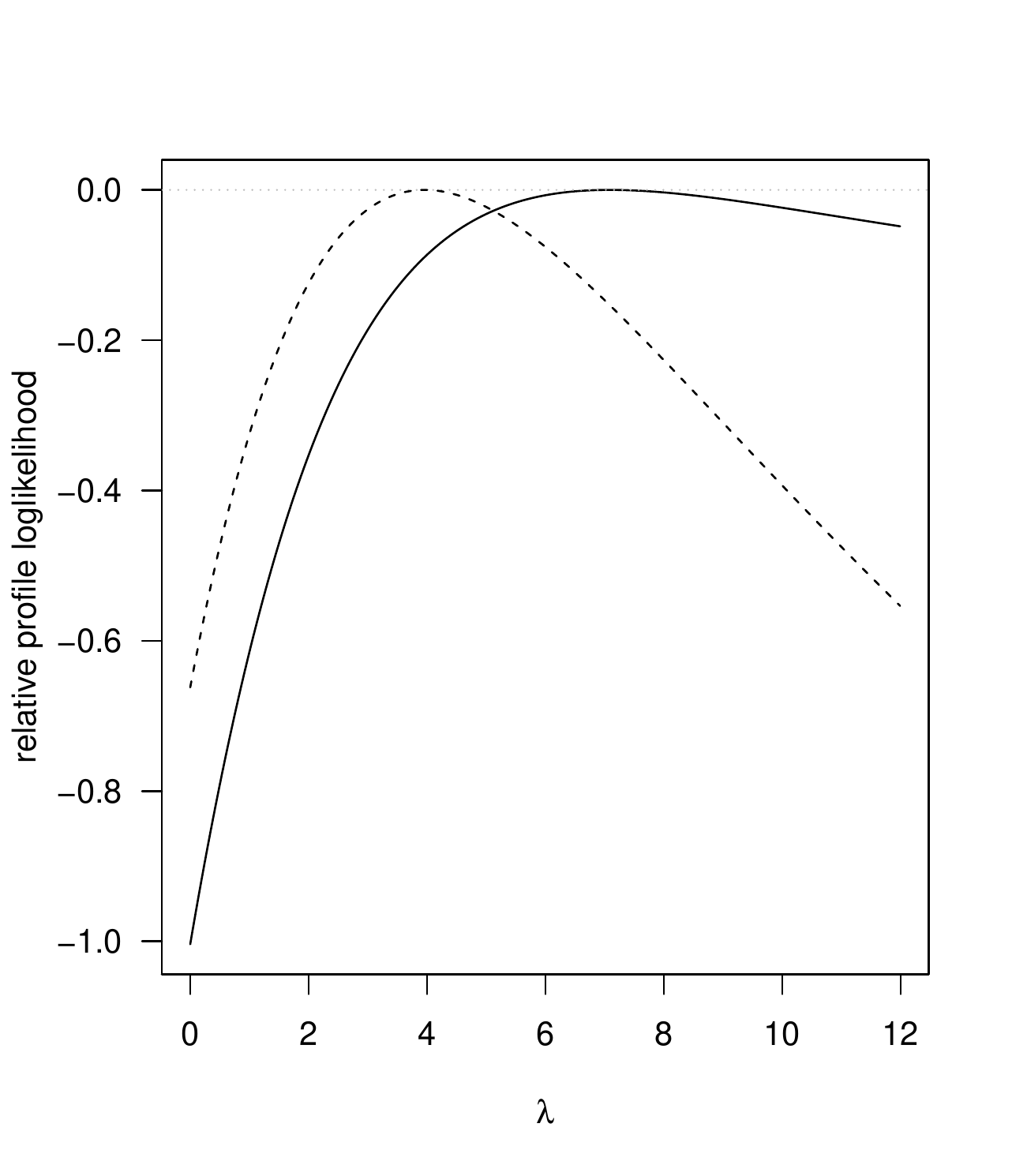}\label{samplePEfraca}}
\subfigure[][]{\includegraphics[scale=0.43]{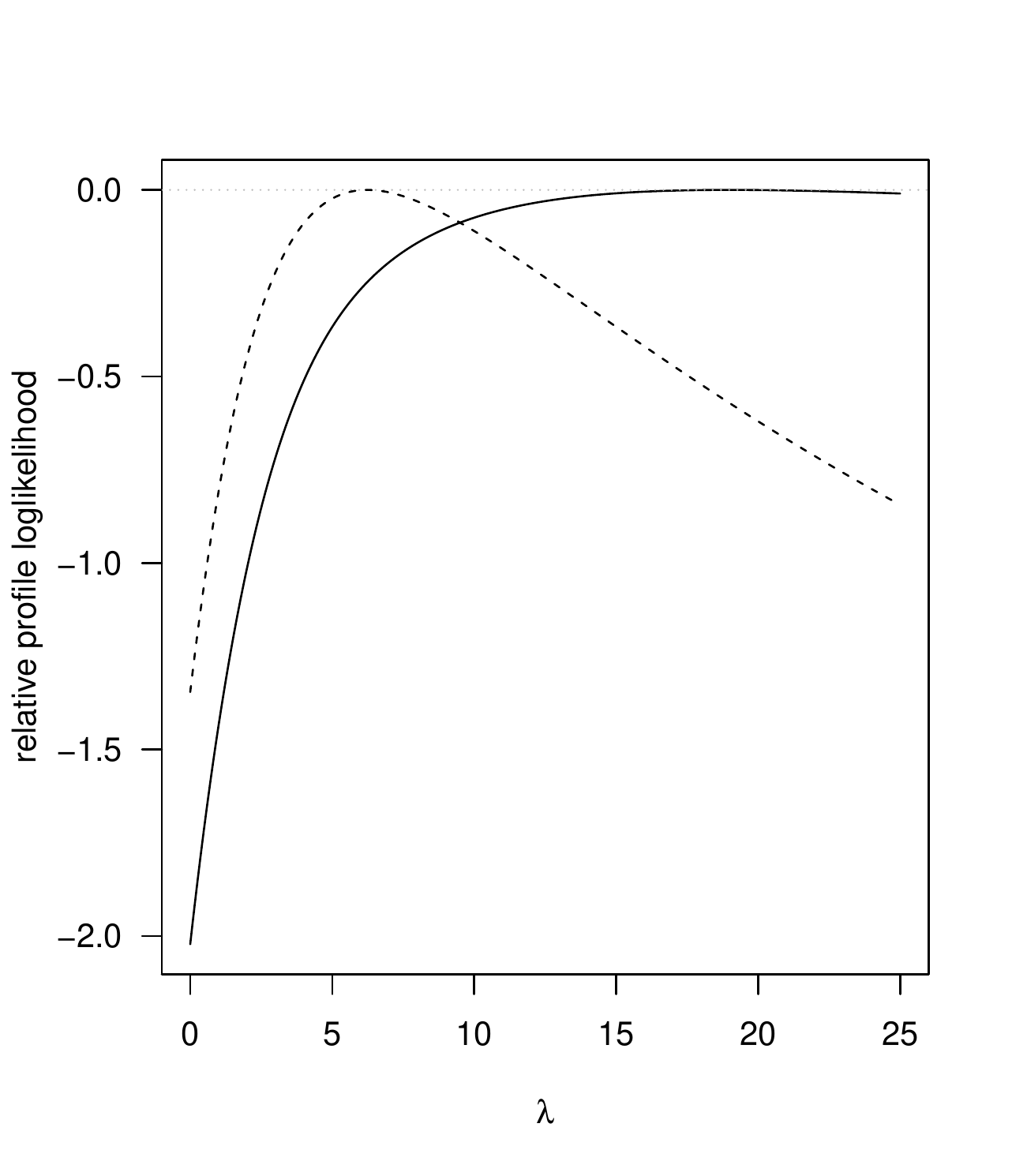}\label{samplePEforte}}
\caption{\small Relative versions of the usual (solid) and penalized (dashed) profile log-likelihood functions of $\lambda$ for three samples.}
\label{sample1T} 
\end{figure}

In applications with simulated data of a small sample size, we noted that the profile log-likelihood for $\lambda$ may display weak concavity. In these cases, the standard non-linear optimization algorithms used for the log-likelihood maximization may produce estimates for $\lambda$ far from its true value. We propose the use of a penalized log-likelihood function for $\lambda$ as it will be described below in detail. We anticipate the expected effect of the penalization in Figure \ref{sample1T}, which presents the plots of the relative profile log-likelihood function of $\lambda$ and the corresponding penalized version, obtained from three random samples of $25$ observations taken from the PL-PE$_{(1.3)}$ distribution with $\mu = 0.7$, $\sigma = 0.5$, and $\lambda = 1$. The profile log-likelihood function for $\lambda$ (solid line) for the first sample (Figure \ref{samplePEok}) is well behaved, but the other two samples give rise to profile log-likelihoods with weak concavity regions (Figures \ref{samplePEfraca}--\ref{samplePEforte}). The corresponding relative profile penalized log-likelihood functions (dashed lines) do not display flat or nearly flat regions for any of the three samples. The estimates of $\lambda$ become closer to the true value when the penalized log-likelihood function is employed.

In the statistical literature there are several reports of monotone likelihoods for different models: Cox regression model \citep{Bryson1981}; logistic regression \citep{Albert1984}; skew normal and skew t distributions \citep{Azzalini2013,Sartori2006}; modified extended Weibull distribution \citep{Lima2019}. To deal with monotone likelihoods, one may consider a modification in the log-likelihood function in order to obtain estimators with better properties.

As in \cite{Sartori2006}, we modify the profile log-likelihood of $\lambda$ instead of the log-likelihood of $\bm{\theta}$, requiring less computational effort. 
We consider a penalization based on the Jeffreys' prior \citep{Jeffreys1946}, with the observed information matrix in place of Fisher's information matrix. 
This approach performed well in simulation experiments. The penalized profile log-likelihood for $\lambda$ is 
\begin{align}\label{penalizedloglik}
\ell_p^* (\lambda)= \ell^* (\lambda) + \dfrac{1}{2} \log \left( \dfrac{  \text{J}_{\lambda \lambda}^*}{n} \right),
\end{align}
in which $\ell^* (\lambda) = \ell(\widehat{\bm{\beta}}_\lambda, \widehat{\bm{\tau}}_\lambda, \lambda )$ is the profile log-likelihood for $\lambda$ and $\text{J}_{\lambda \lambda}^* = \text{J}_{\lambda \lambda}(\widehat{\bm{\beta}}_\lambda, \widehat{\bm{\tau}}_\lambda, \lambda)$, where $\widehat{\bm{\beta}}_\lambda$ and $\widehat{\bm{\tau}}_\lambda$ are the mle for $\bm{\beta}$ and $\bm{\tau}$, respectively, with fixed $\lambda$. The penalized maximum likelihood estimate (pmle) of $\bm{\theta}$, that is, $\widetilde{\bm{\theta}} = (\widetilde{\bm{\beta}}, \widetilde{\bm{\tau}}, \widetilde{\lambda})^\top$, can be computed through numerical optimization following two steps:
\begin{itemize}
\item[i.] Compute $\widetilde{\lambda}$ such that
\[
\widetilde{\lambda} = \underset{\lambda>0}{\mathrm{argmax}}~\ell_p^*(\lambda).
\]
\item[ii.] Find $\widetilde{\bm{\beta}}$ and $\widetilde{\bm{\tau}}$ by maximizing $\ell(\bm{\beta}, \bm{\tau}, \widetilde{\lambda})$.
\end{itemize}

The optimization algorithms require the specification of initial values to be used in the iterative scheme. Our suggestion is to use as an initial point estimate for $\bm{\beta}$ the ordinary least squares estimate of this parameter vector obtained from a linear regression of the transformed responses $d_1(y_1), \ldots, d_1(y_n)$ on $\bm{X}$, that is, $\bm{\beta}^{(0)} = (\bm{X}^\top \bm{X})^{-1} \bm{X}^\top \bm{\upsilon}$, where $\bm{\upsilon} = (d_1(y_1), \ldots, d_1(y_n))^\top$. For $\bm{\tau}$, we suggest $\bm{\tau}^{(0)}= (\tau^{(0)}_1, 0, \ldots, 0)^\top$, where $\tau^{(0)}_1$ is the sample standard deviation of $\log[y/(1-y)]$. Finally, an initial value for $\lambda$ is $\lambda^{(0)} = 1$. In addition, quantities evaluated at $\widetilde{\bm{\theta}}$ ($\widehat{\bm{\theta}}$) will be written with a tilde (circumflex).

%
%

%

The penalization used in the profile log-likelihood for $\lambda$ is $\mathcal{O}_p(1)$ as $n \rightarrow \infty$. Then, both $\ell^* (\lambda)$ and $\ell_p^* (\lambda)$ are $\mathcal{O}_p(n)$ and they differ only by the penalization term, which is $\mathcal{O}_p(1)$. Thus, the first-order asymptotic distribution of
$\widetilde{\bm{\theta}}$ coincides that of $\widehat{\bm{\theta}}$; for more details, see \cite{Azzalini2013}. Then, under suitable regularity conditions, $\widetilde{\bm{\theta}}$ is a consistent estimator of $\bm{\theta}$ and
\[
\sqrt{n} ( \widetilde{\bm{\theta}} - \bm{\theta} ) \stackrel{\mathcal{D}}{\longrightarrow} \mathcal{N}_{p+q+1} (\bm{0}_{p+q+1}, ~\textbf{K}(\bm{\theta})^{-1}),~\mbox{as}~ n \rightarrow \infty,
\]
where $\textbf{K}(\bm{\theta}) = \textbf{K}(\bm{\beta},\bm{\tau}, \lambda)$ is the unit Fisher's information matrix. There is no closed-form expression for $\textbf{K}(\bm{\theta})$, but the asymptotic behavior remains valid if $\textbf{K}(\bm{\theta})$ is approximated by $\textbf{J}_n(\bm{\beta},\bm{\tau}, \lambda)$. The asymptotic normal distribution can be used to construct approximate confidence intervals and confidence regions for the parameters. Also, the usual asymptotic properties of the large sample tests, such as the likelihood ratio and Wald tests, remain valid.


In order to evaluate the performance of the penalized maximum likelihood estimator in the power logit models, we conducted a simulation study for power logit regression models with
$\log(\mu_i/(1-\mu_i) = \beta_1 + \beta_2 x_i$ and  
$\log \sigma_i = \tau_1 + \tau_2 s_i,$ 
where $\beta_1 = 0.5$, $\beta_2 = 1.5$, $\tau_1 =-1$, $\tau_2 = 0.5$, and $\lambda = 1$. The covariates $x_i$ and $s_i$ were generated as independent random draws from a uniform distribution on the unit interval and were kept fixed for all the replicates. We considered four different power logit distributions: PL-t$_{(5)}$, PL-PE$_{(1.5)}$, PL-Hyp$_{(1.2)}$, and PL-slash$_{(1.4)}$. We generated 3,000 Monte Carlo replicates with $n = 40$ and $n = 120$. All simulations were performed using the R software \citep{R2021} with the BFGS algorithm \citep{Press1992}. The empirical bias and the root mean squared error ($\sqrt{\text{MSE}}$) of the maximum likelihood estimates with and without penalization are presented in Table \ref{simulation1}.

\begin{table}[!ht]
\caption{\small Bias and root mean squared error ($\sqrt{\text{MSE}}$) for the mle and pmle.}\label{simulation1}
\scriptsize
\centering
\begin{tabular}{rrrrrrrrrrrrr}
  \hline
 &  & \multicolumn{5}{c}{$n=40$} &  & \multicolumn{5}{c}{$n=120$} \\ 
   \cline{3-7} \cline{9-13}
 &  & \multicolumn{2}{c}{\shortstack{mle}} &  & \multicolumn{2}{c}{\shortstack{pmle}} & & \multicolumn{2}{c}{\shortstack{mle}} & & \multicolumn{2}{c}{\shortstack{pmle}} \\
   \cline{3-4} \cline{6-7} \cline{9-10} \cline{12-13} 
 &  & bias & $\sqrt{\text{MSE}}$ & & bias & $\sqrt{\text{MSE}}$ & & bias & $\sqrt{\text{MSE}}$ & & bias & $\sqrt{\text{MSE}}$ \\
   \hline
\multirow{5}{*}{PL-t$_{(5)}$}  & $\beta_1$  & $-$0.00  & 0.14 & & $-$0.00  & 0.14 &  & $-$0.00  & 0.09 & & $-$0.00 & 0.09 \\ 
  & $\beta_2$  & $-$0.00  & 0.28 & & $-$0.00  & 0.28 &  & $-$0.00  & 0.18 & & $-$0.00 & 0.18 \\ 
  & $\tau_1$  & 0.21   & 0.65 & & 0.05   & 0.48 &  & 0.08   & 0.34 & & 0.04 & 0.31 \\ 
  & $\tau_2$  & $-$0.14  & 0.61 & & $-$0.04  & 0.55 &  & $-$0.06  & 0.36 & & $-$0.03 & 0.35 \\ 
  & $\lambda$  & 2.07   & 5.69 & & 0.79   & 2.67 &  & 0.60   & 1.79 & & 0.36 & 1.49 \\
\hline
\multirow{5}{*}{PL-PE$_{(1.5)}$}   & $\beta_1$  & $-$0.00  & 0.12 & & $-$0.00  & 0.12 &  & $-$0.00  & 0.08 & & $-$0.00 & 0.08 \\ 
  & $\beta_2$  & $-$0.00  & 0.24 & & $-$0.00  & 0.24 &  & $-$0.00  & 0.15 & & $-$0.00 & 0.15 \\ 
  & $\tau_1$   & 0.20   & 0.66 & & 0.05   & 0.48 &  & 0.08   & 0.35 & & 0.04 & 0.32 \\ 
  & $\tau_2$   & $-$0.15  & 0.60 & & $-$0.06  & 0.54 &  & $-$0.06  & 0.36 & & $-$0.04 & 0.34 \\ 
  & $\lambda$   & 2.08   & 5.71 & & 0.80   & 2.60 &  & 0.61   & 1.91 & & 0.41 & 1.60 \\ 
\hline
\multirow{5}{*}{PL-Hyp$_{(1.2)}$}   & $\beta_1$  & $-$0.00  & 0.16 & & 0.00   & 0.16 &  & $-$0.00  & 0.10 & & $-$0.00 & 0.10 \\ 
  & $\beta_2$  & $-$0.01  & 0.33 & & $-$0.01  & 0.33 &  & $-$0.00  & 0.20 & & $-$0.00 & 0.20 \\ 
  & $\tau_1$   & 0.15   & 0.54 & & 0.04   & 0.43 &  & 0.05   & 0.30 & & 0.02 & 0.28 \\ 
  & $\tau_2$   & $-$0.11  & 0.57 & & $-$0.04  & 0.53 &  & $-$0.05  & 0.35 & & $-$0.03 & 0.34 \\ 
  & $\lambda$   & 1.49   & 3.88 & & 0.72   & 2.25 &  & 0.43   & 1.37 & & 0.29 & 1.20 \\ 
\hline
\multirow{5}{*}{PL-slash$_{(1.4)}$}   & $\beta_1$  & $-$0.02  & 0.19 & & $-$0.02  & 0.18 &  & $-$0.01  & 0.12 & & $-$0.00 & 0.12 \\ 
  & $\beta_2$  & $-$0.01  & 0.37 & & $-$0.01  & 0.37 &  & $-$0.00  & 0.23 & & $-$0.00 & 0.23 \\ 
  & $\tau_1$   & $-$0.01  & 0.49 & & $-$0.10  & 0.43 &  & $-$0.01  & 0.29 & & $-$0.04 & 0.28 \\ 
  & $\tau_2$   & 0.01   & 0.57 & & 0.06   & 0.55 &  & 0.01   & 0.35 & & 0.04 & 0.34 \\ 
  & $\lambda$   & 0.50   & 3.79 & & $-$0.08  & 1.70 &  & 0.14   & 1.24 & & $-$0.04 & 1.10 \\ 
   \hline
\end{tabular}
\end{table}

Inspection of Table \ref{simulation1} shows that the maximum likelihood estimators for the $\beta$'s, both usual and penalized, present bias close to zero and small mean squared error for all the investigated models; for the PL-t$_{(5)}$ regression model, the bias of $\widehat{\beta}_2$ is zero up to two decimal places for $n=40$. The penalization seems not to interfere in the estimation of the $\beta$'s. The estimates of the parameters associated with the dispersion submodel, i.e., $\tau_1$ and $\tau_2$, present small bias. The penalized maximum likelihood estimators have smaller bias than the usual maximum likelihood estimators; in the PL-t$_{(5)}$ regression model, for example, the bias of $\widehat{\tau}_1$, when $n=40$, is $0.21$, while the bias of $\widetilde{\tau}_1$ is $0.05$. The corresponding $\sqrt{\text{MSE}}$ are also smaller when the penalization is used; for the PL-PE$_{(1.5)}$ regression model and $n=40$, the $\sqrt{\text{MSE}}$ of $\widehat{\tau}_1$ is $0.66$ and $0.48$ for $\widetilde{\tau}_1$. The penalization is effective in the estimation of $\lambda$. In all the scenarios, when the sample size is small ($n=40$), the bias of the $\widehat{\lambda}$ is considerably large, as well as the mean squared error. The penalized maximum likelihood estimator has much smaller bias and mean squared error. For instance,  in the PL-PE$_{(1.5)}$ regression model with $n=40$, the bias of $\widehat{\lambda}$ is 2.08 and $\sqrt{\text{MSE}} = 5.71$, while the bias and  $\sqrt{\text{MSE}}$ of $\widetilde{\lambda}$ are, respectively, 0.80 and 2.60. As expected, the bias and $\sqrt{\text{MSE}}$ of the estimators decrease with increasing sample size. Simulation studies with other power logit models were carried out and similar results were observed. We recommend using the penalized maximum likelihood estimator when the sample size is small. For large or moderate samples sizes, the usual and penalized maximum likelihood estimators have similar performances.

\subsection{Choosing the extra parameter $\zeta$}\label{upsilom}

As mentioned before, the density generator function $r(\cdot)$ may involve an extra parameter, denoted by $\zeta$.  This parameter is considered fixed in the estimation process. To select a suitable value for $\zeta$, we suggest to choose $\widehat{\zeta}$ such that 
\[
\widehat{\zeta} = \underset{\zeta \in \Theta^\zeta}{\mbox{argmin}} \Upsilon_\zeta,
\]
with 
\begin{align}\label{upsilonmeasure}
\Upsilon_\zeta = n^{-1} \displaystyle \sum_{i=1}^n | \Phi^{-1}[R(\widetilde{z}^{(i)})] - \upsilon^{(i)}|,
\end{align}
where $\Theta^\zeta$ represents the parameter space of $\zeta$, $\widetilde{z}^{(i)}$ is the $i$th order statistic of $\widetilde{z}$, $\upsilon^{(i)}$ is the mean of the $i$th order statistic in a random sample of size $n$ of the standard normal distribution and $\Phi(\cdot)$ is the cdf of the standard normal distribution. Note that $\Phi^{-1}[R(Z)]$ has a standard normal distribution. This measure was proposed by \cite{Vanegas2015} and the idea is to choose the value of $\zeta$ such that $( \Phi^{-1}[R(\widetilde{z}^{(1)})], \ldots, \Phi^{-1}[R(\widetilde{z}^{(n)})] )$ be as close as possible to an ordered sample of the standard normal distribution. Numerical results presented in the Supplementary Material show a good performance of this criterion for the choice of $\zeta$.

\section{Diagnostic tools} \label{diagnostictools}

In this section we present some diagnostic tools for the power logit regression models. First, we define two overall goodness-of-fit measures: the pseudo $R^2$ ($R^2_p$) and the $\Upsilon_\zeta$ measure. Following \cite{FerrariCribariNeto2004}, the pseudo $R^2$ is defined as the square of the sample correlation coefficient between the estimated linear predictor $d_1(\widetilde{\mu})$ and $d_1(y)$. Note that $0 \leq R^2_p \leq 1$ and a perfect agreement between $\widetilde{\mu}$ and $y$ yields $R^2_p = 1$. The $\Upsilon_\zeta$ measure is defined in \eqref{upsilonmeasure}. Small values of $\Upsilon_\zeta$ suggest that the fitted model suitably describes the data. 

Some residuals for the power logit regression models, as well as  influence and leverage measures, are presented below.

\subsection{Residuals}

We propose three residuals for the power logit regression models: the quantile residual, the deviance residual, and the standardized residual. 

\vspace{0.2cm}
\noindent \textit{Quantile residual}: \cite{DunnSmyth1996} defined the quantile residual, which has a standard normal distribution asymptotically if the model is correctly specified and the parameter estimators are consistent. For the power logit regression models, the quantile residual is defined as
\[
r_i^q = \Phi^{-1}[R(\widetilde{z}_i)], \quad i=1, \ldots, n,
\]
where $R(\cdot)$ is the cdf of $Z \sim \text{S}(0, 1; r)$.

\vspace{0.2cm}
\noindent \textit{Deviance residual}: The deviance residual is defined as
\[
\sinal(y_i - \widetilde{\mu}_i)\{ 2[\ell_i(\bar{\mu}_i, \widetilde{\sigma}_i, \widetilde{\lambda}) - \ell_i(\widetilde{\mu}_i, \widetilde{\sigma}_i, \widetilde{\lambda})] \}^{1/2}, \quad i=1, \ldots, n, 
\]
where $\bar{\mu}_i$ is the mle of $\mu_i$ under the saturated model \citep{McCullaghNelder1989}. This residual measures the discrepancy of the fitted model and the data as twice the difference between the maximum log-likelihood achievable and that achieved by the postulated model. For the power logit regression models, we have $\bar{\mu}_i=y_i$ and the deviance residual can be written as
\[
r_i^d = \sinal(\widetilde{z}_i) \left\{ 2 \log \left[\dfrac{r(0)}{r(\widetilde{z}_i^2)}  \right] \right\}^{1/2}, \quad i=1, \ldots, n.
\]
For the PL-N models, the quantile and deviance residuals coincide and $r_i^q=r_i^d=\widetilde{z}_i$.

\vspace{0.2cm}
\noindent \textit{Standardized residual}: The power logit regression models with constant dispersion are equivalent to
\begin{align}\label{symmetricmodel}
y_i^\dag = \mu_i^\dag (\bm{\beta}; \bm{x_i}) + \varepsilon_i, \quad i=1, \ldots,n,
\end{align}
where $\varepsilon_i \sim \mbox{S}(0, \sigma^2; r)$, $y_i^\dagger = \log[y_i^{\lambda}/(1-y_i^{\lambda})]$, and $\mu_i^\dag (\bm{\beta}; \bm{x_i}) = \mu_i^\dagger = \log[ \mu_i^{\lambda}/(1-\mu_i^{\lambda})]$ is a nonlinear function of $\bm{\beta}$ with matrix of derivatives $\bm{\mbox{D}} = \partial \bm{\mu^\dagger}/\partial \bm{\beta} = \sigma \bm{\mu}^* \textbf{T}_1 \textbf{X}$, with $\bm{\mu^\dagger} = (\mu_1^\dag, \ldots, \mu_n^\dag)^\top $. Assuming that $\lambda$ is fixed, model \eqref{symmetricmodel} is a symmetric nonlinear regression model \citep{Cysneiros2008, Galea2005}. Then, an ordinary residual may be defined as $r_i = y_i^\dag - \widetilde{\mu}_i^\dag$, for $i=1, \ldots, n$. Using expansions up to order $n^{-1}$ from \cite{coxSnell1968}, when they exist, $\mathbb{E}(\bm{r}) \approx (\textbf{I}_n - \textbf{H})\bm{\eta}$ and  $\mbox{Var}(\bm{r}) \approx \sigma^2 \xi_r \{ \textbf{I}_n - (d_r \xi_r)^{-1} \textbf{H} \}$ where $\bm{r}=(r_1, \ldots, r_n)^\top$, $d_r = d_r(\zeta) = \mathbb{E}[Z^2 v^2(Z)]$, $\xi_r$ is such that $\text{Var}(Z) = \xi_r$, $\textbf{H} = \bm{\mbox{D}} (\bm{\mbox{D}}^\top \bm{\mbox{D}})^{-1} \bm{\mbox{D}}^\top$, $\textbf{I}_n$ is the identity matrix of order $n$ and $\bm{\eta}$ is the difference between the linear and quadratic approximations of $\mu_i^\dag (\widetilde{\bm{\beta}}; \bm{x_i})$. Therefore, we define a standardized residual for the power logit regression models with constant dispersion based on $r_i$ as
\begin{align}\label{standardizedresidual}
r_i^p = \dfrac{\widetilde{y}_i^\dag - \widetilde{\mu}_i^\dag}{\widetilde{\sigma} \sqrt{\widetilde{\xi_r} \{ 1 - (\widetilde{d}_r \widetilde{\xi_r})^{-1} \widetilde{h}_{ii} \} }}, \quad i=1, \ldots, n,
\end{align}
where $\widetilde{y}_i^\dag = \log (y_i^{\widetilde{\lambda}}/(1-y_i^{\widetilde{\lambda}}))$, $\widetilde{h}_{ii}$ is the $i$th diagonal element of $\textbf{H}$ evaluated in $(\widetilde{\bm{\beta}}, \widetilde{\bm{\tau}}, \widetilde{\lambda})$. For power logit regression models with varying dispersion, the standardized residual takes the form \eqref{standardizedresidual}, with $\widetilde{\sigma}$ replaced by $\widetilde{\sigma}_i$ and $\widetilde{h}_{ii}$ being the $i$th diagonal element of $\textbf{H} = \bm{\Sigma}^{-1/2} \bm{\mbox{D}} (\bm{\mbox{D}}^\top \bm{\Sigma}^{-1} \bm{\mbox{D}})^{-1} \bm{\mbox{D}}^\top \bm{\Sigma}^{-1/2}$ evaluated at $(\widetilde{\bm{\beta}}, \widetilde{\bm{\tau}}, \widetilde{\lambda})$, where $\bm{\Sigma} = \diag\{ \sigma^2_1, \ldots, \sigma^2_n \}$.


Table \ref{funcaoxidr} presents $\xi_r$ and $d_r$ for some power logit models. In Table \ref{funcaoxidr}, $K_2(\cdot)$ denotes the modified Bessel function of third order and index 2, $h_1(\zeta) = \int_1^\infty (\sqrt{x^2-1}/x) \exp\{-\zeta x \} \text{d}x$, $\text{erf}(x) = (2/\sqrt{\pi}) \int_0^z \exp\{-t^2 \} \text{d}t$ is the error function, and
$$
q(\zeta) = \begin{cases}
   (\zeta^2/4) \left( 1 - \zeta^2/4 \right), & \text{for } 0 < \zeta < 2, \\
    2.197543451-1.963510026 \log(\zeta \sqrt{2}) + [\log(\zeta \sqrt{2})]^2, & \text{for } \zeta\geq2.
  \end{cases}
  $$
\begin{table}[!ht]
\centering
\caption{\small $\xi_r$ and $d_r$ for some power logit models.} \label{funcaoxidr}
\begin{tabular}{l|lll}
  \hline
    Model & $\xi_r$& & $d_r$ \\
    \hline
PL-N  & $1$ & & 1 \\
PL-t$_{(\zeta)}$  & $\zeta/(\zeta -2)$, $\zeta>2$ && $(\zeta+1)/(\zeta+3)$ \\
PL-PE$_{(\zeta)}$  & 1& & $\zeta^2 \Gamma\left(1/\zeta \right)^{-2} \Gamma\left(3/\zeta \right)\Gamma\left((2\zeta -1)/\zeta \right)  $, $\zeta> 1/2$   \\
PL-LOI & $\approx$ 0.79569 & & $\approx$ 1.47724 \\
PL-LOII &  $\pi^2/3$ & & $1/3$ \\
PL-slash$_{(\zeta)}$  & $\dfrac{\zeta}{\zeta-1}$, $\zeta>1$ & & $ \approx \dfrac{4 \zeta (\zeta +1/2) [(\zeta+3/2)(\zeta+5/2) + \zeta + 1] }{(\zeta +1)(\zeta + 3/2)^2 (\zeta + 5/2)} $ \\
PL-Hyp$_{(\zeta)}$  & $K_2(\zeta)/[\zeta K_1(\zeta)]$ & & $\zeta^2 h_1(\zeta)/K_1(\zeta)$ \\
PL-SN$_{(\zeta)}$ & $\approx q(\zeta)$ & &$ \approx 2+4/\zeta^2 - (\sqrt{2 \pi}/\zeta) (1- \text{erf}( \sqrt{2}/\zeta )  ) \exp \left\{ 2/\zeta^2 \right\} $  \\
   \hline
\end{tabular}
\end{table}

\begin{figure}[!htb]
\centering
\includegraphics[scale=0.32]{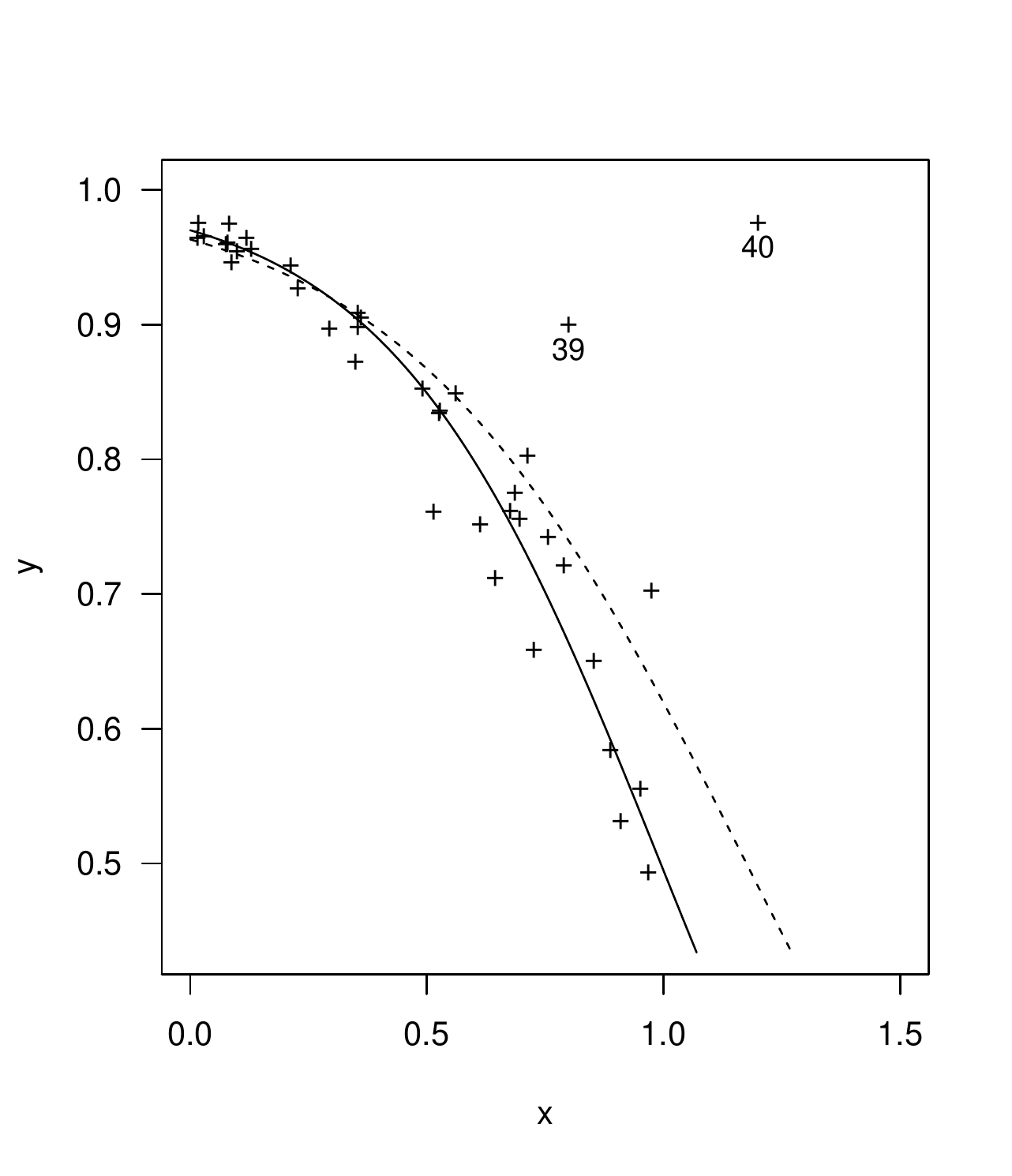}
\includegraphics[scale=0.32]{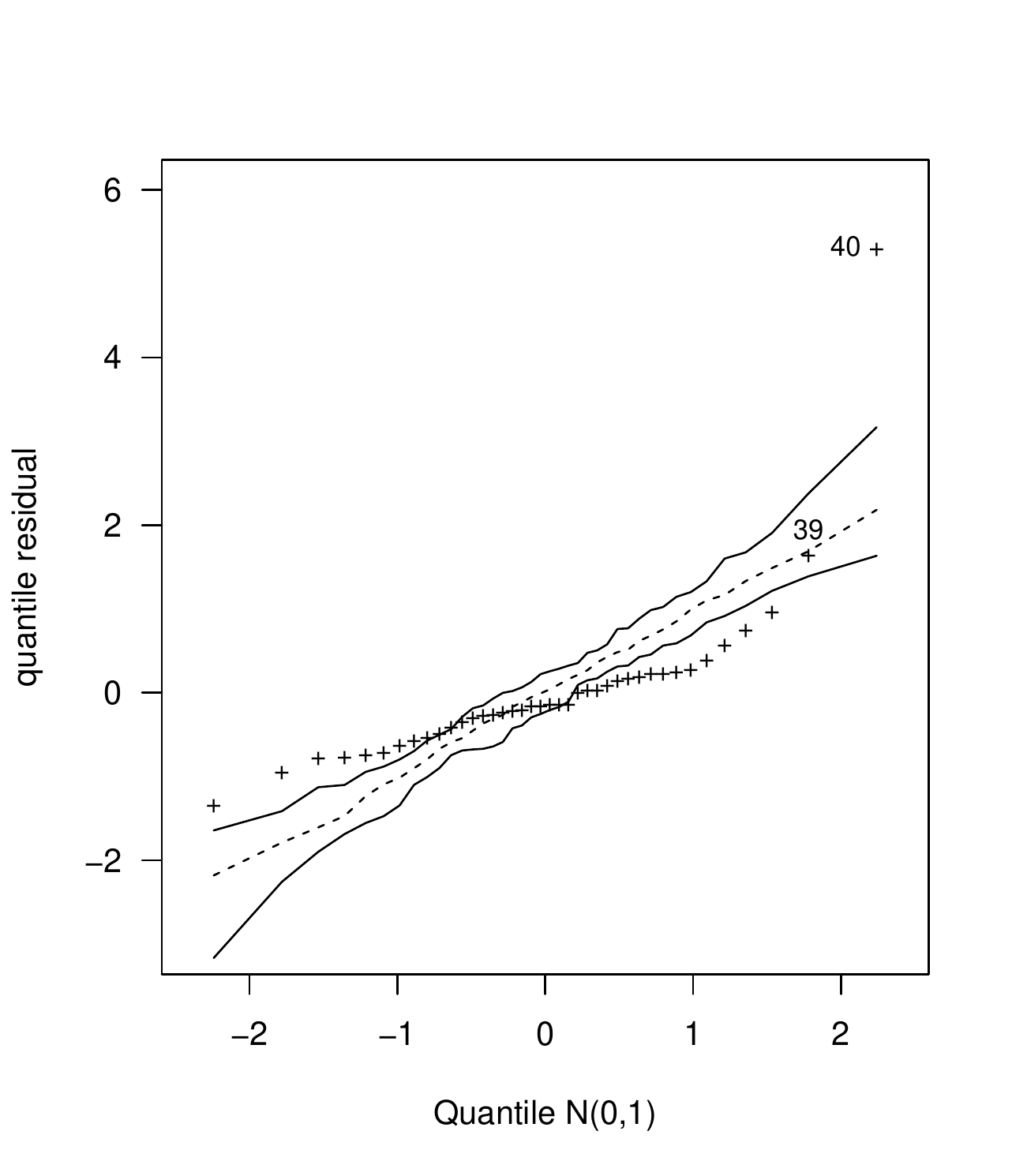}
\includegraphics[scale=0.32]{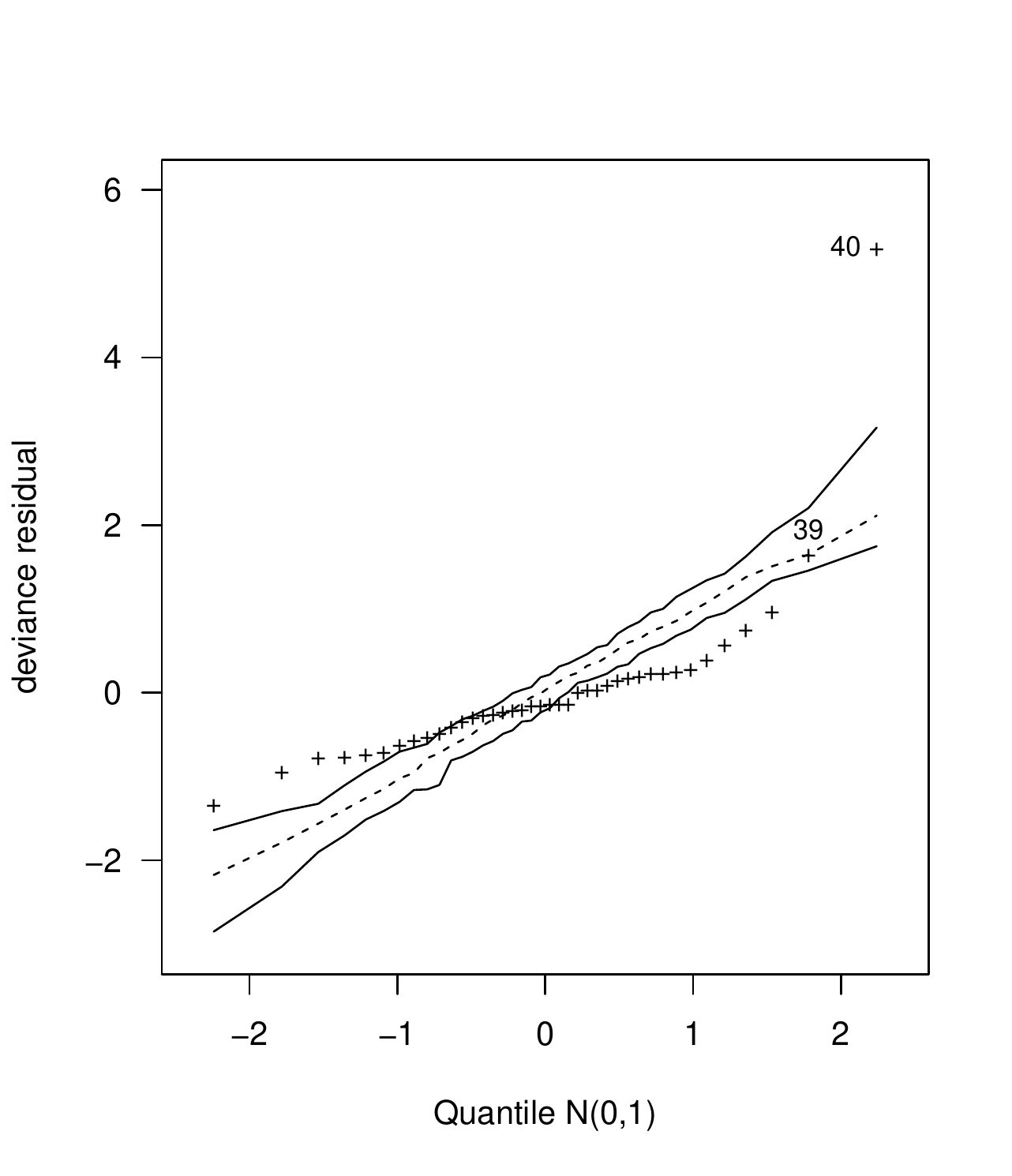}
\includegraphics[scale=0.32]{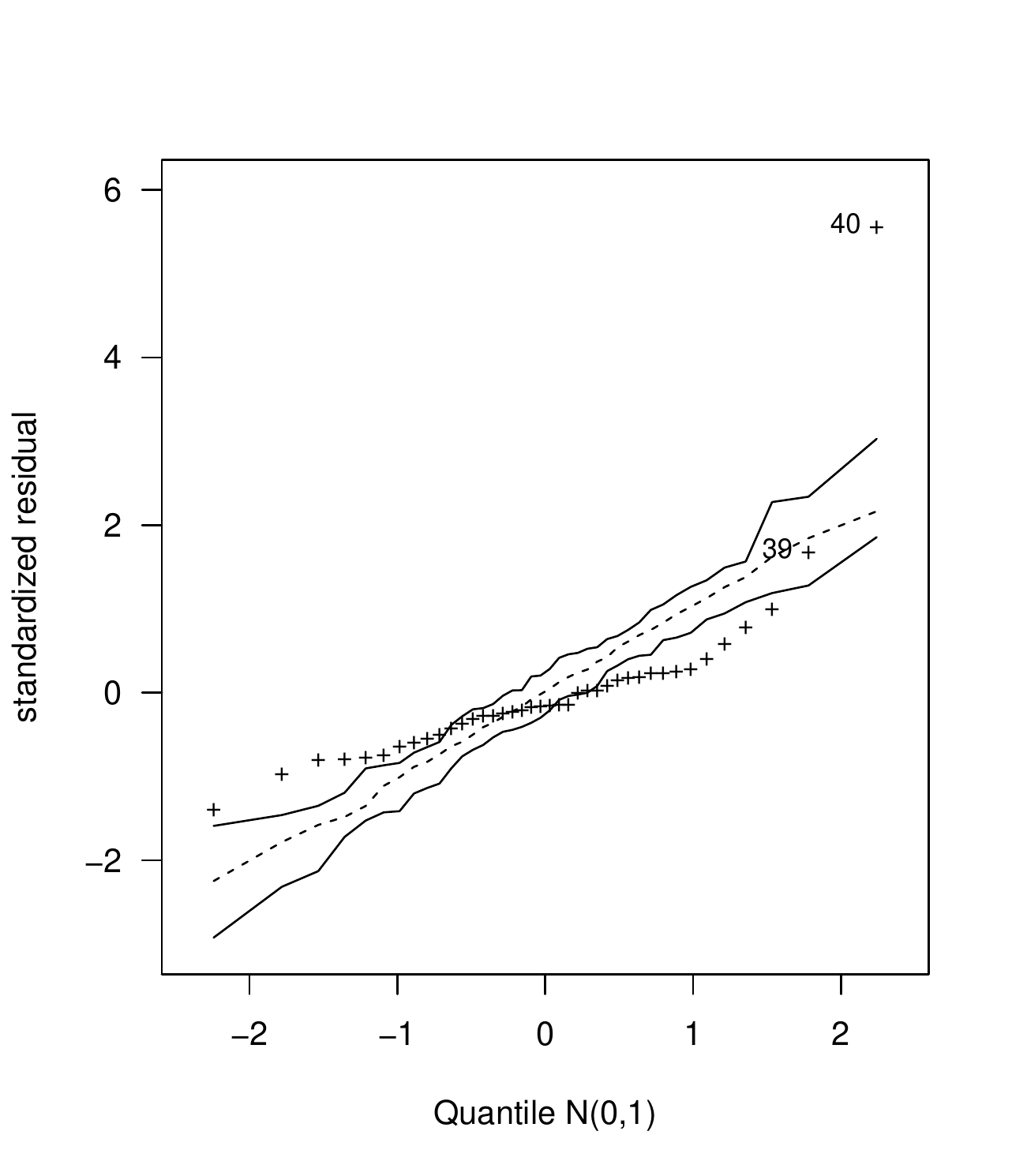}
\includegraphics[scale=0.32]{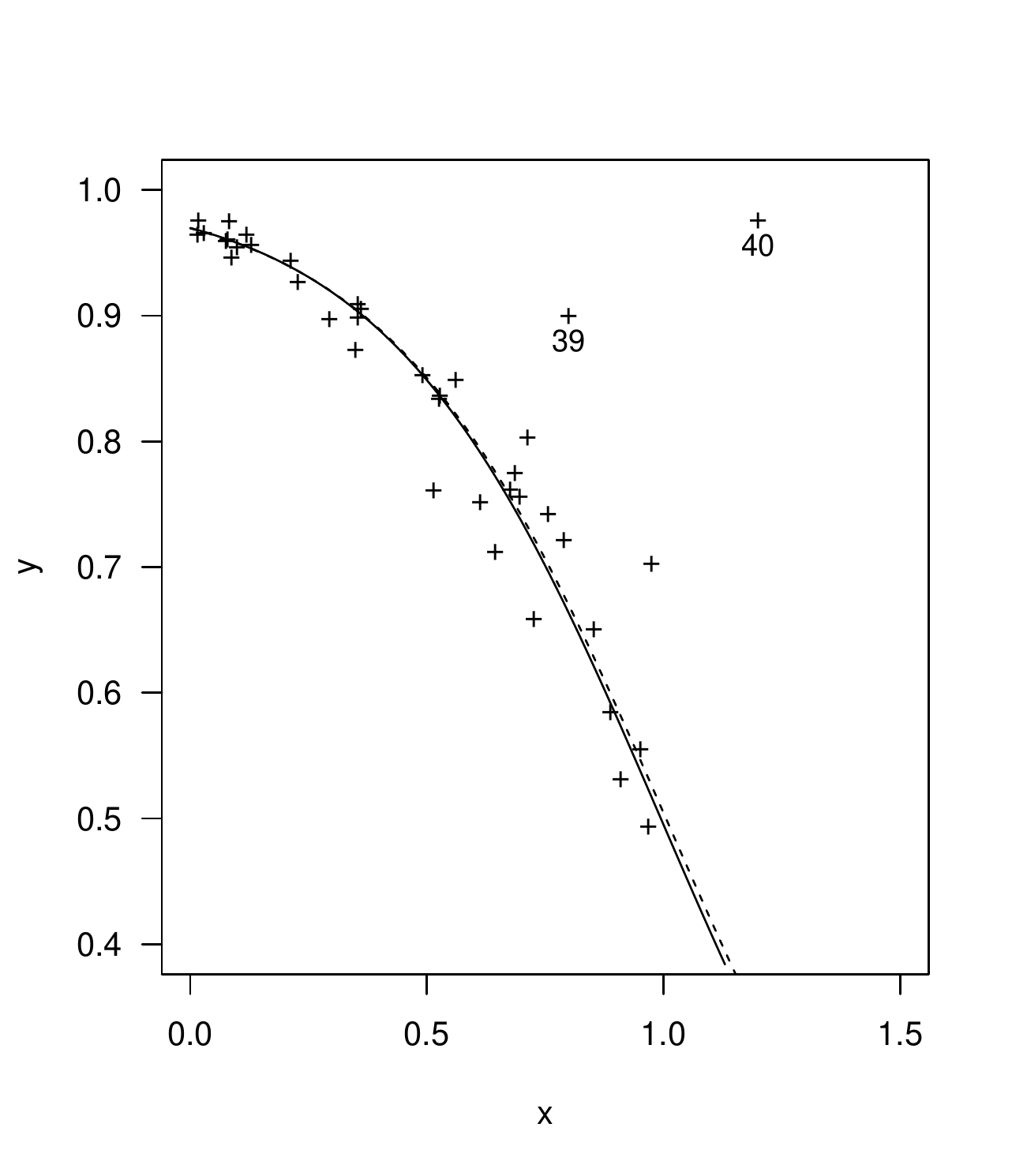}
\includegraphics[scale=0.32]{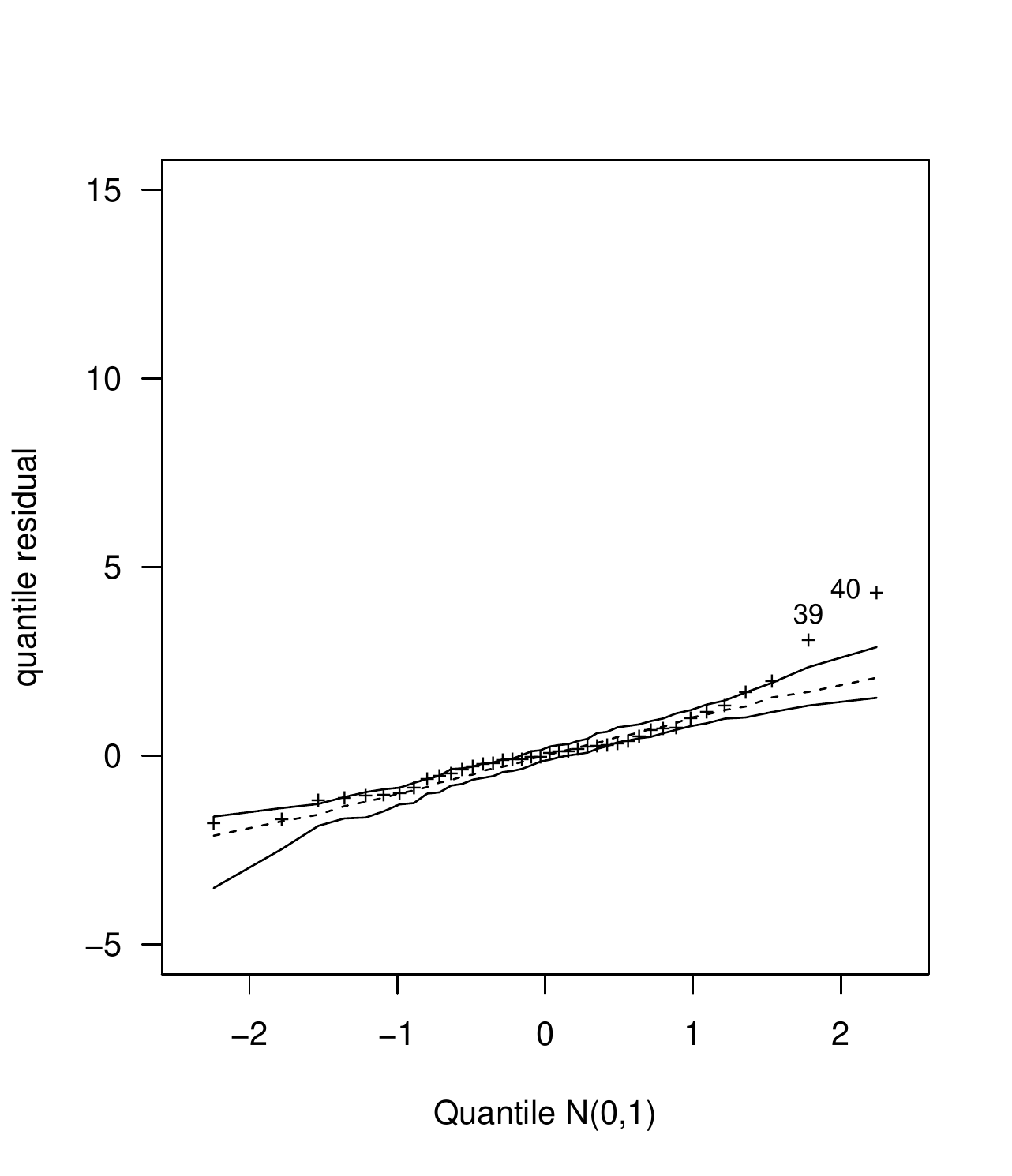}
\includegraphics[scale=0.32]{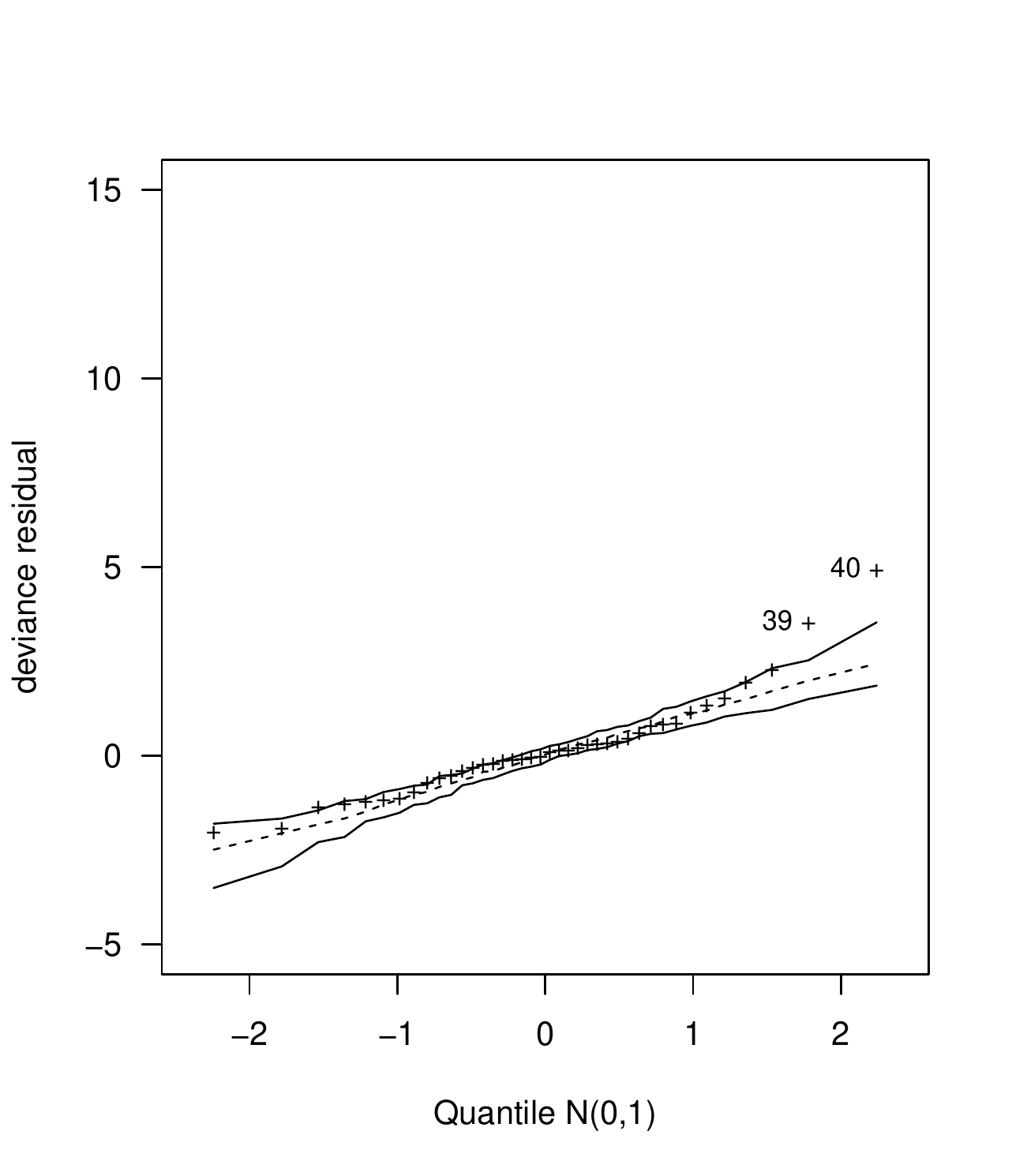}
\includegraphics[scale=0.32]{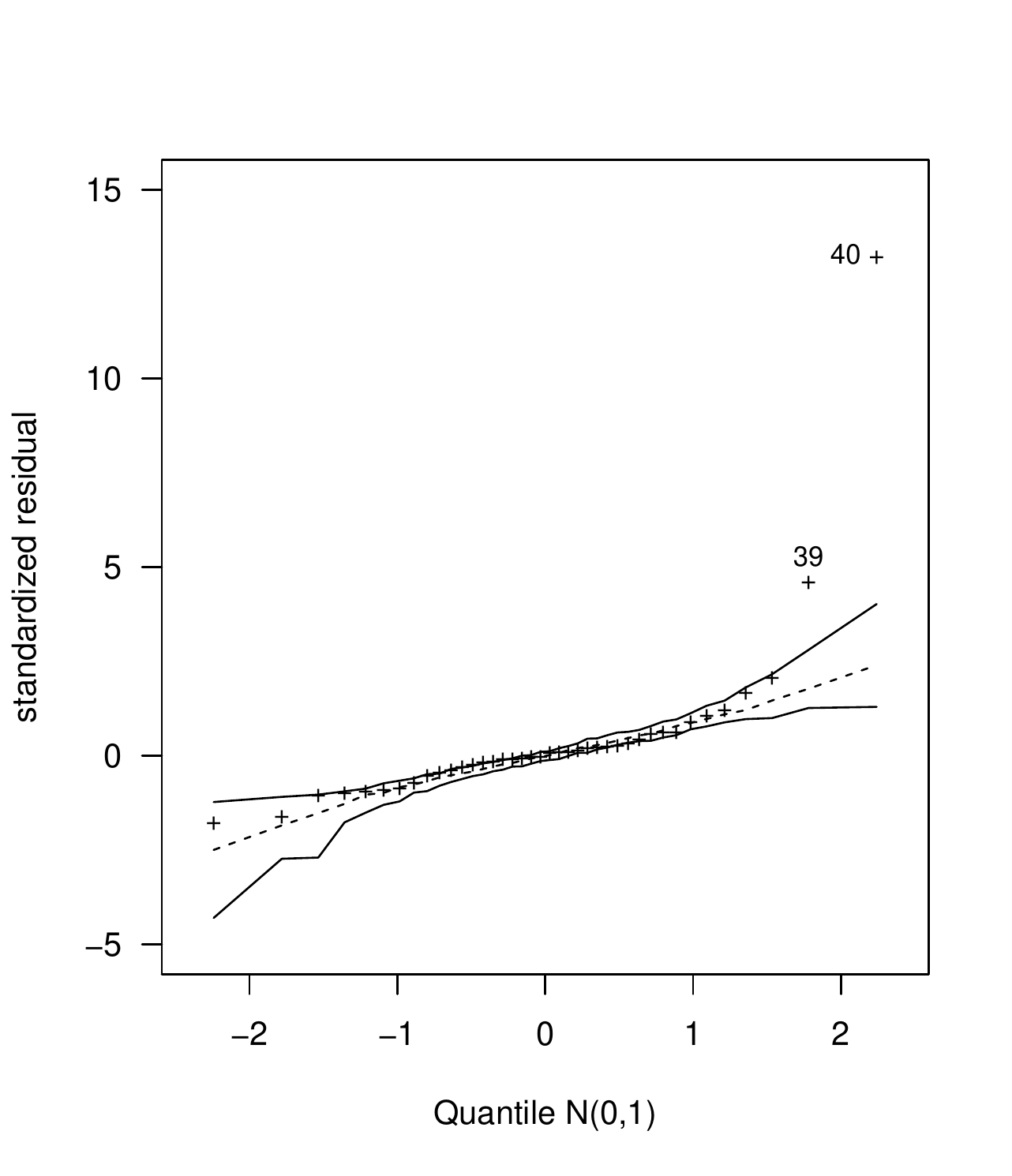}
\caption{\small Scatter plots with the fitted lines for the uncontaminated (solid line) and contaminated (dashed line) data and normal probability plots of the quantile, deviance and standardized residuals with simulated envelopes for the contaminated data; PL-N (top line) and PL-t$_{(5)}$ (bottom line).}\label{Scatterplotappp}
\end{figure}

In order to study the performance of the proposed residuals in detecting outlier observations, we conducted an application in simulated data. We generated a sample of size $n=40$ of a power logit normal regression model with constant dispersion and $\log(\mu_i/(1-\mu_i)) = 3.5 - 3.5 x_i$, for $i=1,\ldots, 40$, where $x_i$ is taken from a uniform distribution on the unit interval, for  $i=1,\ldots, 38$, and $x_{39} = 0.8$, $x_{40} = 1.2$, $\log \sigma = -1.5$ and $\lambda = 0.5$. We contaminated the data as follows: we replaced $y_{39}$ and $y_{40}$ by $y^*_{39}=0.9$ and $y^*_{40} = \max y_i$. These data sets, contaminated and uncontaminated, were analysed using the PL-N and PL-t$_{(5)}$ regression models. The results are presented in Figure \ref{Scatterplotappp}. The plots suggest that the proposed residuals are efficient in identifying atypical observations. For the PL-t$_{(5)}$ regression model, the standardized residual highlights more clearly observation $\#40$, that is a leverage point. Furthermore, we observe that the PL-t$_{(5)}$ regression model is less sensitive to outliers than the PL-N regression model, as expected. This is because $v(z)$ returns small weights for observations with large residuals in the PL-t$_{(\zeta)}$ models, when $\zeta$ is not large. This experiment was performed for other power logit regression models and similar results were observed. 
Other applications in simulated data are presented in the Supplementary Material in order to verify different types of perturbations in the model.


\subsection{Local influence}

The local influence analysis is a diagnostic method proposed by \cite{Cook1986} that evaluates the effect of small perturbations in the data or the model. For the power logit regression models, we are interested in evaluating the influence of small perturbations on the estimation of the parameters associated with the median and dispersion. Thus, we consider $\lambda$ as a fixed constant. In practice, this parameter is replaced by its (penalized) maximum likelihood estimate. Therefore, let $\bm{\theta}=(\bm{\beta}, \bm{\tau})$. To assess the local influence, we use the likelihood displacement $LD_{\bm{\omega}} = 2[\ell (\bm{\widehat{\theta}}) - \ell (\bm{\widehat{\theta}}_\omega)]$, where $\bm{\widehat{\theta}}_\omega$ denotes the mle under the perturbed model. Thus, the normal curvature for $\bm{\theta}$ at the direction $\bm{h}$, $\norm{\bm{h}} = 1 $, is given by $C_{\bm{h}} =  2 \vert \bm{h}^\top \bm{\Delta}^\top \ddot{\bm{\ell}}_{\widehat{\theta} \widehat{\theta}}^{-1} \bm{\Delta}\bm{h} \vert$, where $\ddot{\bm{\ell}}_{\theta \theta} = -\textbf{J}_n(\bm{\beta},\bm{\tau}, \lambda)$ and $\bm{\Delta}$ is a $(p+q) \times n$ matrix that depends on the perturbation scheme and is defined as $\bm{\Delta} = \partial^2 \ell(\bm{\theta} \vert \bm{\omega})/\partial \bm{\theta} \partial \bm{\omega}^\top$ evaluated at $\widehat{\bm{\theta}}$ and $\bm{\omega}_0$, the no perturbation vector. The index graph of $\bm{h}_{\mathrm{max}}$, the eigenvector corresponding to the highest absolute eigenvalue of $-\bm{\Delta}^\top \ddot{\bm{\ell}}_{\widehat{\theta} \widehat{\theta}}^{-1} \bm{\Delta}$ , can reveal the most influential observations in $\bm{\widehat{\theta}}$. Another possibility was proposed by \cite{LesaffreVerbeke1998}, named total local influence, which consists of the construction of the index plot of $C_i =  2 \vert \bm{\Delta}_i^\top \ddot{\bm{\ell}}_{\bm{\theta} \bm{\theta}}^{-1} \bm{\Delta}_i \vert$, where $\bm{\Delta}_i$ is the $i$th column of $\bm{\Delta}$. 

In this work, we consider two perturbation schemes: case-weights and covariate perturbation. The structure of $\bm{\Delta}$ for
each perturbation scheme is given in the following.

\vspace{0.2cm}
\noindent \textit{Case-weights perturbation}: Here, $\bm{\omega} = (\omega_1, \ldots, \omega_n)^\top$ is an $n \times 1$ vector of weights, with $0 \leq w_i \leq 1$, for $i=1, \ldots, n$, and $\bm{\omega}_0 = (1, \ldots, 1)^\top$. The perturbed log-likelihood function is $ \ell (\bm{\theta} \vert \bm{\omega}) = \sum_{i=1}^{n} \omega_i \ell_i(\mu_i, \sigma_i, \lambda)$. For this perturbation scheme, $\bm{\Delta} = ( \bm{\Delta}_{\bm{\beta}}^\top, \bm{\Delta}_{\bm{\tau}}^\top)^\top$, where
\[
\bm{\Delta}_{\bm{\beta}} =  \textbf{X}^{\top} \widehat{\textbf{W}} \widehat{\textbf{T}}_1 \widehat{\textbf{D}}_\beta \quad \text{and} \quad \bm{\Delta}_{\bm{\tau}} = \textbf{S}^{\top} \widehat{\textbf{T}}_2 \widehat{\textbf{D}}_\tau,
\]
where $\textbf{D}_\beta = \diag\{ \mu_1^*, \ldots, \mu_n^*\}$ and $\textbf{D}_\tau = \diag\{ \sigma_1^*, \ldots, \sigma_n^*\}$.

\vspace{0.2cm}
\noindent \textit{Median covariate perturbation}: Now, we perturb a continuous covariate in the median submodel, say $\bm{x}_j$. Following \cite{thomas1989}, we replace $x_{ij}$ by $x_{ij\omega}= x_{ij}+\sigma_x \omega_i$, where $\sigma_x $ is the sample standard deviation of $\bm{x}_j$. 
The perturbed linear predictor for the median submodel is
\begin{align*}
\eta_{1i\omega}=\beta_1 x_{i1} + \cdots + \beta_j (x_{ij}+\sigma_x \omega_i) + \cdots + \beta_p x_{ip} = \bm{x}_{i\omega}^\top \bm{\beta},
\end{align*}
\sloppy where $\bm{x}_{i\omega} = (x_{i1}, \ldots, x_{ij\omega}, \ldots, x_{ip})^\top$. The perturbed log-likelihood function is $ \ell (\bm{\theta} \vert \bm{\omega}) = \sum_{i=1}^{n} \ell_i(\mu_{i\omega}, \sigma_i, \lambda)$, where $\mu_{i\omega} = d_1^{-1}(\eta_{1i\omega})$ and $\bm{\omega}_0 = (0, \ldots, 0)^\top$. Assuming that $\textbf{X}$ and $\textbf{S}$ are functionally independent, the components of $\bm{\Delta}$ are 
\[
\bm{\Delta}_{\bm{\beta}} =  - \sigma_x \widehat{\beta}_j \textbf{X}^\top \widehat{\textbf{W}}_1 \widehat{\textbf{T}}_1 + \sigma_x \bm{c}_{\bm{\beta}}^j \widehat{\bm{\mu}}^{*\top} \widehat{\textbf{W}}\widehat{\textbf{T}}_1 \quad \text{and} \quad \bm{\Delta}_{\bm{\tau}} = - \sigma_x \widehat{\beta}_j \textbf{S}^\top \widehat{\textbf{W}}_4 \widehat{\textbf{T}}_1 \widehat{\textbf{T}}_2,
\]
where $\bm{c}_{\bm{\beta}}^j$ denotes a $p \times 1$ vector with one at the $j$th position and zero elsewhere, and $\widehat{\beta}_j$ denotes the $j$th element of $\bm{\widehat{\beta}}$.

\vspace{0.2cm}
\noindent \textit{Dispersion covariate perturbation}: Here, we perturb a continuous covariate in the dispersion submodel, say $\bm{s}_k$. We replace $s_{ik}$ by $s_{ik\omega}= s_{ik}+\sigma_s \omega_i$, where $\sigma_s $ is the sample standard deviation of $\bm{s}_k$. 
The perturbed linear predictor for the dispersion submodel is
\begin{align*}
\eta_{2i\omega}=\tau_1 s_{i1} + \cdots + \tau_k (s_{ik}+\sigma_s \omega_i) + \cdots + \tau_q s_{iq} = \bm{s}_{i\omega}^\top \bm{\tau},
\end{align*}
\sloppy where $\bm{s}_{i\omega} = (s_{i1}, \ldots, s_{ik\omega}, \ldots, s_{iq})^\top$. The perturbed log-likelihood function is $\ell (\bm{\theta} \vert \bm{\omega}) = \sum_{i=1}^{n} \ell_i(\mu_{i}, \sigma_{i\omega}, \lambda)$, where $\sigma_{i\omega} = d_2^{-1}(\eta_{2i\omega})$, and $\bm{\omega}_0 = (0, \ldots, 0)^\top$. Assuming that $\textbf{X}$ and $\textbf{S}$ are functionally independent, the components of $\bm{\Delta}$ are 
\[
\bm{\Delta}_{\bm{\beta}} =  - \sigma_s \widehat{\tau}_k \textbf{X}^\top \widehat{\textbf{W}}_4 \widehat{\textbf{T}}_1 \widehat{\textbf{T}}_2 \quad \text{and} \quad \bm{\Delta}_{\bm{\tau}} = - \sigma_s \widehat{\tau}_k \textbf{S}^\top \widehat{\textbf{W}}_2 \widehat{\textbf{T}}_2 + \sigma_s \bm{c}_{\bm{\tau}}^k \widehat{\bm{\sigma}}^{*\top} \widehat{\textbf{T}}_2,
\]
where $\bm{c}_{\bm{\tau}}^k$  denotes a $q \times 1$ vector with one at the $k$th position and zero elsewhere, and $\widehat{\tau}_k$ denotes the $k$th element of $\bm{\widehat{\tau}}$.

\vspace{0.2cm}
\noindent \textit{Simultaneous (median and dispersion) covariate perturbation}: Now, we perturb a continuous covariate in the median and in the dispersion submodels, say $\bm{x}_j$ and $\bm{s}_k$. We replace $x_{ij}$ and $s_{ik}$ by $x_{ij\omega} = x_{ij}+\sigma_x \omega_i$ and $s_{ik\omega} = s_{ik}+\sigma_s \omega_i$, respectively. Then
\begin{align*}
\begin{split}
\eta_{1i\omega}& =\beta_1 x_{i1} + \cdots + \beta_j (x_{ij}+\sigma_x \omega_i) + \cdots + \beta_p x_{ip} = \bm{x}_{i\omega}^\top \bm{\beta},\\
\eta_{2i\omega}& =\tau_1 s_{i1} + \cdots + \tau_k (s_{ik}+\sigma_s \omega_i) + \cdots + \tau_q s_{iq} = \bm{s}_{i\omega}^\top \bm{\tau},
\end{split}
\end{align*}
The perturbed log-likelihood function is $\ell (\bm{\theta} \vert \bm{\omega}) = \sum_{i=1}^{n} \ell_i(\mu_{i\omega}, \sigma_{i\omega}, \lambda)$, and $\bm{\omega}_0 = (0, \ldots, 0)^\top$ and 
\[
\bm{\Delta}= \left[
\begin{array}{ccc}
- \sigma_x \widehat{\beta}_j \textbf{X}^\top \widehat{\textbf{W}}_1 \widehat{\textbf{T}}_1 + \sigma_x \bm{c}_{\bm{\beta}}^j \widehat{\bm{\mu}}^{*\top} \widehat{\textbf{W}} \widehat{\textbf{T}}_1 - \sigma_s \widehat{\tau}_k \textbf{X}^\top \widehat{\textbf{W}}_4 \widehat{\textbf{T}}_1 \widehat{\textbf{T}}_2 \\
- \sigma_s \widehat{\tau}_k \textbf{S}^\top \widehat{\textbf{W}}_2 \widehat{\textbf{T}}_2 + \sigma_s \bm{c}_{\bm{\tau}}^k \widehat{\bm{\sigma}}^{*\top} \widehat{\textbf{T}}_2 - \sigma_x \widehat{\beta}_j \textbf{S}^\top \widehat{\textbf{W}}_4 \widehat{\textbf{T}}_1 \widehat{\textbf{T}}_2. \\
\end{array}
\right].
\]
 
\subsection{Generalized leverage} 

The generalized leverage measure is defined by $\text{GL}_{ij} = \partial \widehat{y_i}/\partial y_j$ and reflects the rate of instantaneous change in the $i$th predicted value, when the $j$th response variable is increased by an infinitesimal amount \citep{Wei1998}. The idea is to use $\text{GL}_{ii}$ to evaluate the influence of $y_i$ on its own predicted value. A high leverage value suggests that the observation may be an outlier on the covariates.  \cite{Wei1998} showed that the generalized leverage matrix can be expressed as
\begin{equation*}\label{alavancageneralizada}
\text{GL}(\bm{\theta}) = \dot{\textbf{L}}_{\bm{\theta}} \textbf{J}_n^{-1} \ddot{\textbf{L}}_{\bm{\theta} \bm{y}},
\end{equation*}
where $\dot{\textbf{L}}_{\bm{\theta}} = \partial \bm{\mu}/ \partial \bm{\theta}$ and $\ddot{\textbf{L}}_{\bm{\theta} \bm{y}}  = \partial \ell(\bm{\theta})/ \partial \bm{\theta} \partial \bm{y}^\top$, with $\bm{\theta}$ evaluated at $\widehat{\bm{\theta}}$. For the power logit regression models, we consider the penalized maximum likelihood estimator, which has better small sample properties. After some algebra, we can show that $\dot{\textbf{L}}_{\bm{\theta}} = [\textbf{T}_1 \textbf{X}, ~~ \bm{0}_{n, q}, ~~ \bm{0}_{n, 1}]$ and 
\[
\ddot{\textbf{L}}_{\bm{\theta} \bm{y}} =
\begin{bmatrix}
\textbf{X}^{\top} \textbf{T}_1 \textbf{D}_\beta \textbf{D}_y \textbf{W}_\beta\\
\textbf{S}^{\top}\textbf{T}_2 \textbf{D}_y \textbf{W}_\tau\\
\bm{1}_n^{\top} \textbf{D}_y \textbf{W}_\lambda
\end{bmatrix},
\]
where $\bm{0}_{a, b}$ is an $a \times b$ matrix of zeros, $\textbf{D}_\beta = \diag\{ \mu_1^*, \ldots, \mu_n^*\}$, $\textbf{D}_y = \diag\{ y_1^*, \ldots, y_n^*\}$, $\textbf{W}_\beta = \diag\{ v(z_1) + z_1 v'(z_1), \ldots, v(z_n) + z_n v'(z_n)\}$, $\textbf{W}_\tau = \diag\{ \sigma_1^{-1} z_1 [2 v(z_1) + z_1 v'(z_1)], \ldots, \sigma_n^{-1} z_n [2 v(z_n) + z_n v'(z_n)]\}$, $\textbf{W}_\lambda = \diag\{ w^\lambda_1, \ldots, w^\lambda_n\}$, $y_i^* = \lambda/[\sigma_i y_i (1- y_i^\lambda)]$, and 
\begin{align*}
w^\lambda_i & = \dfrac{\sigma_j y_i^\lambda (1+\lambda \log y_j - y_j^\lambda)}{\lambda (1-y_j^\lambda)} - \dfrac{1}{\sigma_j} \left( \dfrac{\log y_j}{ 1-y_j^\lambda} - \dfrac{\log \mu_j}{ 1-\mu_j^\lambda} \right) [v(z_j) + z_j v'(z_j)] \\
 & - z_j v(z_j) \dfrac{y_i^\lambda (\lambda \log y_j - 1) + 1}{\lambda (1-y_j^\lambda)},
\end{align*}
for $i=1, \ldots, n$. The index plot of the diagonal elements of $\text{GL}(\widetilde{\bm{\theta}})$ may be used to assess observations with high influence on their own predicted values.

\section{R implementation: \texttt{PLreg} package} \label{package}

To facilitate the practical use of the proposed models, we developed the new R package \texttt{PLreg}. It allows fitting power logit regression models, in which the density generator may be normal, Student-t, power exponential, slash, hyperbolic, sinh-normal, or type II logistic. Diagnostic tools associated with the fitted model, such as the residuals, local influence measures, leverage measures, and goodness-of-fit statistics, are implemented. The estimation process follows the maximum likelihood approach and, currently, the package supports two types of estimators: the usual maximum likelihood estimator and the penalized maximum likelihood estimator. The skewness parameter $\lambda$ may be fixed, so the package also allows fitting GJS ($\lambda=1$) and log-log ($\lambda=0$) regression models.

The main function of the \texttt{PLreg} package is \texttt{PLreg()}, which follows the standard approach for implementing regression models in \texttt{R}. Once the fitting process has been accomplished, an object of the \texttt{S2} class ``\texttt{PLreg}'' is produced for which several methods are available. The arguments of the \texttt{PLreg()} function are similar to those of functions in other packages for regression models in \texttt{R}, such as the \texttt{betareg()} function. 

The package \texttt{PLreg} and the codes for the applications in the next section are available at the GitHub repository, respectively, at \url{https://github.com/ffqueiroz/PLreg} and \url{https://github.com/ffqueiroz/PowerLogitRegression}.

\section{Applications} \label{app}

\subsection{Employment in non-agricultural sectors}\label{agricul_app}

The data set refers to the employment in non-agricultural sectors in $200$ randomly selected Brazilian municipal districts of the state of S\~ao Paulo in the year $2010$. The data were extracted from the Atlas of Brazil Human Development database, available at \url{https://www.pnud.org.br/atlas/}. The response variable is the proportion of people aged 18 or over who are employed in non-agricultural activities ($y$). 

We fitted different distributions in the power logit class, including some GJS distributions. The extra parameter, if any, was chosen as described in Section \ref{upsilom}. For the sake of comparison we also fitted the beta distribution because it is the most employed distribution to model continuous proportions. The parameterization of the beta law uses the mean and a precision parameter \citep{FerrariCribariNeto2004}, denoted here by $\mu$ and $\sigma$. The results are presented in Table \ref{AP11AC1} \footnote{The $\Upsilon$ measure for the beta distribution is computed similarly to \eqref{upsilonmeasure}.}. The fitted models with the smallest values of $\Upsilon$ and AIC are those of the power logit distributions; for instance, the $\Upsilon$ value of the GJS-SN$_{(0.53)}$ model is almost four times that of the PL-SN$_{(0.97)}$ model. Based on these measures, all the PL models fit the data similarly and better than the beta and the GJS distributions. In addition, the estimates of $\lambda$ are large (around 9), indicating that the GJS distributions ($\lambda=1$) may not be suitable to represent the distribution of the data.

\begin{table}[ht]
\centering
\scriptsize
\caption{\small Estimates, standard errors, asymptotic 95\% confidence intervals for $\mu$, $\Upsilon$ measure, and AIC for the beta, GJS and PL models - Employment in non-agricultural sectors data.} \label{AP11AC1}
\begin{tabular}{ccccccc}
\hline
& \multicolumn{3}{c}{Est. (s.e.)} &  &  &  \\ 
& $\mu$ & $\sigma$ & $\lambda$ & CI & $\Upsilon$ & AIC \\ 
\hline
beta                & $0.82 ~(0.07)$ & $5.43 ~ (0.54)$ & $            $ & $(0.80, 0.84)$ & $0.15$ & $-286.00$ \\ 
GJS-N               & $0.88 ~(0.01)$ & $1.42 ~ (0.07)$ & $            $ & $(0.86, 0.90)$ & $0.22$ & $-255.20$ \\ 
GJS-t$_{(4.56)}$    & $0.85 ~(0.01)$ & $1.19 ~ (0.08)$ & $            $ & $(0.83, 0.88)$ & $0.21$ & $-248.63$ \\ 
GJS-PE$_{(1.44)}$   & $0.85 ~(0.01)$ & $1.46 ~ (0.09)$ & $            $ & $(0.83, 0.88)$ & $0.21$ & $-249.88$ \\ 
GJS-SN$_{(0.53)}$   & $0.88 ~(0.01)$ & $5.54 ~ (0.26)$ & $            $ & $(0.86, 0.90)$ & $0.23$ & $-255.05$ \\ 
PL-N                & $0.84 ~(0.01)$ & $2.43 ~ (0.12)$ & $9.41 ~ (0.75)$ & $(0.81, 0.86)$ & $0.07$ & $-306.54$ \\ 
PL-t$_{(100)}$     & $0.84 ~(0.01)$ & $2.41 ~ (0.12)$ & $9.44 ~ (0.76)$ & $(0.81, 0.86)$ & $0.07$ & $-305.98$ \\ 
PL-PE$_{(2.69)}$    & $0.84 ~(0.01)$ & $2.38 ~ (0.10)$ & $9.16 ~ (0.65)$ & $(0.81, 0.86)$ & $0.05$ & $-310.81$ \\ 
PL-SN$_{(0.97)}$    & $0.84 ~(0.01)$ & $5.31 ~ (0.23)$ & $9.06 ~ (0.63)$ & $(0.81, 0.86)$ & $0.06$ & $-310.63$ \\ 
\hline
\end{tabular}
\end{table}

\begin{figure}[!htb]
\centering
\subfigure[][]{\includegraphics[scale=0.43]{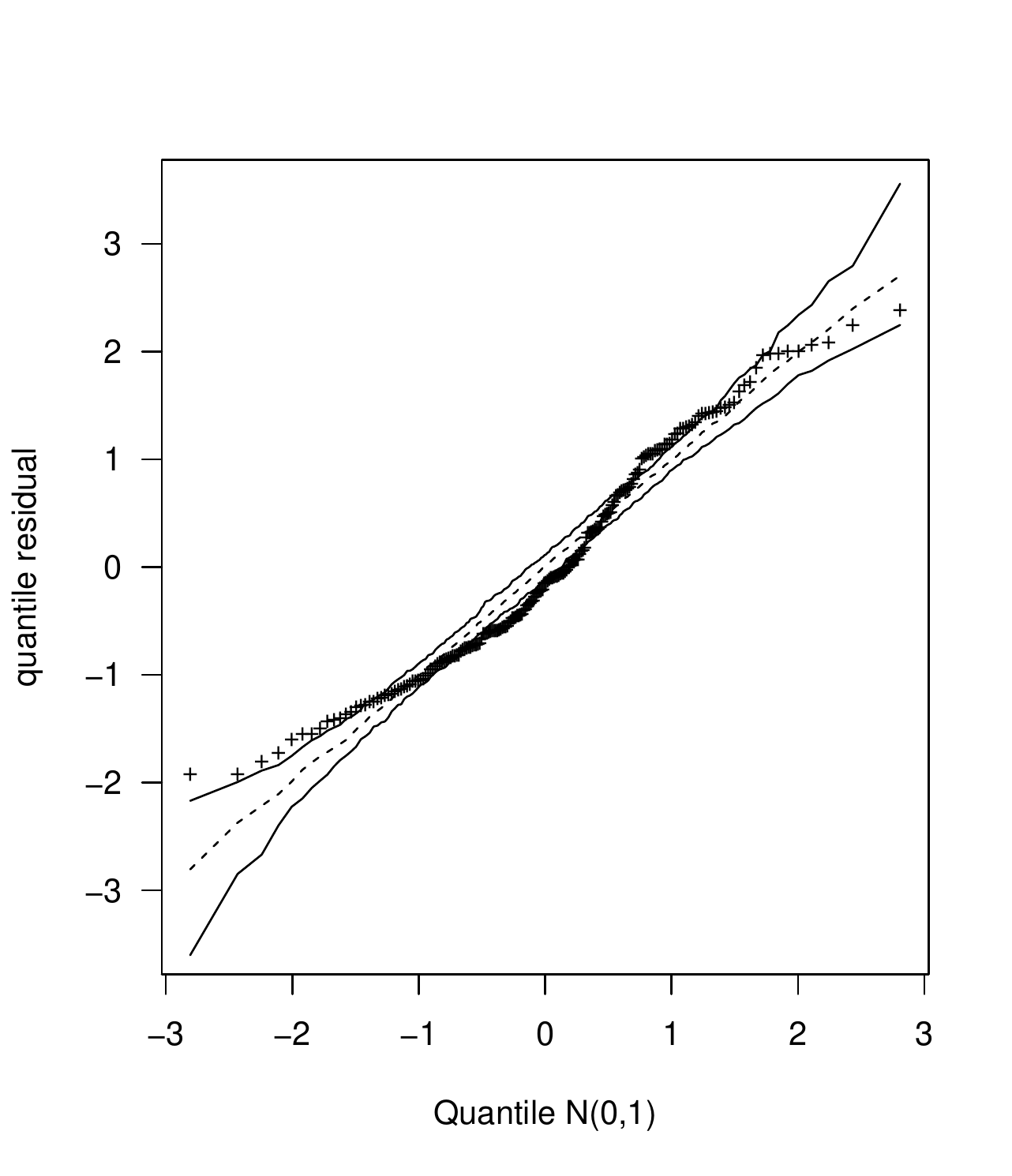}\label{AP1_env_beta}}
\subfigure[][]{\includegraphics[scale=0.43]{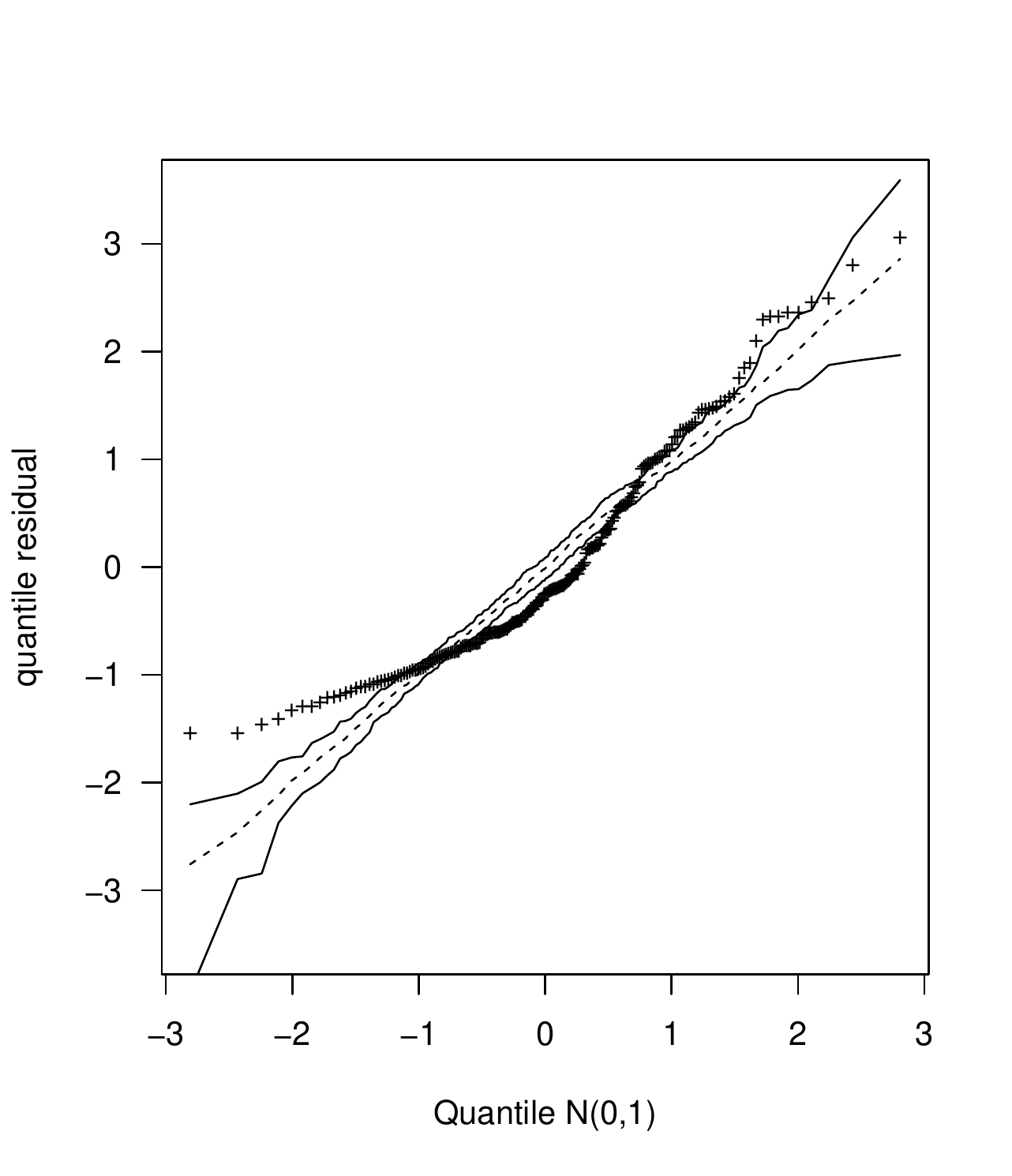}\label{AP1_env_GJSN}}
\subfigure[][]{\includegraphics[scale=0.43]{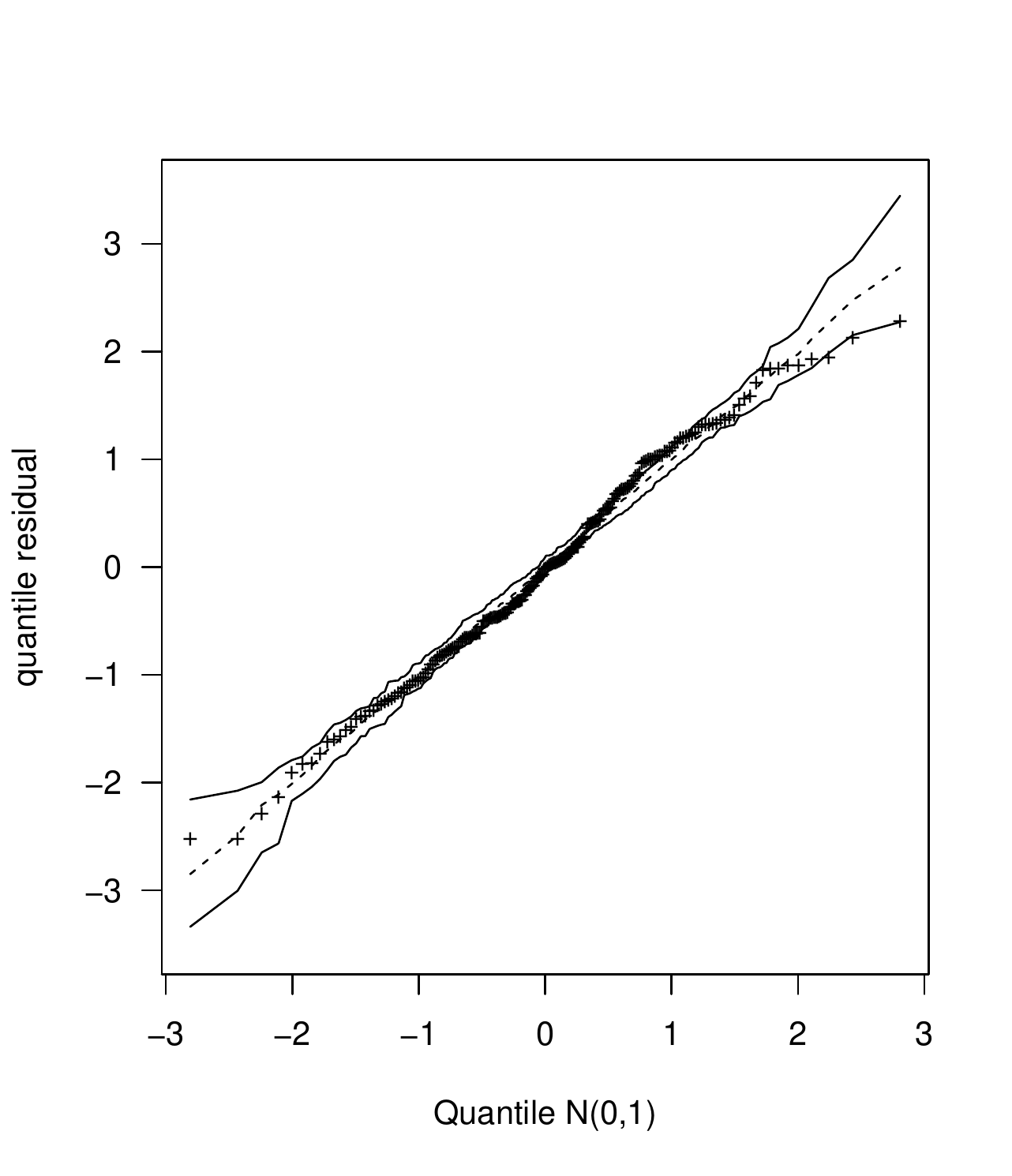}\label{AP1_env_PLN}}
\subfigure[][]{\includegraphics[scale=0.43]{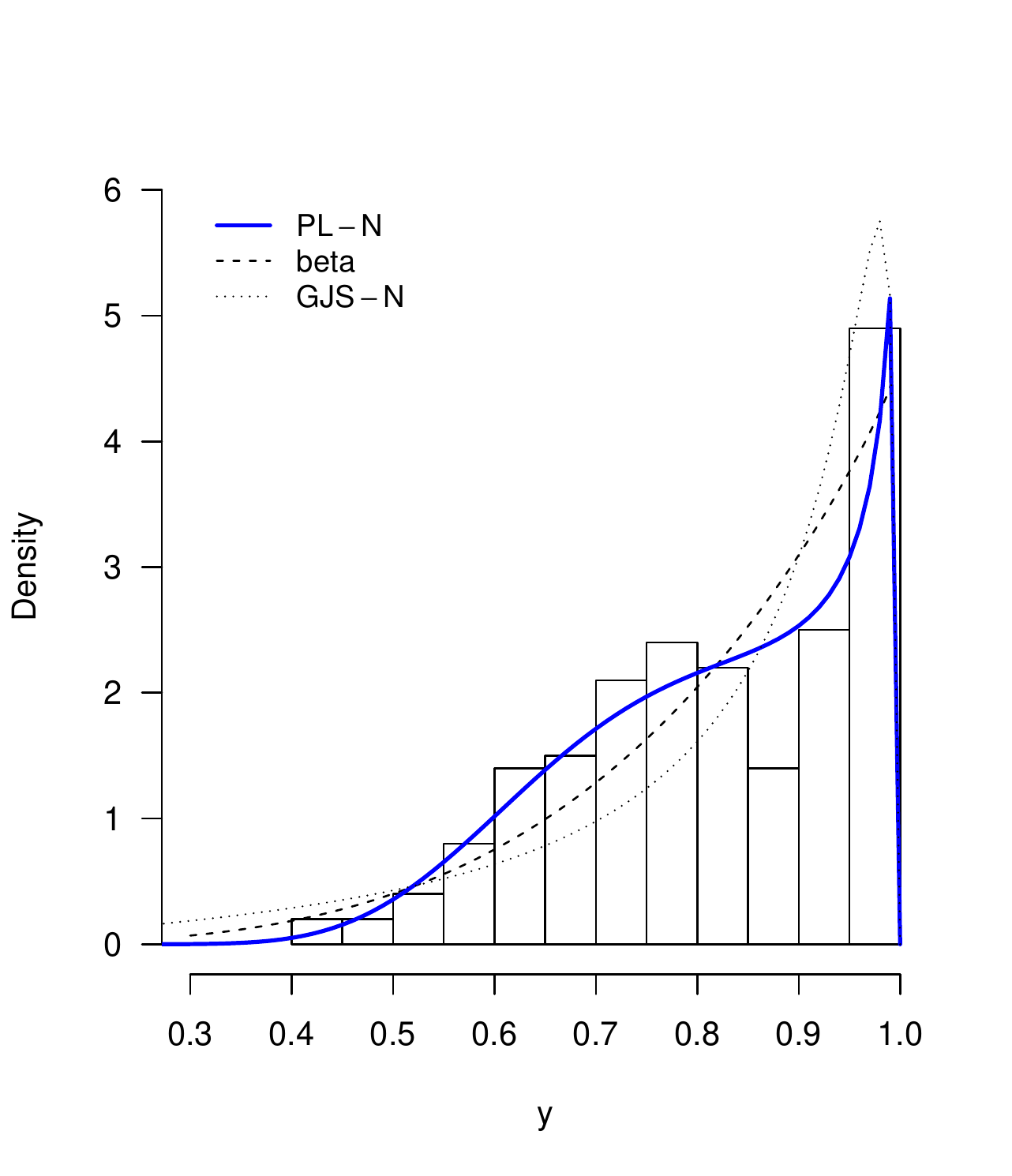}\label{AP1_hist}}
\subfigure[][]{\includegraphics[scale=0.43]{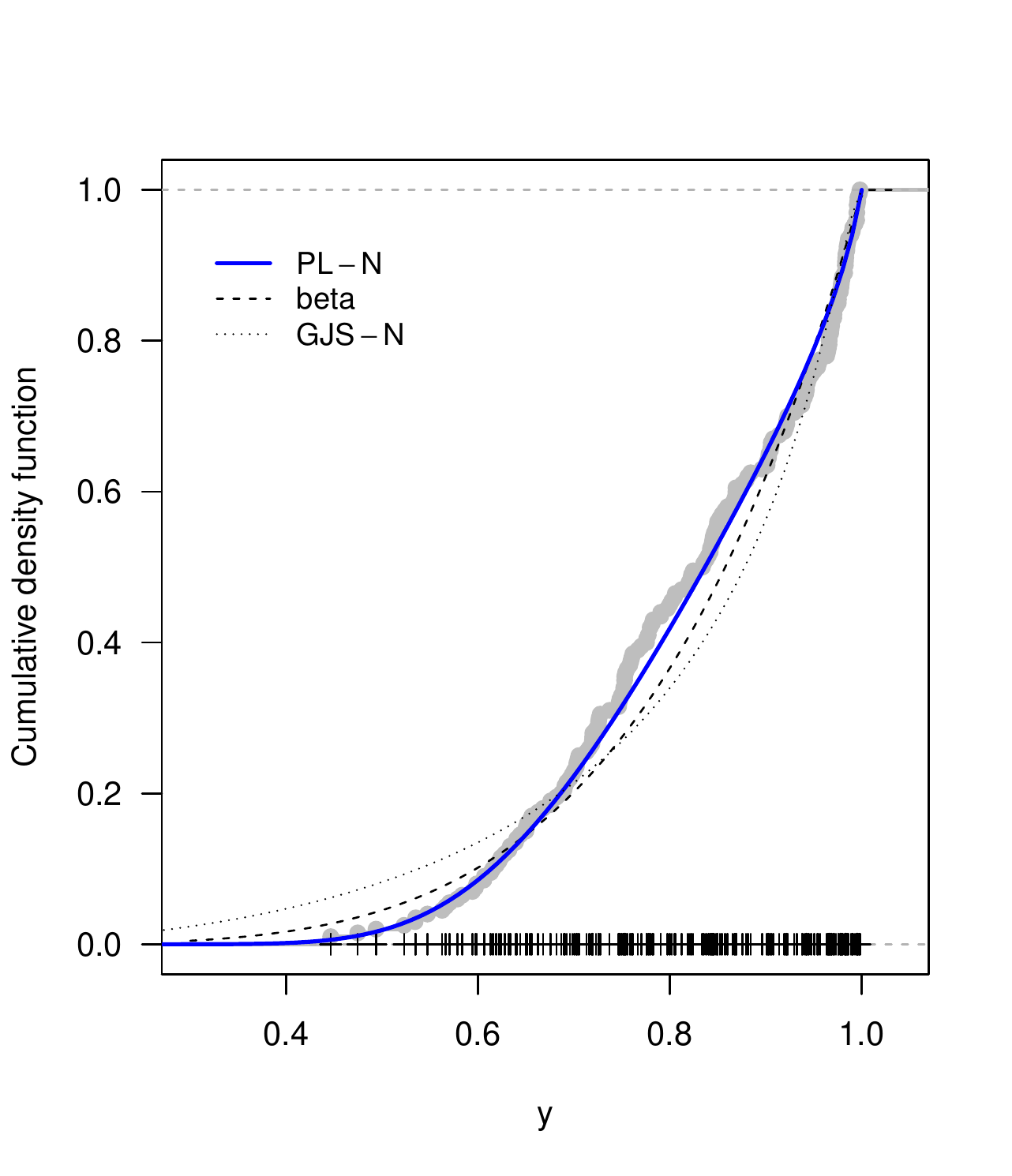}\label{AP1_cdf}}
\subfigure[][]{\includegraphics[scale=0.43]{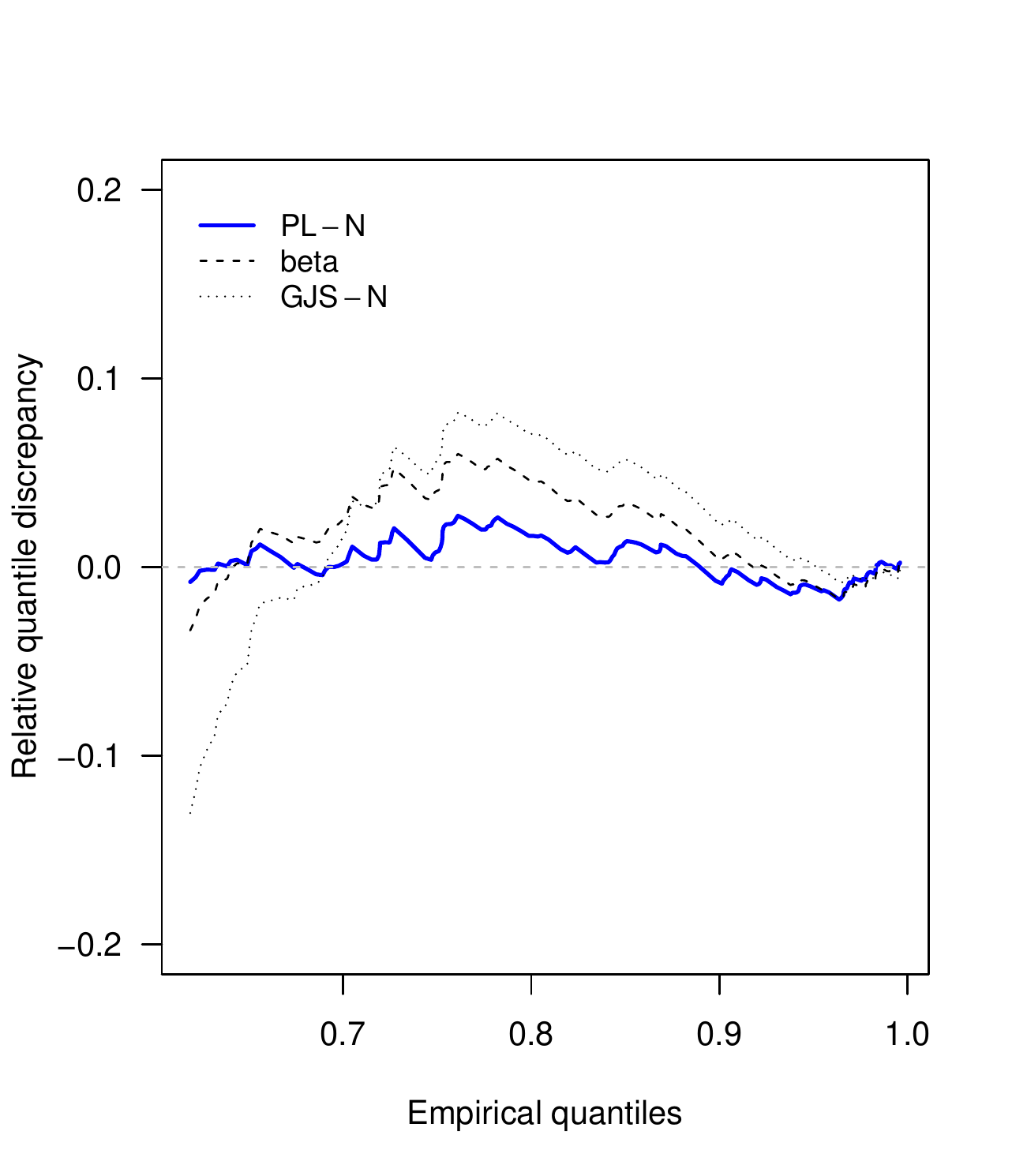}\label{AP1_dqr}}
\caption{\small Normal probability plots of the quantile residual for the beta (a), GJS-N (b) and PL-N (c) models, histogram of $y$ with fitted pdfs (d), fitted cdfs (e) and quantile relative discrepancies (f) for the beta, GJS-N and PL-N models - Employment in non-agricultural sectors data.}\label{AP11_diag}
\end{figure}

Figure \ref{AP11_diag} displays some diagnostic plots. Since the $\Upsilon$ measure and the AIC are similar for all of the power logit distributions, we chose the PL-N distribution to summarize the data, because of its simplicity. Figures \ref{AP1_env_beta}-\ref{AP1_env_PLN} present the normal probability plots with simulated envelope of the quantile residual for the beta, GJS-N and PL-N. These plots show that the  beta and GJS-N models do not fit the data well, but the PL-N distribution seems to be suitable for modeling the data. This is confirmed in Figures \ref{AP1_hist}-\ref{AP1_dqr}\footnote{The relative quantile discrepancy in Figure \ref{AP1_dqr} is the difference between estimated quantiles and empirical quantiles divided by the latter.}. 
Diagnostic plots for the other GJS distributions were done and showed similar behavior. Based on the PL-N model, the estimated median proportion of adults who are employed in non-agricultural sectors in the cities of the state of S\~ao Paulo is $0.84$. 
The 95\% confidence interval is $(0.81, 0.86)$. 

\subsection{Firm cost data}

This application is from a questionnaire sent to risk managers of large corporations in the USA. The data set was introduced by \cite{schmit1990} (see also \cite{RibeiroFerrari2020, gomezSordo2014}) and is available in the personal web page of Professor E. Frees (Wisconsin School of Business Research) \footnote[1]{Available at: \url{https://instruction.bus.wisc.edu/jfrees/jfreesbooks/Regression\%20Modeling/BookWebDec2010/CSVData/RiskSurvey.csv}.}. The response variable is \texttt{firmcost}, defined as premiums plus uninsured losses as a percentage of the total assets and is a measure of the firm's risk management cost effectiveness. A good risk management performance corresponds to smaller values of \texttt{firmcost}. Following \cite{RibeiroFerrari2020}, we considered two covariates: \texttt{sizelog}, the logarithm of total assets, and \texttt{indcost}, a measure of the firm's industry risk. The data set contains information on 73 firms. The postulated initial model is a power logit slash regression model with 
\begin{align*}
\begin{split}
\log \left( \dfrac{\mu_i}{1-\mu_i} \right) &= \beta_1 + \beta_2 \times \texttt{sizelog}_i  + \beta_3 \times \texttt{indcost}_i, \\
\log \sigma_i &= \tau_1 + \tau_2 \times \texttt{sizelog}_i + \tau_3 \times \texttt{indcost}_i, 
\end{split} 
\end{align*}
for $i=1, \ldots, 73$. The results are presented in Table \ref{tab:AP2_est_vary}. Note that all the covariates are statistically significant for the median submodel. For the dispersion submodel, both covariates are not significant. Removing each covariate at a time, the other covariate remains non significant, indicating that a constant dispersion model should be considered; see Table \ref{tab:AP2_est_const}. The fitted model indicates that the median \texttt{firmcost} is positively related with \texttt{indcost} and negatively related with \texttt{sizelog}. The estimate of $\lambda$ is $1.788$ and the asymptotic 95\% confidence interval is $(1.441, 2.135)$, indicating that the GJS-slash regression model ($\lambda=1$) may not be suitable.


\begin{table}[!ht]
    \centering
		\scriptsize
\begin{minipage}[t]{0.48\linewidth}\centering
\caption{\small Estimates, standard errors, and p-values for the PL-slash regression model with varying dispersion - Firm cost data.}\label{tab:AP2_est_vary}
\begin{tabular}{rrrr}
\hline
& Est. & s.e. & p-value \\
\hline
$\mu$                 &           &       &             \\ 
\texttt{intercept}    &  3.822    & 0.987 & $<0.001$ \\
\texttt{indcost}      &  2.312    & 0.806 & 0.005 \\ 
\texttt{sizelog}      & $-$0.908  & 0.119 & $<0.001$ \\ 
\\
$\sigma$               &           &       &             \\ 
\texttt{intercept}     &  $-$0.569 & 0.755 & 0.451 \\ 
\texttt{indcost}       &  0.366    & 0.541 & 0.498 \\ 
\texttt{sizelog}       &  0.074    & 0.092 & 0.416 \\ 
\\    
$\lambda$              &     2.035 &  0.203 &         \\ 
$\zeta$                &     1.88  &        &     \\                     
\hline
\end{tabular}
\end{minipage}\hfill%
\begin{minipage}[t]{0.48\linewidth}\centering
\caption{\small Estimates, standard errors, and p-values for the PL-slash regression model with constant dispersion - Firm cost data.}
\label{tab:AP2_est_const}
\begin{tabular}{rrrr}
\hline
& Est. & s.e. & p-value \\
\hline
$\mu$                &        &       &             \\ 
\texttt{intercept}   &  3.867  & 0.983 & $<0.001$ \\
\texttt{indcost}     &  2.133 & 0.569 & $<0.001$ \\
\texttt{sizelog}     &  $-$0.905 & 0.111  & $<0.001$ \\ 
\\
$\log \sigma$        &    0.133 & 0.093 &  0.155   \\ 
$\lambda$            &     1.788 &  0.177 &          \\ 
$\zeta$              &      2.29 &       &             \\                     
\hline
\end{tabular}
\end{minipage}
\end{table}

\begin{figure}[!htb]
\centering
\subfigure[][]{\includegraphics[scale=0.43]{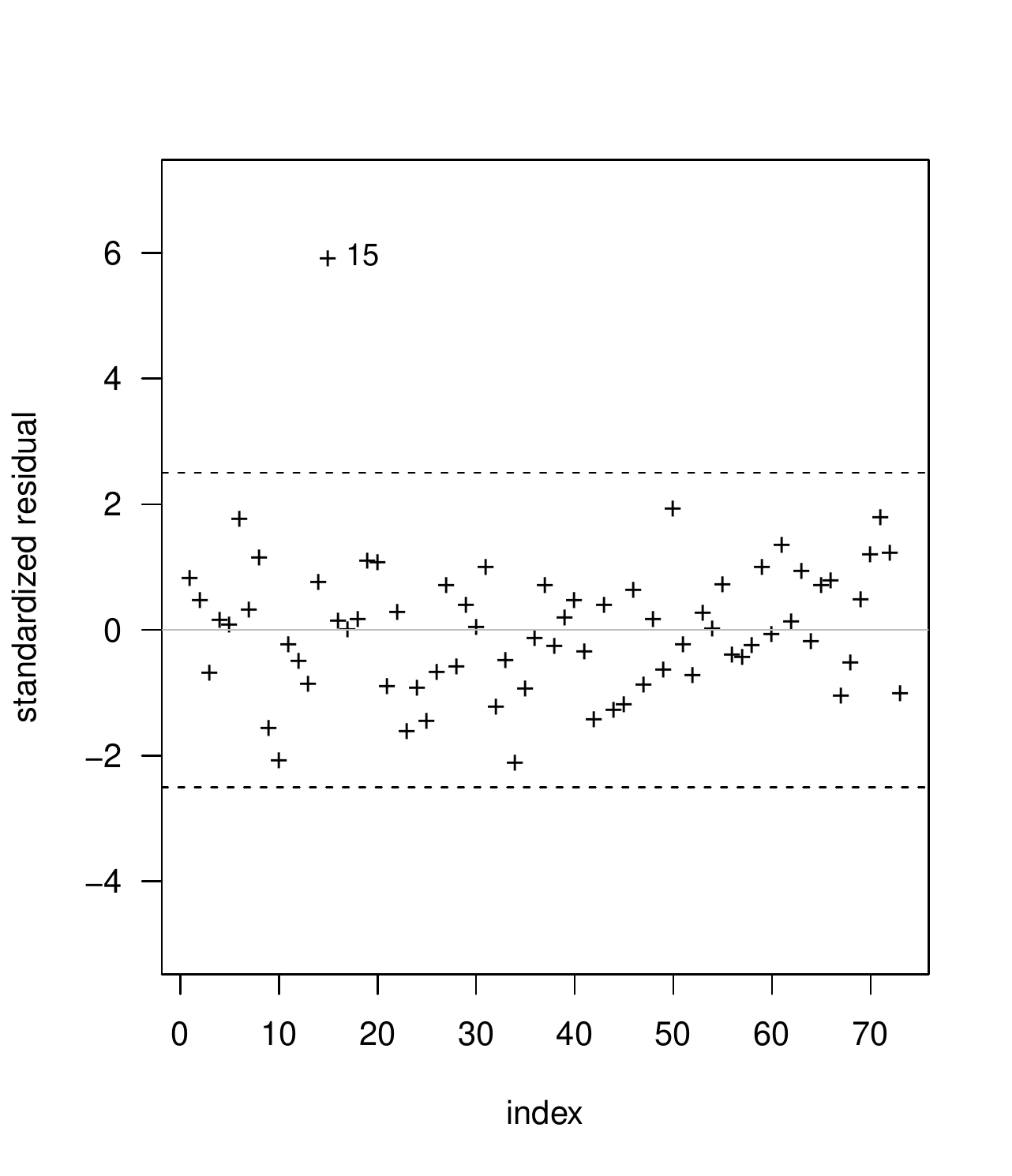}\label{Diag2PL13}}
\subfigure[][]{\includegraphics[scale=0.43]{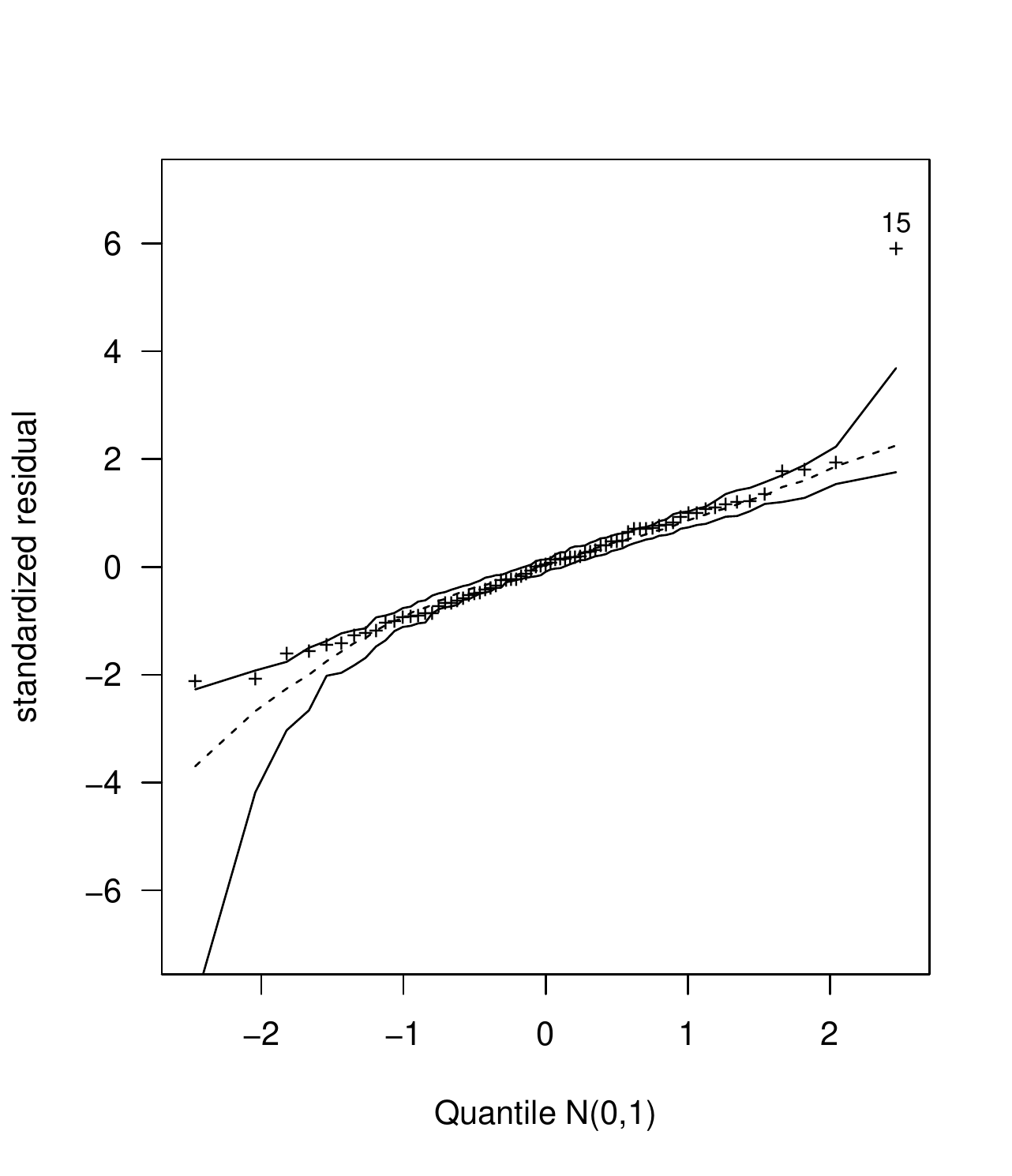}\label{Diag2PL23}}
\subfigure[][]{\includegraphics[scale=0.43]{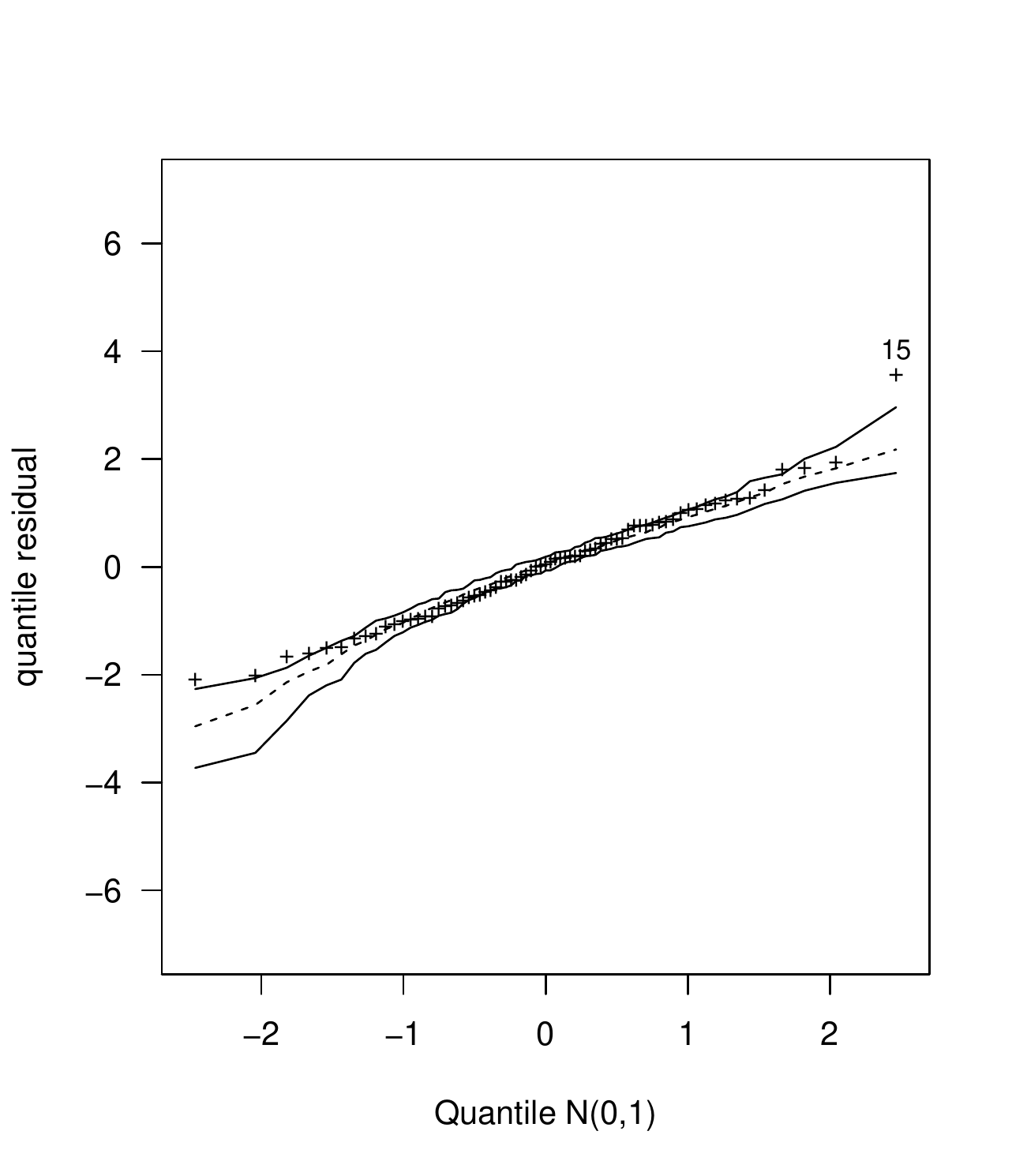}\label{Diag2PL53}}
\subfigure[][]{\includegraphics[scale=0.43]{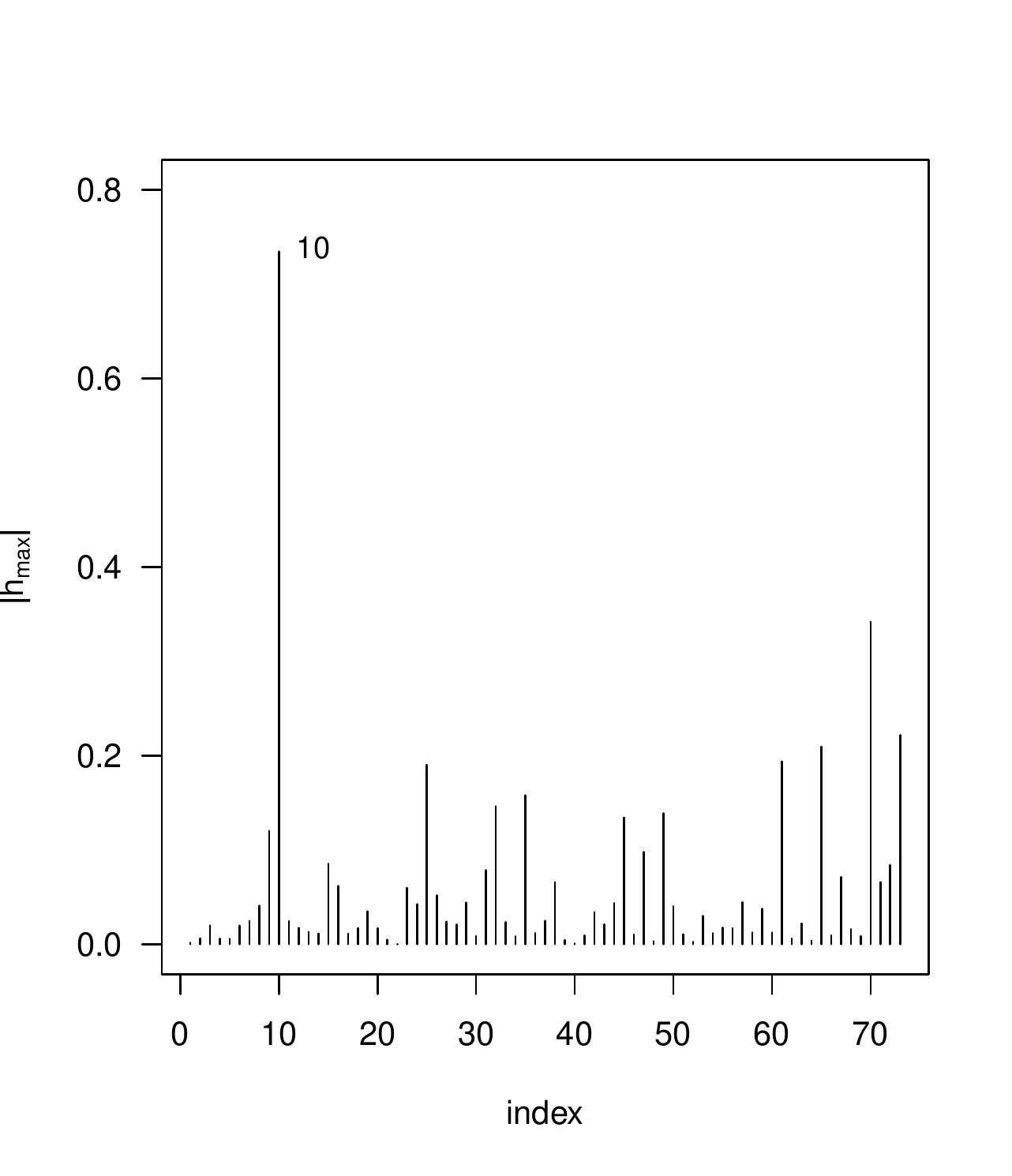}\label{Diag2PL33}}
\subfigure[][]{\includegraphics[scale=0.43]{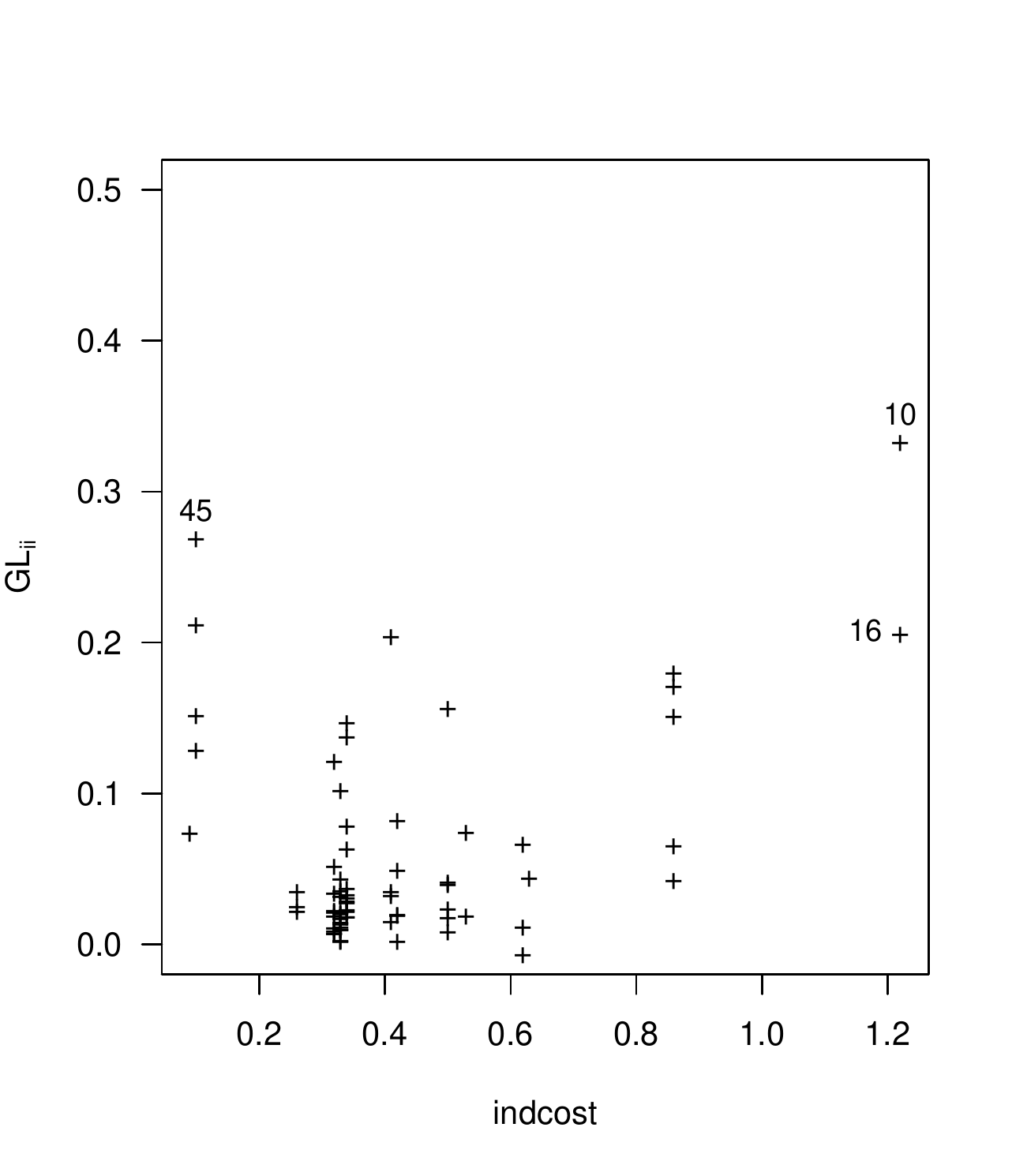}\label{Diag2PL43}}
\subfigure[][]{\includegraphics[scale=0.43]{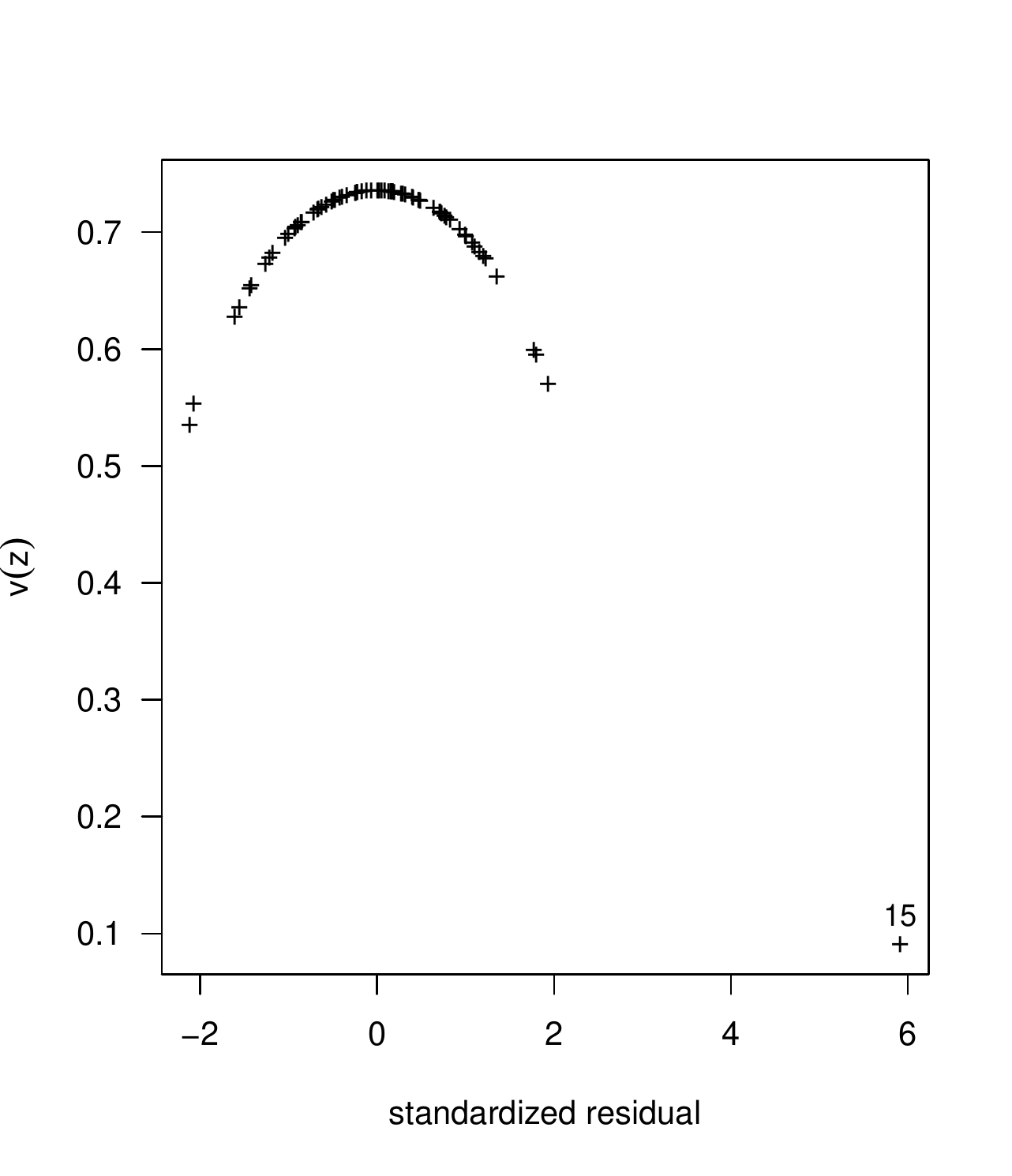}\label{Diagvzfunction3}}
\caption{\small Scatter plot of the standardized residual against index of the observations (a), normal probability plots of the standardized residual (b) and quantile residual (c), index plot of $\vert h_{\text{max}}\vert$ under case-weight perturbation (d), index plot of $GL_{ii}$ (e), and scatter plot of $v(z)$ against the standardized residual (f) for the PL-slash regression model with constant dispersion -  Firm cost data.}\label{AP3PL1_diag}
\end{figure}

Diagnostic plots are presented in Figure \ref{AP3PL1_diag}. Figures \ref{Diag2PL13}-\ref{Diag2PL53} indicate that the postulated model suitably fits the data. Case \#15 is highlighted in almost all the graphics; this observation corresponds to a firm with the highest value of \texttt{firmcost}. On the other hand, this case does not appear in Figure \ref{Diag2PL33} as an influential observation; in fact, the weight for this case in the estimation process is small (see Figure \ref{Diagvzfunction3}). Additionally, case \#10 is highlighted in the influence plot (Figure \ref{Diag2PL33}) and in the generalized leverage plot (Figure \ref{Diag2PL43}), but the exclusion of this case from the data set does not substantially change the fitted model. Overall, we conclude that the PL-slash regression model fits the data well.

\begin{figure}[!htb]
\centering
\subfigure[][]{\includegraphics[scale=0.45]{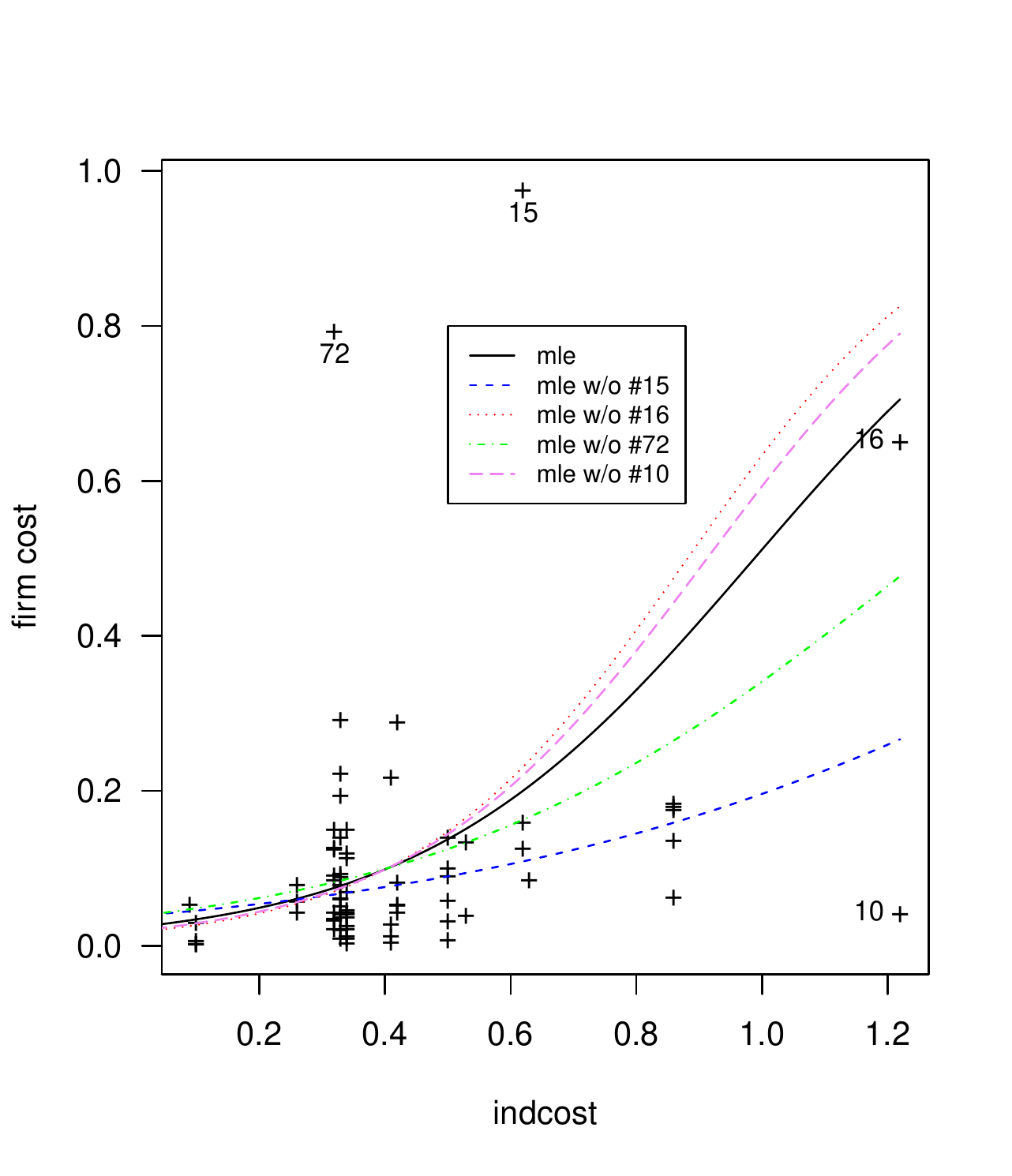}\label{AP3_betafig}}
\subfigure[][]{\includegraphics[scale=0.45]{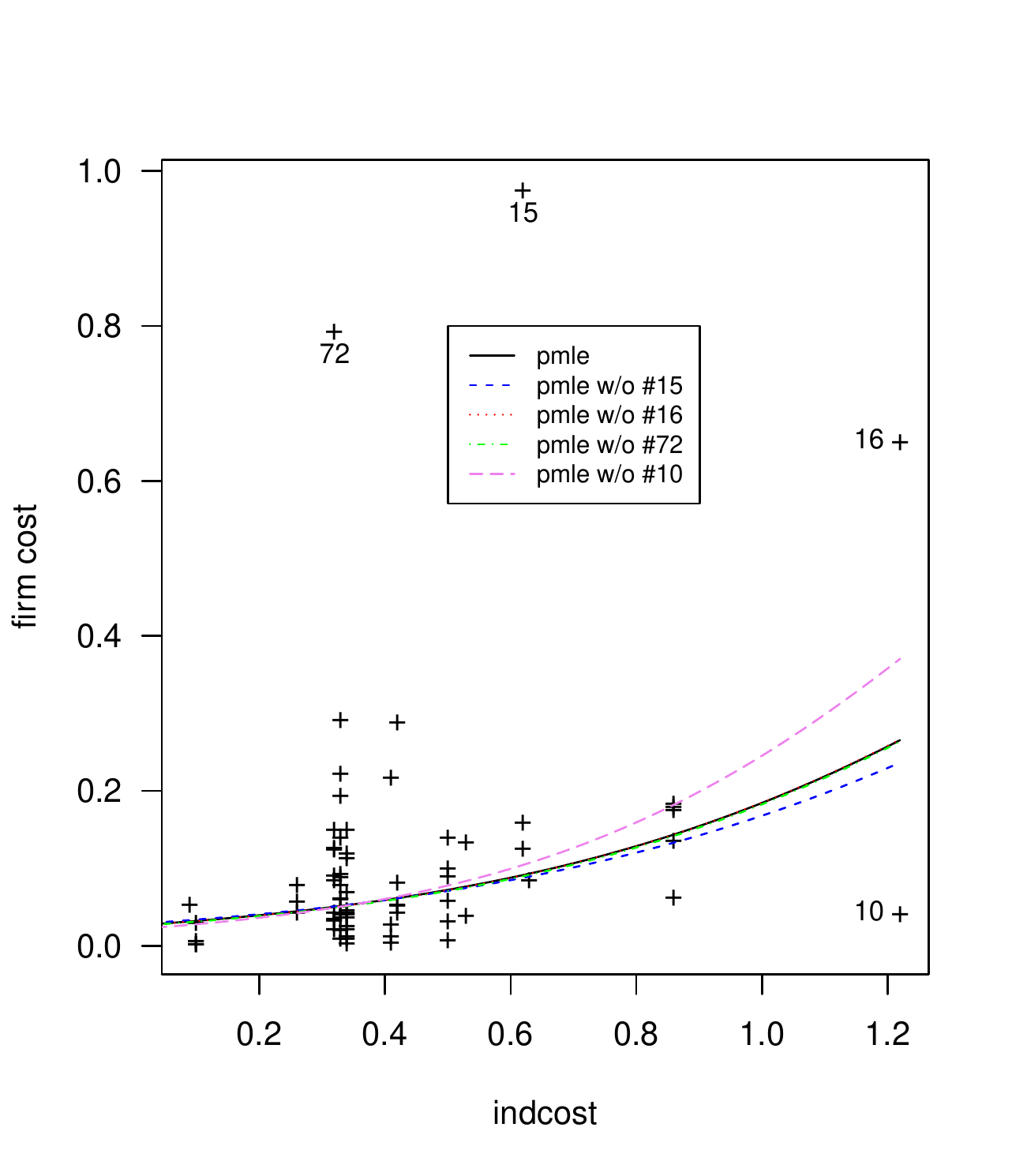}\label{AP3_PLfig}}
\caption{\small Scatter plots of firm cost versus \texttt{indcost} with the fitted lines based on the beta regression model with varying precision (a) and PL-slash with constant dispersion (b) for the full data and the data without outliers -  Firm cost data.}\label{AP2_graphPL3}
\end{figure}

\cite{RibeiroFerrari2020} fittted a beta regression model with varying precision (with covariates \texttt{indcost} and \texttt{sizelog}) to this data set using the maximum likelihood approach. Observations \#15, \#16 and \#72 were detected as atypical observations. Figure \ref{AP3_betafig} displays the scatter plot of \texttt{firmcost} versus \texttt{indcost} with the fitted lines based on the beta regression model for the full data and the data without outliers; the scatter plots were produced by setting the value of \texttt{sizelog} at its sample median. The exclusion of the outliers changes substantially the fitted lines. Case \#15 causes the largest change. In comparison, the fitted lines change much less for the PL-slash regression model (Figure \ref{AP3_PLfig}), suggesting a robust fit. In \cite{RibeiroFerrari2020} the lack of robustness of the maximum likelihood estimation in beta regression is remedied by a modification in the estimation procedure. Here, we replaced the beta distribution by the PL-slash law, a highly flexible distribution. This application suggests that the PL-slash model induced robustness in the likelihood-based estimation process.

\subsection{Body fat of little brown bats}

\begin{figure}[!htb]
\centering
\subfigure[][]{\includegraphics[scale=0.43]{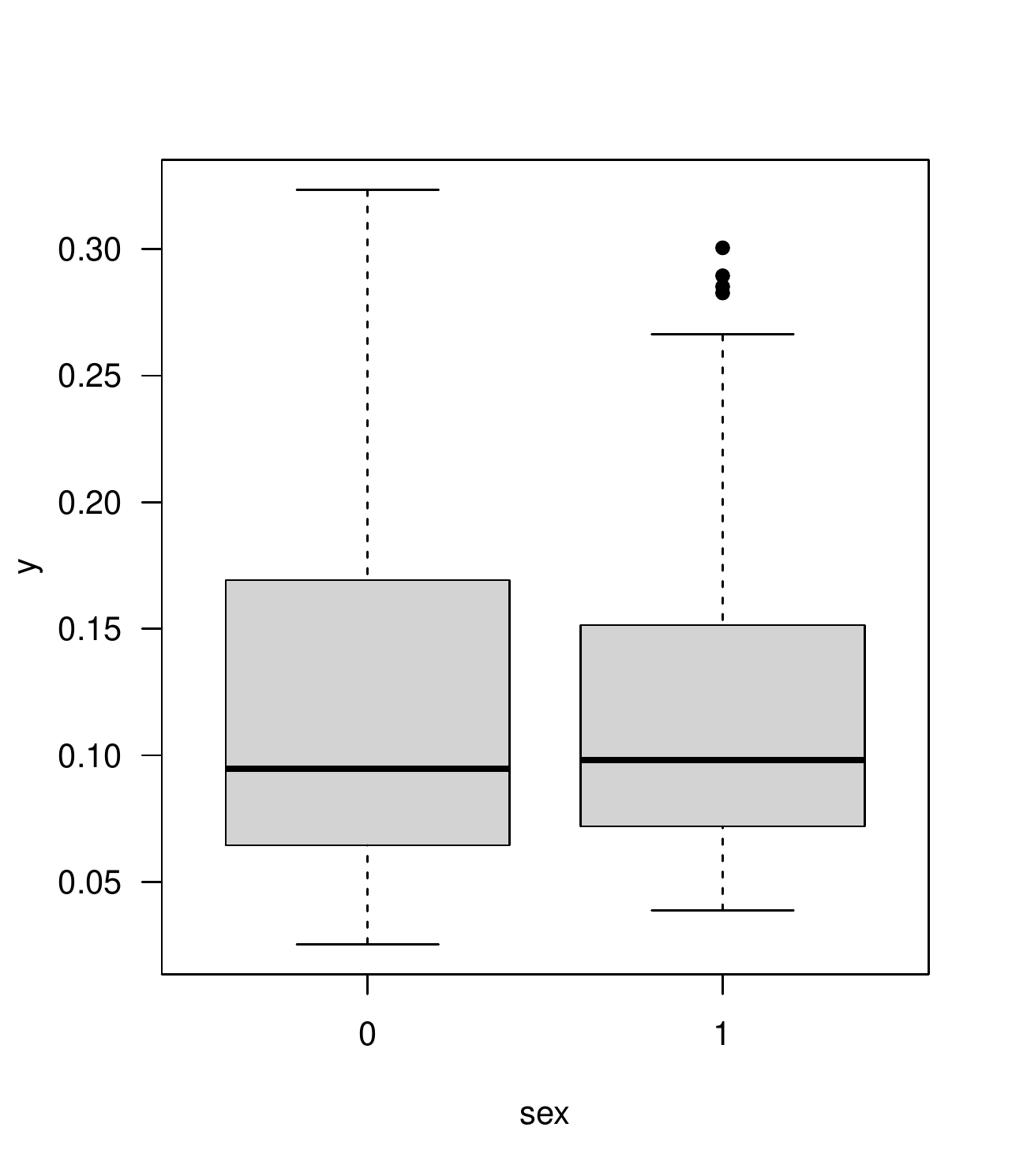}\label{Diag2summary1}}
\subfigure[][]{\includegraphics[scale=0.43]{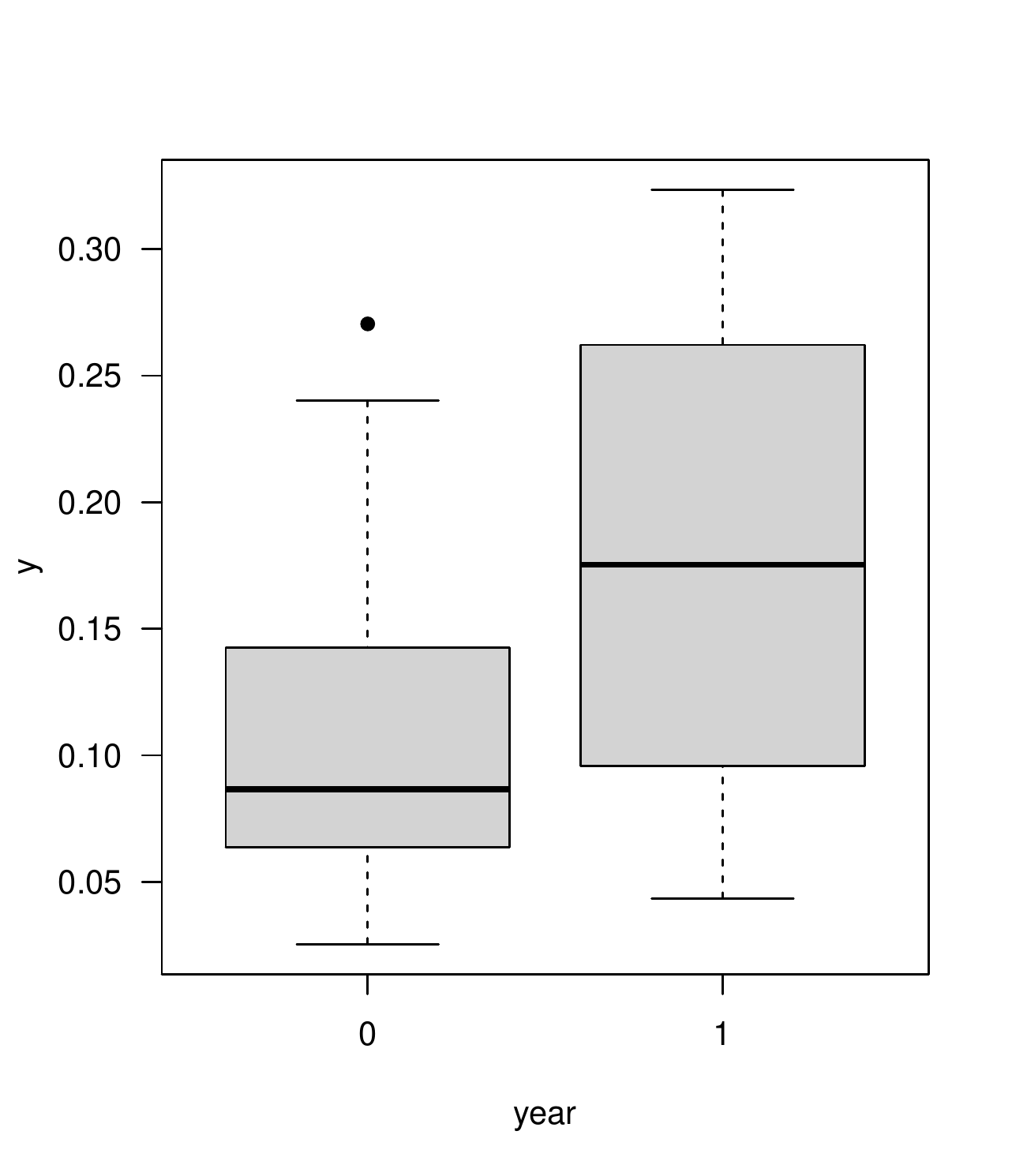}\label{Diag2summary2}}
\subfigure[][]{\includegraphics[scale=0.43]{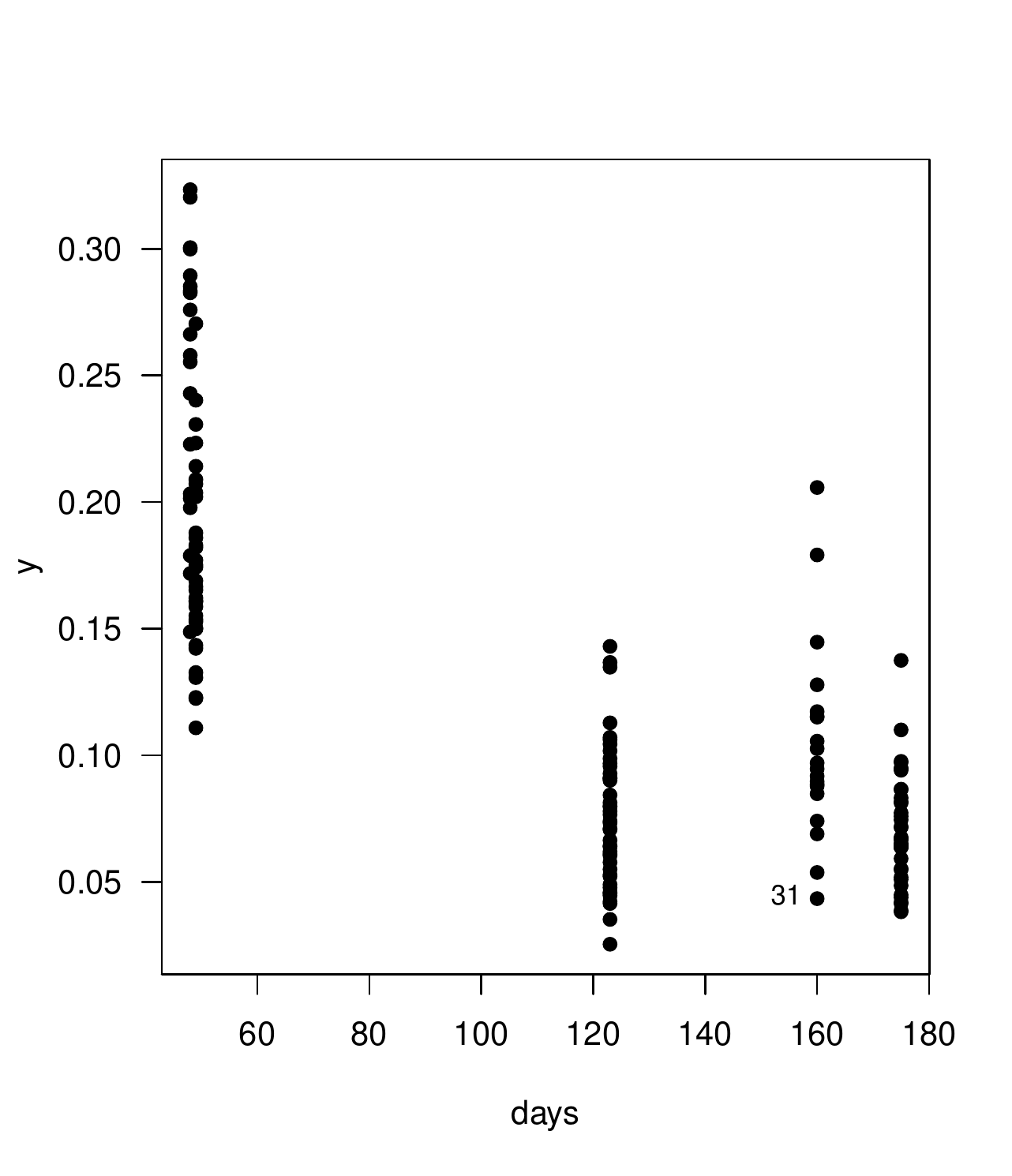}\label{Diag2summary3}}
\caption{\small Boxplot of $y$ against \texttt{sex} (a) and \texttt{year} (b), and scatter plot of $y$ against \texttt{days} (c) -  Body fat of little brown bat data.}\label{AP2PL1_summary}
\end{figure}

We now consider a data set reported in \cite{Cheng2019}. The response variable is the proportion of body fat of little brown bats. The data set used here was collected in Aeolus Cave, located in East Dorset, Vermont, in the USA. The bats were sampled during the winter of 2009 (covering the winter season from October 2008 to April 2009) and 2016 (October 2015 to April 2016). Here, the interest lies in modeling the proportion of body fat of little brown bats ($y$) as a function of the year (\texttt{year}, 1 for 2016 and 0 for 2009), sex of the sampled bat (\texttt{sex}, 1 for male and 0 for female) and the hibernation time (\texttt{days}), defined as the number of days since the fall equinox. Some plots of the data are presented in Figure \ref{AP2PL1_summary}. 

We consider power logit regression models with
\begin{align*}
\begin{split}
\log \left( \dfrac{\mu_i}{1-\mu_i} \right) &= \beta_1 + \beta_2 \times \texttt{year}_i + \beta_3 \times \texttt{sex}_i + \beta_4 \times \texttt{days}_i, \\
\log \sigma_i &= \tau_1 + \tau_2 \times \texttt{year}_i + \tau_3 \times \texttt{sex}_i + \tau_4 \times \texttt{days}_i, 
\end{split} 
\end{align*}
for $i=1, \ldots, 159$, with different distributions, namely PL-N, PL-PE$_{(\zeta)}$, PL-Hyp$_{(\zeta)}$ and PL-SN$_{(\zeta)}$. Table \ref{AP2AC1} presents the estimates and asymptotic p-values for each fitted model, as well as the AIC and the $\Upsilon$ measure. The pseudo $R^2$ of all the estimated regression models is approximately 0.68. Note that \texttt{year} and \texttt{days} are significant in the median component and \texttt{sex} is significant in the dispersion component for all the considered models. The AIC and $\Upsilon$ value are similar for all the fitted models. In addition, the estimate of $\lambda$ is close to zero, indicating that the corresponding limiting models (log-log regression models) should be considered. We then fit different models in the log-log class; see Table \ref{AP2AC1_loglog}. As expected, all estimates, p-values and goodness-of-fit measures are close to those of the associated power logit regression model. Moreover, the AIC and the $\Upsilon$ measure for the log-log models are similar. We will continue the analysis only with the log-log normal regression because it is the simplest model.

\begin{table}[ht]
\centering
\tiny
\caption{\small Estimates and p-values for power logit regression models - Body fat of little brown bat data.} \label{AP2AC1}
\begin{tabular}{rrrrrrrrrrrrrrrr}
\hline
& \multicolumn{3}{c}{ $\text{PL-N}$} & & \multicolumn{3}{c}{ $\text{PL-PE}_{(1.6)}$} & & \multicolumn{3}{c}{ $\text{PL-Hyp}_{(5.5)}$} & & \multicolumn{3}{c}{ $\text{PL-SN}_{(0.52)}$} \\
\hline
& Est. & s.e. & p-value & & Est. & s.e. &  p-value & & Est. &  s.e. & p-value & & Est. & s.e. &  p-value \\ 
\hline
$\mu$              &           &         &         & &          &         &         & &            &         &           & &          &         &       \\ 
\texttt{intercept} & $-1.153 $ & $0.067$ & $0.000$ & & $-1.154$ & $0.063$ & $0.000$ & &   $-1.152$ & $0.065$ & $0.000$   & & $-1.153$ & $0.067$ & $0.000$ \\ 
\texttt{days}      & $-0.009 $ & $0.001$ & $0.000$ & & $-0.009$ & $0.001$ & $0.000$ & &   $-0.009$ & $0.001$ & $0.000$   & & $-0.009$ & $0.001$ & $0.000$ \\ 
\texttt{sex}       & $-0.032 $ & $0.053$ & $0.542$ & & $-0.049$ & $0.054$ & $0.368$ & &   $-0.039$ & $0.053$ & $0.460$   & & $-0.028$ & $0.053$ & $0.595$ \\ 
\texttt{year}      & $0.504  $ & $0.058$ & $0.000$ & & $0.517 $ & $0.061$ & $0.000$ & &   $0.511 $ & $0.059$ & $0.000$   & & $0.499 $ & $0.058$ & $0.000$ \\ 
$\sigma$              &        &       &       & &        &       &       & &        &         &       & &        &         &       \\ 
\texttt{intercept} & $-1.966 $ & $0.180$ & $0.000$ & & $-1.931$ & $0.199$ & $0.000$ & &   $-1.206$ & $0.183$ & $0.000$   & & $-0.600$ & $0.158$ & $0.000$ \\ 
\texttt{days}      & $0.001  $ & $0.001$ & $0.546$ & & $0.001 $ & $0.001$ & $0.654$ & &   $0.000 $ & $0.001$ & $0.662$   & & $0.001 $ & $0.001$ & $0.459$ \\ 
\texttt{sex}       & $-0.287 $ & $0.115$ & $0.013$ & & $-0.296$ & $0.129$ & $0.022$ & &   $-0.298$ & $0.126$ & $0.018$   & & $-0.283$ & $0.108$ & $0.009$ \\ 
\texttt{year}      & $0.109  $ & $0.132$ & $0.408$ & & $0.119 $ & $0.148$ & $0.419$ & &   $0.118 $ & $0.144$ & $0.413$   & & $0.102 $ & $0.124$ & $0.411$ \\ 
$\lambda$          & $0.001  $ & $0.065$ &         & & $0.008 $ & $0.072$ &       & &     $0.000 $ & $0.056$ &           & & $0.001 $ & $0.046$ &       \\ 
\\ 
$\Upsilon$         &           & $0.053$ &   & & & $0.052$ &    & & & $0.050$ &     & & & $0.057$ &      \\ 
AIC             & & $-628.8$ & & & & $-627.4$ &  & & & $-627.7$ &  & & &$-628.9 $&   \\ 
\hline
\end{tabular}
\end{table}

\begin{table}[ht]
\centering
\tiny
\caption{\small Estimates and p-values for log-log regression models - Body fat of little brown bat data.} \label{AP2AC1_loglog}
\begin{tabular}{rrrrrrrrrrrrrrrr}
\hline
& \multicolumn{3}{c}{ $\text{log-log-N}$} & & \multicolumn{3}{c}{ $\text{log-log-PE}_{(1.6)}$} & & \multicolumn{3}{c}{ $\text{log-log-Hyp}_{(5.5)}$} & & \multicolumn{3}{c}{ $\text{log-log-SN}_{(0.52)}$} \\
\hline
& Est.      & s.e.    & p-value & & Est.       & s.e.    & p-value & & Est.      &  s.e.   & p-value & & Est.     & s.e.    & p-value \\ \hline
$\mu$              &           &         &         & &            &         &         & &           &         &         & &          &         &         \\ 
\texttt{intercept} & $ -1.153$ & $0.067$ & $0.000$ & & $-1.152 $  & $0.058$ & $0.000$ & &    $-1.152$  & $0.065$ & $0.000$    & & $-1.153$ & $0.067$ & $0.000$ \\ 
\texttt{days}      & $ -0.009$ & $0.001$ & $0.000$ & & $-0.009 $  & $0.001$ & $0.000$ & &    $-0.009$  & $0.001$ & $0.000$    & & $-0.009$ & $0.001$ & $0.000$ \\ 
\texttt{sex}       & $ -0.032$ & $0.053$ & $0.542$ & & $-0.049 $  & $0.054$ & $0.356$ & &    $-0.039$  & $0.054$ & $0.460$    & & $-0.028$ & $0.053$ & $0.594$ \\ 
\texttt{year}      & $ 0.504 $ & $0.058$ & $0.000$ & & $0.517  $  & $0.060$ & $0.000$ & &    $0.511 $  & $0.059$ & $0.000$    & & $0.499 $ & $0.058$ & $0.000$ \\ 
$\sigma$           &           &         &         & &            &         &         & &              &         &            & &          &         &         \\ 
\texttt{intercept} & $ -1.967$ & $0.166$ & $0.000$ & & $-1.940 $  & $0.185$ & $0.000$ & &    $-1.206$  & $0.181$ & $0.000$    & & $-0.601$ & $0.157$ & $0.000$ \\ 
\texttt{days}      & $ 0.001 $ & $0.001$ & $0.541$ & & $0.001  $  & $0.001$ & $0.658$ & &    $0.000 $  & $0.001$ & $0.662$    & & $0.001 $ & $0.001$ & $0.460$ \\ 
\texttt{sex}       & $ -0.287$ & $0.115$ & $0.012$ & & $-0.295 $  & $0.128$ & $0.021$ & &    $-0.298$  & $0.126$ & $0.018$    & & $-0.283$ & $0.108$ & $0.009$ \\ 
\texttt{year}      & $ 0.109 $ & $0.131$ & $0.408$ & & $0.120  $  & $0.148$ & $0.418$ & &    $0.118 $  & $0.144$ & $0.413$    & & $0.102 $ & $0.124$ & $0.410$ \\ 
\\ 
$\Upsilon_\zeta$   &           & $0.054$ &         & &            & $0.052$ &         & &           & $0.050$ &         & &          & $0.057$ &         \\ 
AIC                &           &$-630.8$ &         & &            & $-629.4$&         & &           & $-629.7$&         & &          & $-630.9$&         \\ 
\hline
\end{tabular}
\end{table}

Table \ref{AP2AC1_loglog} shows that \texttt{sex} and both \texttt{days} and \texttt{year} are not marginally significant, respectively for the median and dispersion submodels. The likelihood ratio statistic for testing the reduced model against the full model is $1.4$ (p-value $=$ $0.29$). That is, the reduced model is not rejected at the usual significance levels, and hence, those covariates can be excluded from the model. Table \ref{AP2PL1} presents the results. The fitted model indicates a negative relationship between the median proportion of body fat and the hibernation time (\texttt{days}) irrespectively of the year. Also, for constant hibernation time, the median proportion of body fat is higher in 2016 than in 2009. Additionally, the dispersion of the proportion of body fat in male bats is estimated to be smaller than in female bats since the coefficient of \texttt{sex} in the dispersion submodel is negative.

\begin{table}[ht]
\centering
\scriptsize
\caption{\small Estimates, standard errors, and p-values for the log-log-N regression model - Body fat of little brown bat data.} \label{AP2PL1}
\begin{tabular}{rccccccc}
  \hline
& Est. & s.e. & p-value \\ 
\hline 
$\mu$                 &           &       &   \\
\texttt{intercept}    & $-$1.170  & 0.063 &  $<0.001$ \\ 
\texttt{days}         & $-$0.009  & 0.001 &  $<0.001$ \\ 
\texttt{year}         &  0.496    & 0.056 & $<0.001$\\ 
$\sigma$              &           &       &        \\
\texttt{intercept}    &  $-$1.849 & 0.073 & 0.000 \\ 
\texttt{sex}          &  $-$0.288 & 0.115 & 0.012 \\ 
\hline
\end{tabular}
\end{table}

\begin{figure}[!htb]
\centering
\subfigure[][]{\includegraphics[scale=0.44]{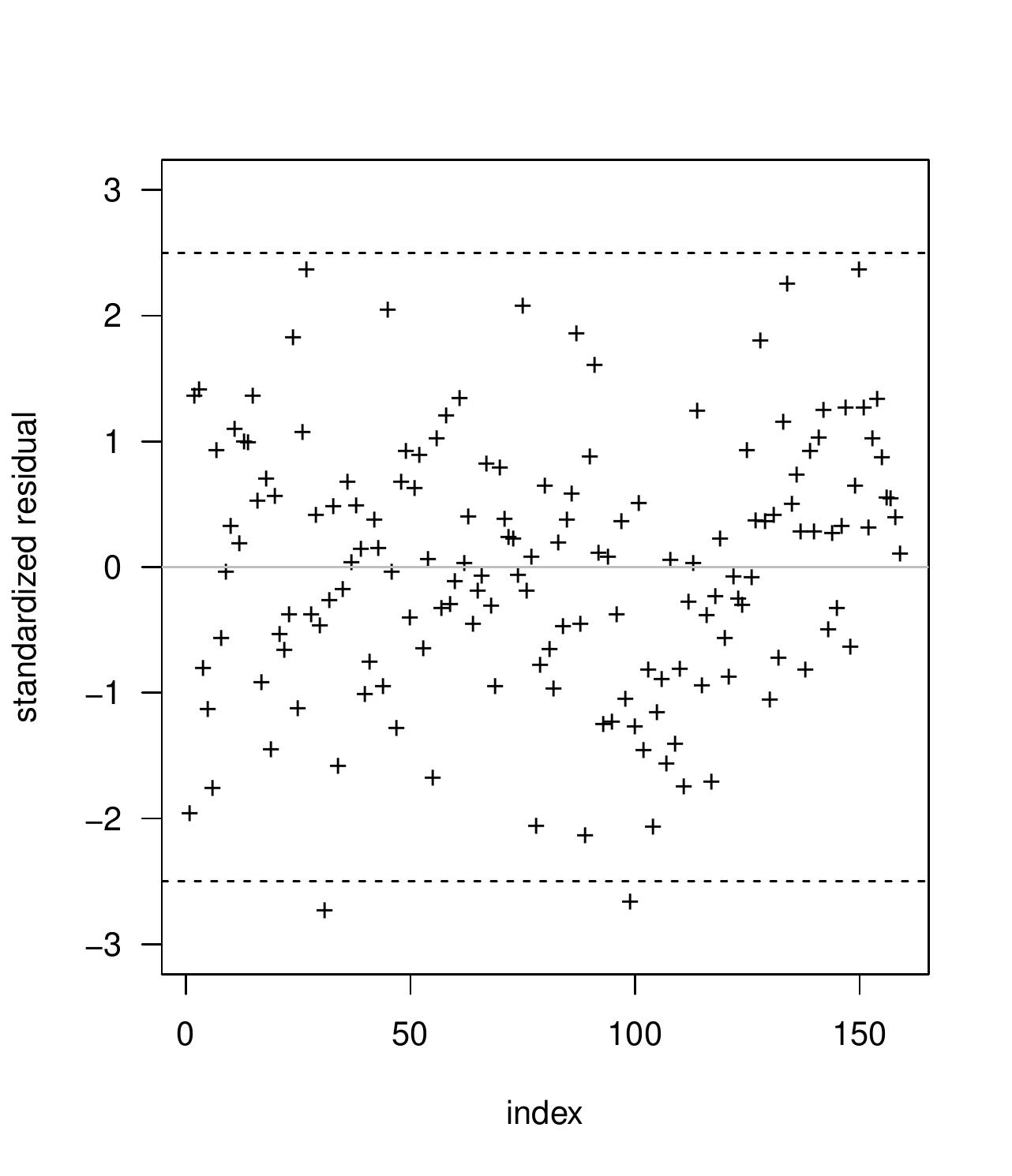}\label{Diag2loglog1}}
\subfigure[][]{\includegraphics[scale=0.44]{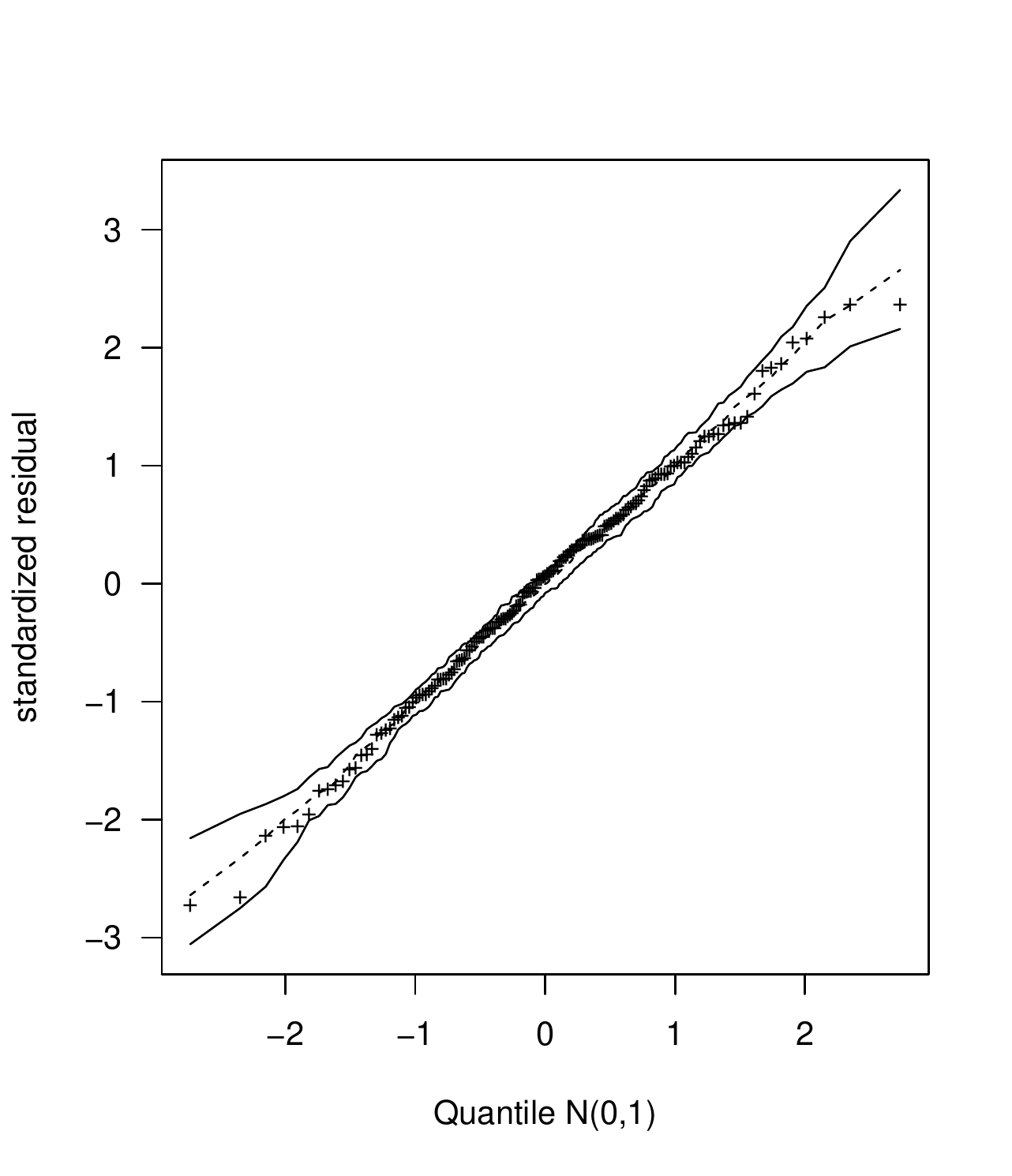}\label{Diag2loglog2}}
\subfigure[][]{\includegraphics[scale=0.44]{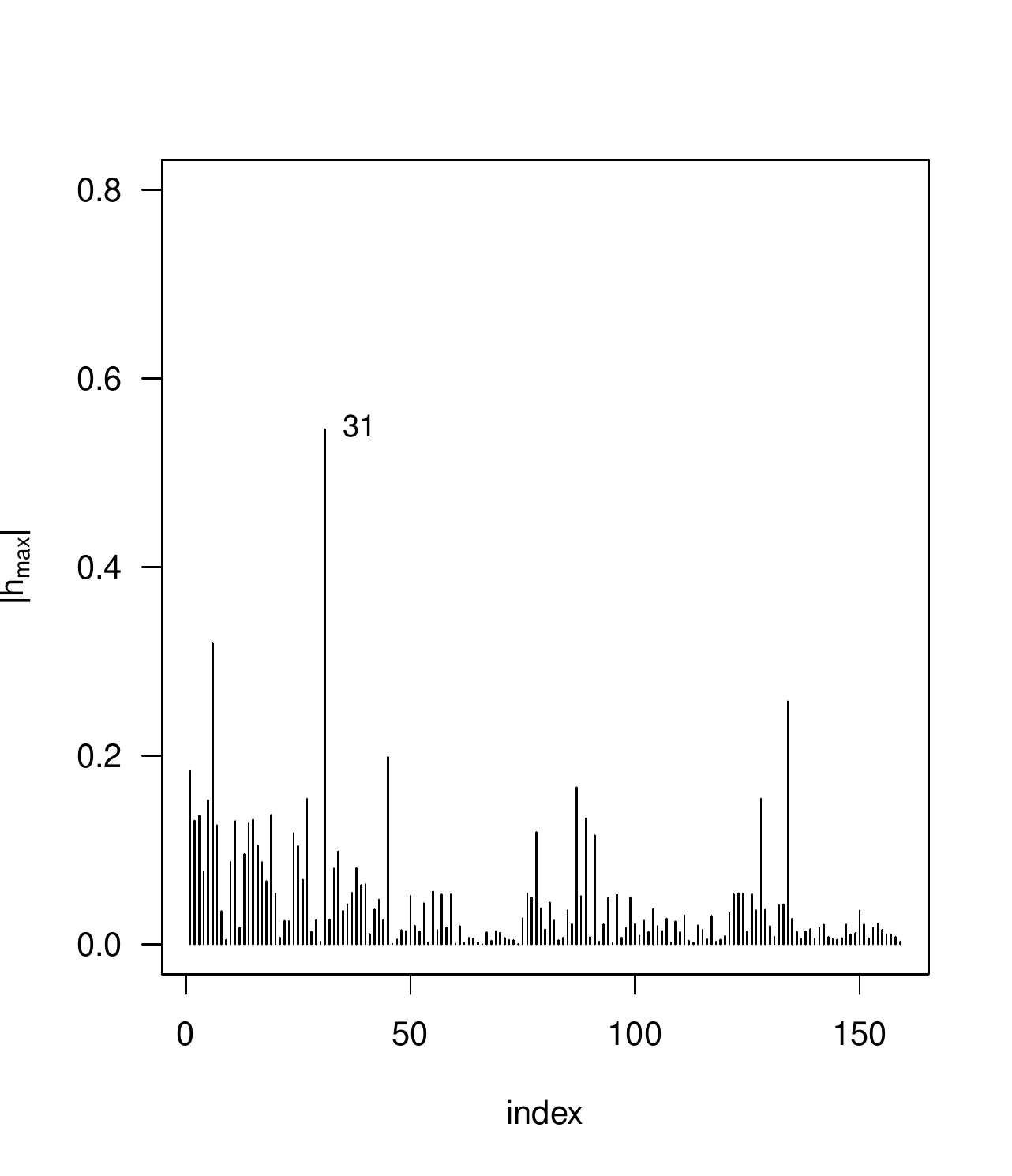}\label{Diag2loglog3}}
\subfigure[][]{\includegraphics[scale=0.43]{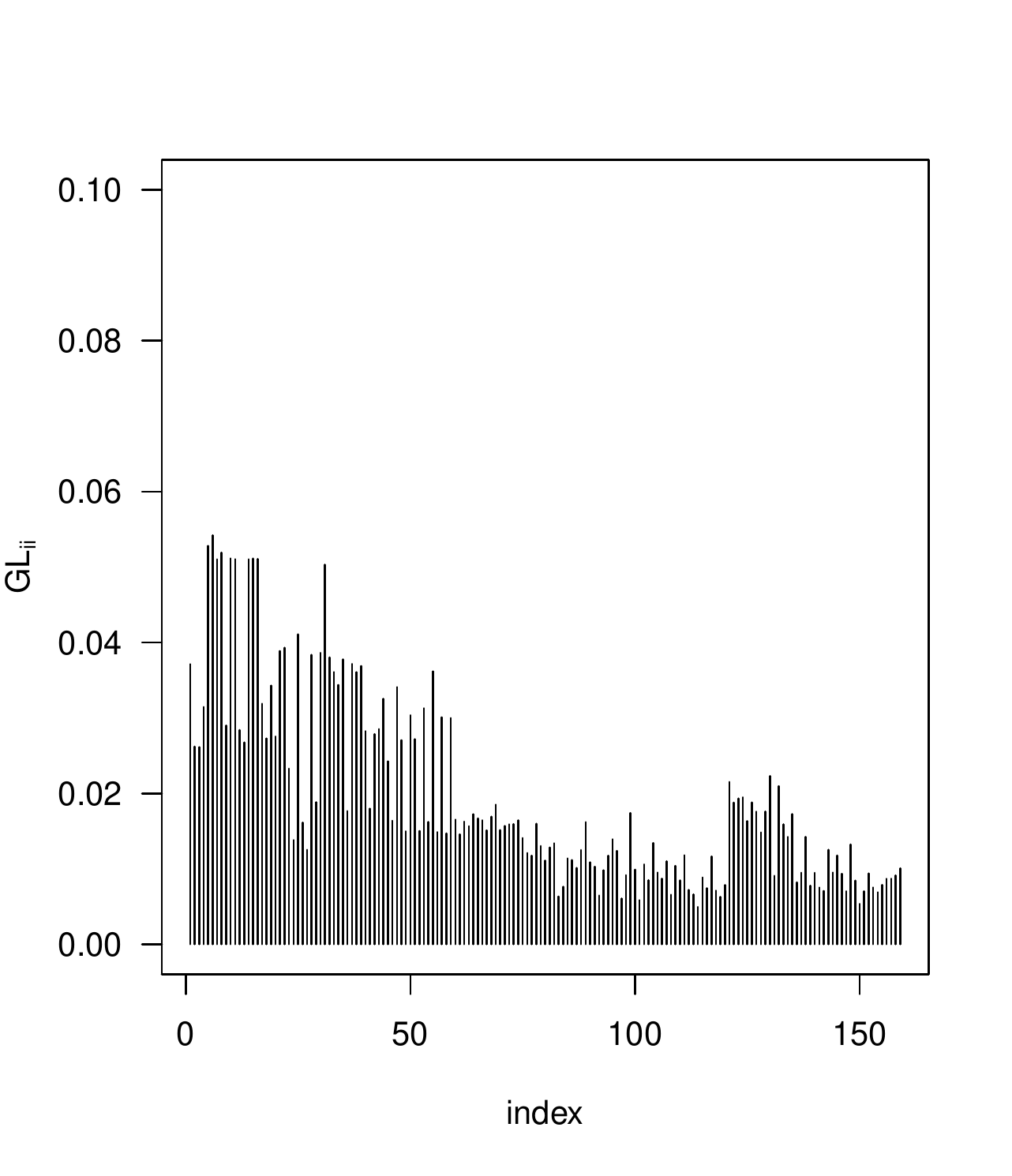}\label{Diag2loglog4}}
\caption{\small Scatter plot of the standardized residual against index of the observations (a), normal probability plots of the standardized residual (b), index plot of $\vert h_{\text{max}}\vert$ under case-weight perturbation (c) and index plot of $GL_{ii}$ for the log-log-N regression model -  Body fat of little brown bat data.}\label{loglogAP2PL1_diag}
\end{figure}

The diagnostic plots in Figure \ref{loglogAP2PL1_diag} suggest that the fit is adequate. Figure \ref{Diag2loglog3} highlights \#31 as a possible influential observation. This case corresponds to a bat with the smallest percentage of body fat in the year 2016 (approximately 4\%) after a long period of hibernation (160 days). The predicted value for this observation is approximately 10\%, much higher than the observed value. The exclusion of this observation causes only small changes in the estimated parameters. Thus, we may conclude that the postulated log-log normal regression model provides an adequate fit to the data.

%

%
%
%

\section{Concluding Remarks}

The power logit distributions generalize the GJS distributions at the cost of introducing an additional parameter. The application in Section \ref{agricul_app} reveals that the GJS distributions are not flexible enough to produce an adequate fit to the data. The use of the corresponding power logit distributions improves the fit considerably. An interesting feature of the power logit distributions is that their three parameters are interpretable as the median, dispersion and skewness parameters. Additionally, they tend to be naturally more flexible than two-parameter distributions such as the beta, simplex, and Kumaraswamy distributions. They may also depend on an extra parameter that index the underlying symmetric distribution. The extra parameter, if any, may be seen as a tuning constant that is chosen to promote even more flexibility to suitably fit different configurations of data, particularly with outlying observations.

The paper introduces regression models in which the response variable is assumed to follow a distribution in the power logit class. The real data applications in Section \ref{app} suggest that the proposed models are useful for modeling continuous bounded data. We emphasize that the paper offers a broad set of tools for likelihood inference and diagnostics. Three types of residuals are defined. We draw attention to the standardized residual, which can more clearly highlight leverage observations than the deviance and quantile residuals. Applications in simulated data in Section \ref{diagnostictools} and the Supplementary Material show that the proposed residuals can identify different deviations from the postulated model and detect the presence of atypical observations. 


The new \texttt{R} package \texttt{PLreg} enables practitioners to employ the power logit regression models in their own data analyses. The package is introduced in this paper and provides a comprehensive range of facilities for fitting both PL and GJS regression models and performing diagnostic analysis. Moreover, \texttt{PLreg} provides procedures for choosing the extra parameter, when needed.

%

We conclude this paper by outlining some interesting directions for further research. The power logit regression models may be extended to accomodate situations in which the data include observations at the boundaries. The authors have been working on zero-and/or-one inflated
power logit regression models and the findings will be reported elsewhere. Different regression structures, such as nonlinear, spatial, time series, mixed components, and random forest regressions, also deserve investigation and computing implementation.

\section*{Acknowledgments}
This study was financed in part by the Coordena\c{c}\~ao de Aperfei\c{c}oamento de Pessoal de N\'ivel Superior - Brazil (CAPES) - Finance Code 001 and by the Conselho Nacional de Desenvolvimento Cient\'ifico e Tecnol\'ogico - Brazil (CNPq). The second author gratefully acknowledges funding provided by CNPq (Grant No. 305963-2018-0). This paper is based on part of the first author's Ph.D. thesis which was supervised by the second author at the University of S\~ao Paulo, Brazil.


\end{document}